\documentclass[useAMS,usenatbib]{mn2e}

\usepackage{color}
\usepackage{txfonts}
\usepackage{rotating}
\usepackage{supertabular}
\def\Htwo{\mbox{H$_2$}}
\def\twelveCO{\mbox{$^{12}$CO}}
\def\thirteenCO{\mbox{$^{13}$CO}}
\def\CeighteenO{\mbox{C$^{18}$O}}
\def\HthirteenCOplus{\mbox{H$^{13}$CO$^+$}}
\def\HCOplus{\mbox{HCO$^+$}}
\def\Msun{M$_{\odot}$}
\def\Lsun{L$_{\odot}$}
\def\cmcube{cm$^{-3}$}
\def\arcsecs{$^{\prime\prime}$}
\def\arcmin{$^{\prime}$}
\def\antemp{$T_{\rm A}^*$}
\def\kms{km~s$^{-1}$}

\long\def\changed#1{\textcolor{black}{#1}}

\def\spectra#1#2#3{
\centerline{\includegraphics[width=4.5cm,angle=-90]{#1_compspec-eps-converted-to.pdf}\quad
	    \includegraphics[width=4.5cm,angle=-90]{#2_compspec-eps-converted-to.pdf}\quad
	    \includegraphics[width=4.5cm,angle=-90]{#3_compspec-eps-converted-to.pdf}
\smallskip}}

\title[Dense gas survey in Perseus]{The JCMT dense gas survey of the Perseus Molecular Cloud}
\author[S. L. Walker-Smith et al.]{S. L. Walker-Smith$^{1}$\thanks{E-mail:
sw547@mrao.cam.ac.uk}, J. S. Richer$^{1,2}$, J. V. Buckle$^{1,2}$, J. Hatchell$^{3}$ and E. Drabek-Maunder$^{3,4}$\\
$^{1}$Astrophysics Group, Cavendish Laboratory, J J Thomson Avenue, Cambridge, CB3 0HE\\
$^{2}$Kavli Institute for Cosmology, Institute of Astronomy, University of Cambridge, Madingley Road, Cambridge, CB3 0HA\\
$^{3}$University of Exeter, School of Physics, Stocker Rd, Exeter, EX4 4QL\\
$^{4}$Imperial College London, Blackett Laboratory, Prince Consort Rd, London, SW7 2AZ}

\begin{document}

\date{}

\pagerange{\pageref{firstpage}--\pageref{lastpage}} \pubyear{2013}

\maketitle

\label{firstpage}

\begin{abstract}
We present the results of a large-scale survey of the very dense ($n$ \textgreater $10^6$~\cmcube) gas in the Perseus molecular cloud using \HCOplus\ and HCN ($J = 4\to3$) transitions. We have used this emission to trace the structure and kinematics of gas found in pre- and protostellar cores, as well as in outflows. We compare the \HCOplus/HCN data, highlighting regions where there is a marked discrepancy in the spectra of the two emission lines. We use the \HCOplus\ to identify positively protostellar outflows and their driving sources, and present a statistical analysis of the outflow properties that we derive from this tracer. We find that the relations we calculate between the \HCOplus\ outflow driving force and the $M_{\rm env}$ and $L_{\rm bol}$ of the driving source are comparable to those obtained from similar outflow analyses using \twelveCO, indicating that the two molecules give reliable estimates of outflow properties. We also compare the \HCOplus\ and the HCN in the outflows, and find that the HCN traces only the most energetic outflows, the majority of which are driven by young Class 0 sources. We analyse the abundances of HCN and \HCOplus\ in the particular case of the IRAS~2A outflows, and find that the HCN is much more enhanced than the \HCOplus\ in the outflow lobes. We suggest that this is indicative of shock-enhancement of HCN along the length of the outflow; this process is not so evident for \HCOplus, which is largely confined to the outflow base.
\end{abstract}

\section{Introduction}

\subsection{Outflows and chemistry}
High-velocity molecular outflows were first discovered in 1976 \citep{1976ApJ...209L.137Z}, and are thought to be one of the earliest observable signatures of star formation --- nearly all core-collapse candidate protostars are found to have outflows. Indeed outflows are thought to be integral to the star formation process, as they remove excess angular momentum \citep{2007prpl.conf..245A}. 

However despite the ubiquity of outflows, there is still much that we do not understand about them. One example is the outflow driving mechanism: they are thought to be powered by the gravitational energy of the infalling material in a contracting core \citep{1980ApJ...239L..17S}. However, as their source is usually deeply embedded in a dense infalling envelope, it is difficult to observe directly the outflow region and investigate the outflow driving mechanism.

Molecular outflows are thought to develop from the entrainment of low-velocity flows by well-collimated (optical) jets --- the primary jet injects its momentum into the surrounding gas, resulting in a molecular outflow that traces the interaction between the jet and the surrounding environment \citep{1987ARA&A..25...23S}. Wide-angled winds sweeping up a shell of gas have also been invoked to explain outflows such as HH~211 \citep{2006ApJ...636L.137P}. In these outflow models, emission from molecular outflows originates from ambient gas that has been accelerated by a supersonic wind, and has been shock-processed. Observations support this as outflows tend to be associated with other objects such as H{\sc ii} regions, HH objects, H$_2$ jets and H$_2$O masers \citep{2004A&A...426..503W}. Outflows and shock chemistry are therefore inextricably linked.

Some outflows show a greater degree of molecular richness than others --- exciting a wide range of molecules and enhancing the abundance of some to many orders of magnitude over their standard molecular cloud abundances. These `chemically-active' outflows are not necessarily distinguishable using CO emission (although they do tend to be young Class 0 sources), and their activity is also highly transient \citep{2011IAUS..280...88T}.

Although SiO is the canonical outflow tracer, molecules such as \HCOplus\ and HCN are increasingly used to study outflows: the higher transitions (e.g. $J = 4\to3$) trace both the dense gas associated with the protostellar objects, as well as any outflowing/infalling gas \citep{1997ApJ...484..256G}. This allows a comparison of the physical and chemical properties in the source envelopes, with those in the outflow lobes further away from the central sources. In addition, both these molecules have similar critical densities and excitation energies (see Table \ref{mol_trans}), and are found to have highly-correlated spatial distributions. However, recent studies have shown that the abundances of the two molecules are enhanced to different degrees in particular outflows --- \citet{2010A&A...522A..91T} have shown that the HCN abundance is enhanced by $\sim 10-100$ times more than the \HCOplus\ abundance in the outflow linewings. 

One theory by \citet{2000ApJ...536..857H} is that the differences between HCN and \HCOplus\ in the linewings can be attributed to the effect of magnetic field lines that are not aligned with the turbulent flow direction. The ions are trapped on the magnetic field lines while the neutrals follow the turbulent flow, resulting in the ionic species having narrower lines and suppressed linewings compared to the neutral species \citep{2000ApJ...537..245H}. This theory assumes that the two molecules coexist and sample the same parts of the molecular cloud, exposing them to the same dynamical processes.

This is not necessarily the case, as \HCOplus\ and HCN have similar critical densities and excitation conditions but follow different chemical networks \citep{1990MNRAS.244..668P}, making it likely that differences in abundances between the two molecules are due to chemical effects. This difference in enhancements between the two molecules is therefore more commonly attributed to shock-chemistry, as HCN is known to be enhanced in shocks \citep{2004A&A...413..993J}.

\citet{2011IAUS..280...88T} have suggested that there exists an incomplete knowledge of outflows and their chemistry due to a too-focused approach on a few selected objects, and a lack of a statistically-significant number of outflows explored using molecules other than CO. This dense gas survey of \HCOplus\ and HCN $J = 4\to3$ transitions, over multiple sub-regions in the Perseus molecular cloud, is thus very timely. These sub-regions have previously been observed in \twelveCO/\thirteenCO/\CeighteenO\ by \citet{2010MNRAS.408.1516C}, tracing the large-scale structure and kinematics of Perseus, as well as the protostellar activity occurring there. We have performed follow-up observations using \HCOplus\ and HCN ($J = 4\to3$ transitions) to characterise the denser ($n$ \textgreater $10^6$~\cmcube) gas present, and to investigate a large number of previously-mapped outflows using our molecules for comparison.

\subsection{The Perseus molecular cloud}
The Perseus molecular cloud is a well-studied region of low to intermediate-mass star formation. Distance estimates for Perseus range from 230 to 350~pc \citep{2011A&A...535A..44F}, \changed{and \citet {2011PASJ...63....1H} have recently calculated the distance to Perseus to be 235~pc by astrometry}; but for the purposes of this paper we assume the value of 250~pc used by \citet{2010MNRAS.408.1516C}, with whom we will be comparing our results.

We investigate four sub-regions in the Perseus molecular cloud with differing degrees of turbulence, clustering and star formation activity. A comparison of these sub-regions allows us to investigate the effects of the environment on the dense gas structure and outflow properties. The four sub-regions are: 
\begin{enumerate}
\item NGC~1333: This is thought to be the most active and clustered region of star formation in Perseus, and contains a large concentration of TT stars, HH objects, bipolar jets and outflows, all of which will affect the outflow properties that we calculate. It is fairly young --- possessing cores at an age of $\sim 1$~Myr \citep{2005A&A...440..151H}; it also has a lumpy and filamentary structure --- dust ridges extend between clumps of YSOs, and there are several cavities filled with high-velocity outflow gas. \citet{2001ApJ...546L..49S} found that the strongest submillimetre emission originates in the south, and is associated with the YSOs IRAS~2, 3 and 4.
\item IC~348: This is a slightly older region than NGC~1333, but is still undergoing active star formation. The most well-known feature in the region is HH~211 --- a highly-collimated bipolar outflow driven by a Class 0 protostar, which has been the subject of several interferometric studies (e.g. Chandler \& Richer, 2001).
\item L~1448: This region contains a large number of young Class 0 protostars, and has been found to be dominated by outflow activity, which argues for it being relatively young. In particular, the outflow L~1448-C has been studied in great detail \citep{2013A&A...549A..16N,2000AJ....120.1467W}.
\item L~1455: This is the smallest and faintest of the four regions. It has the highest proportion of Class I sources (compared with Class 0s), which points to it being older than L~1448. There are still quite a few prominent outflows, several of which have H$_2$ objects associated with them \citep{2008MNRAS.387..954D}.
\end{enumerate}
%


\subsection{Outline}
We present \HCOplus\ and HCN molecular data that we observed in the Perseus molecular cloud. Section \ref{sec:obs} presents an overview of the observations and data reduction procedure, and Section \ref{sec:structure} compares the spatial and velocity structures of the \HCOplus\ and HCN emission. We present an analysis of outflow properties in the 4 Perseus sub-regions, in both \HCOplus\ (Section \ref{sec:hcooutflows}) and in HCN (Section \ref{sec:hcnoutflows}), comparing the results obtained for the two molecules. Section \ref{sec:abundances} presents a comparison of the relative abundances of \HCOplus/HCN in IRAS~2 (NGC~1333) to investigate the chemistry within the outflow. 

\section{Observations and reduction}\label{sec:obs}

\subsection{Description of observations}
Emission from \HCOplus\ and HCN was observed and details of the transitions and frequencies are given in Table \ref{mol_trans}. The frequencies and energies quoted are taken from LAMDA, the Leiden Atomic and Molecular Database\footnote{http://home.strw.leidenuniv.nl/{\char'176}moldata/}. \changed{The critical densities of each transition were calculated using the Einstein-A coefficients from LAMDA, and the collisional rate coefficients from \citet{1999MNRAS.305..651F} for \HCOplus\ and \citet{2010MNRAS.406.2488D} for HCN.} The data were observed in single sub-band mode, splitting the 250~MHz bandwidth into 8192 channels, giving an initial velocity resolution of 0.026~\kms. 

\begin{table}
\caption{Molecules observed in the Perseus dense gas survey, with their transitions (Column 2), frequencies (Column 3) and energies of the upper level above ground (Column 4). \changed{The critical densities of each transition (Column 5) are calculated using Einstein-A coefficients from LAMDA$^1$, and the collisional rate coefficients are obtained from \citet{1999MNRAS.305..651F} for \HCOplus\ and \citet{2010MNRAS.406.2488D} for HCN.}}
\begin{center}
\begin{tabular}{ccccc}
\hline
Molecule & $\Delta J$ & $\nu_{\rm trans}$/~GHz & $E_{\rm u}$/~K & $n_{\rm crit}$/~\cmcube \\ \hline
\HCOplus\ & 4 -- 3 & 356.734 & 42.8 & 4.4 $\times$ 10$^6$ \\
HCN & 4 -- 3 & 354.506 & 42.5 & 2.4 $\times$ 10$^7$ \\ \hline
\end{tabular}
\end{center}
\label{mol_trans}
\end{table}

The \HCOplus\ observations were taken over a total of 13 nights between 3 August and 8 December 2011; the HCN observations were taken over a total of 6 nights between 9 November and 11 December 2012, and all data were taken using HARP at the JCMT \citep{2009MNRAS.399.1026B}. The sizes and centres of the sub-regions mapped in each molecule are presented in Table \ref{regions}. The areas mapped in HCN differ from those mapped in \HCOplus\ due to time constraints: HCN maps were only made where there was at least a 1$\sigma$ detection of \HCOplus\ emission.  


\begin{table*}
\caption{Columns 2--5 show the centre positions and sizes of the areas mapped for the dense gas survey in Perseus; the sixth column gives the total area in a sub-region mapped in that particular molecule; the final column shows the rms noise level reached (at a velocity resolution of 0.2~\kms) for each map.}
\centering
\begin{tabular}{ccccccc}
\hline
Region & RA (J2000)  & Dec (J2000) & Width & Height & Area & RMS noise\\
       & (h m s) & ($^{\circ}$ \arcmin\  \arcsecs) & (arcsec) & (arcsec) & (arcmin)$^2$ & (K) \\ \hline
NGC1333 (\HCOplus) & 03:28:57 & 31:18:10 & 760 & 900 & 190.0 & 0.23\\
IC348 (\HCOplus) & 03:44:14 & 32:01:50 & 1150 & 700 & 223.6 & 0.20\\
L1448 (\HCOplus) & 03:25:34 & 30:43:45 & 700 & 500 & 97.2 & 0.16\\
L1455 (\HCOplus) & 03:27:37 & 30:14:10 & 580 & 520 & 83.8 & 0.15\\ \hline
NGC1333 (HCN) & 03:28:57 & 31:18:10 & 700 & 900 & 175.0 & 0.18\\
IC348-A (HCN) & 03:43:54 & 32:02:05 & 450 & 300 & 37.5 & 0.15\\
IC348-B (HCN) & 03:44:47 & 32:01:08 & 160 & 150 & 6.7 & 0.14\\
L1448 (HCN) & 03:25:35 & 30:44:31 & 350 & 300 & 29.2 & 0.14\\
L1455 (HCN) & 03:27:43 & 30:12:30 & 350 & 300 & 29.2 & 0.14\\ \hline
\end{tabular}
\label{regions}
\end{table*}

All the standard telescope observing procedures were followed for each night of observations, with regular pointings and focusing of the JCMT's secondary mirror. Standard spectra were also taken towards various well-known calibration sources, to verify that the intensity in the tracking receptor matched recorded standards to within \changed{calibration tolerance\footnote{The JCMT guidelines give an absolute calibration tolerance of between 20 and 30\%, but the relative flux scale of our \HCOplus/HCN data is estimated to be accurate to within 5 to 10\%.}} before subsequent observations were allowed to continue. The fully-sampled maps were taken in raster position-switched mode, an `on-the-fly' data collection method where the HARP array continuously scans in a direction parallel to the sides of the map to produce a fully-sampled map of the area \citep{2009MNRAS.399.1026B}. The telescope is pointed at an off-position after every row in the map to obtain background values that are subtracted from the raw data. Data are presented in units of corrected antenna temperature \antemp, which is related to the main beam temperature ($T_{\rm mb}$) using $T_{\rm mb}$ = \antemp/$\eta_{\rm mb}$. A value of $\eta_{\rm mb}$ = 0.66 was used, following \citet{2009MNRAS.399.1026B}.

The HCN $J = 4\to3$ transition is known to exhibit hyperfine splitting. However, due to the thermal and turbulent broadening of the lines, we are \changed{unable to distinguish between the three hyperfine lines with the greatest intensities (having a velocity range of 0.13~\kms) at the velocity resolution of our observations (0.2~\kms); the other lines are undetectable above the noise levels.} We therefore do not consider the hyperfine structure of HCN to be significant in our subsequent analyses.

\subsection{Data reduction}
The data were reduced using the {\sc oracdr} pipeline, part of the Starlink project\footnote{http://starlink.jach.hawaii.edu/starlink} software. The pipeline utilises {\sc kappa} routines to remove poorly-performing detectors (e.g. those that are particularly noisy or exhibit oscillatory behaviour) and spectra with bad baselines/overly-high rms noise. Linear baselines are removed, and the pipeline then uses {\sc smurf makecube} routines to convert from time-series to spectral (RA--Dec--velocity) 3D cubes. The final cubes were sampled onto a 6 arcsec grid using a Gaussian gridding kernel with a FWHM of 9 arcsec, resulting in an equivalent FWHM beam size of 16.8 arcsec (taking into account the JCMT beamsize of 14 arcsec at 345~GHz). The individual spectral cubes for each observation are then co-added together using {\sc wcsmosaic}, and regridded to a velocity resolution of 0.2~\kms\ using {\sc sqorst}. Almost all the data reached or surpassed the original target of 0.2~K mean r.m.s noise at a velocity resolution of 0.2~\kms, and the noise for each sub-region is shown in Table \ref{regions}.

\section{Overall spatial and velocity structure}\label{sec:structure}

\subsection{Integrated intensity structure}

\begin{figure}
\centerline{\includegraphics[width=8cm]{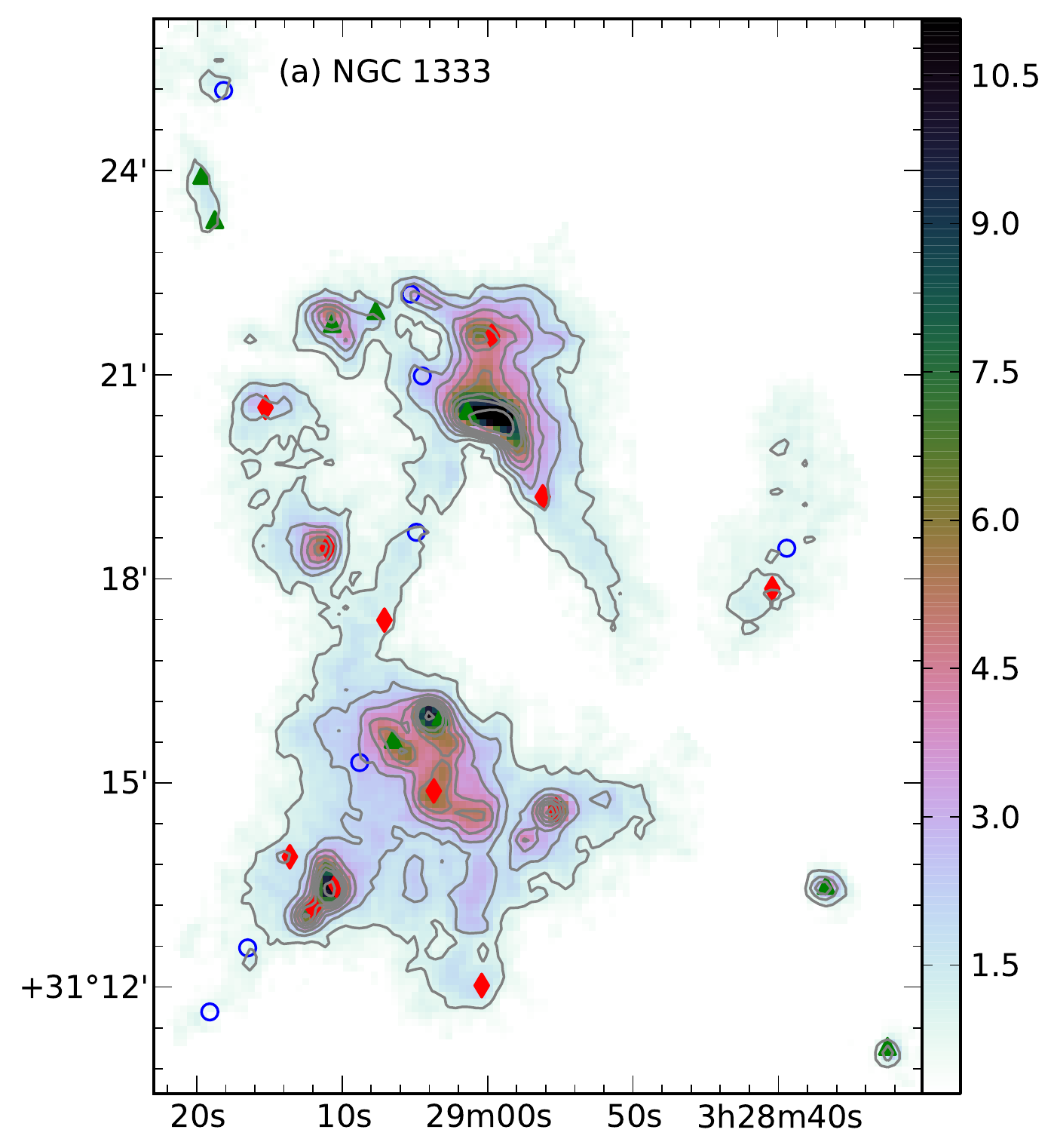}}
\smallskip
\centerline{\includegraphics[width=8cm]{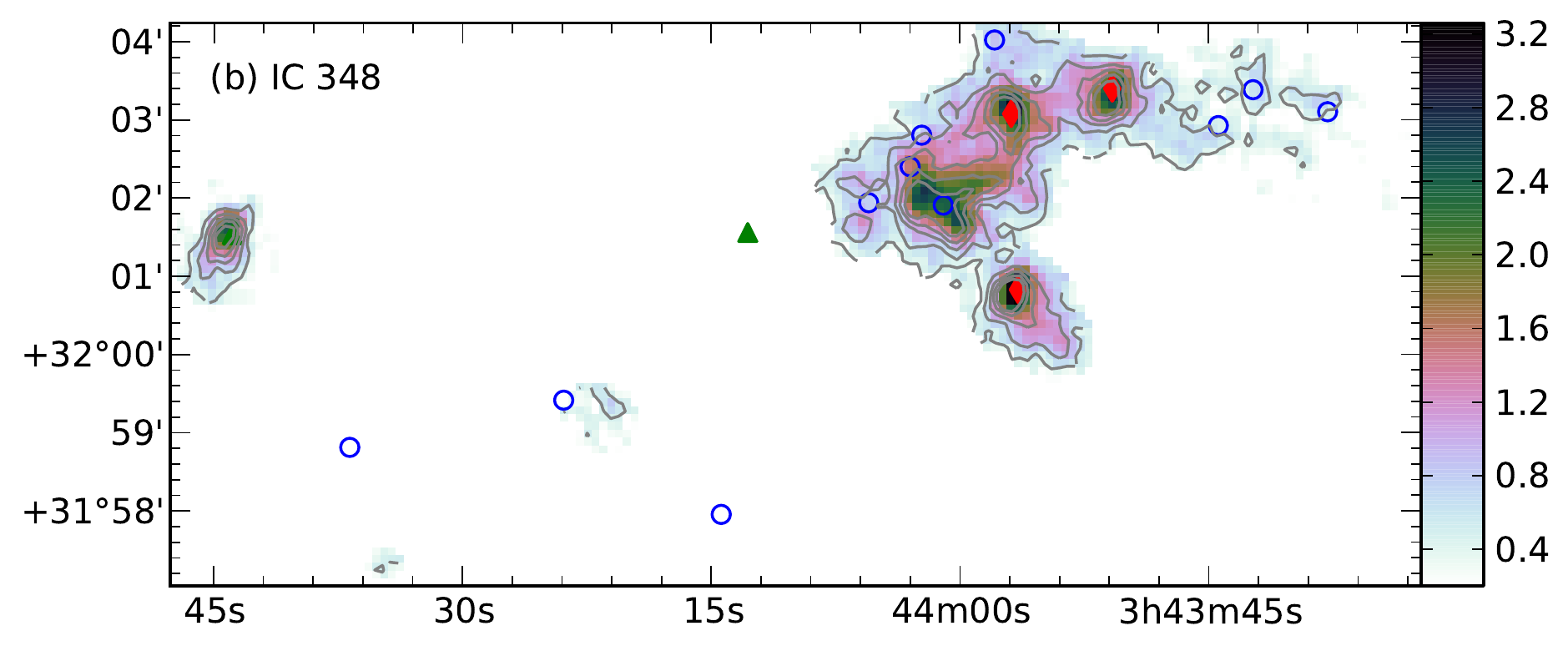}}
\smallskip
\centerline{\includegraphics[width=7cm]{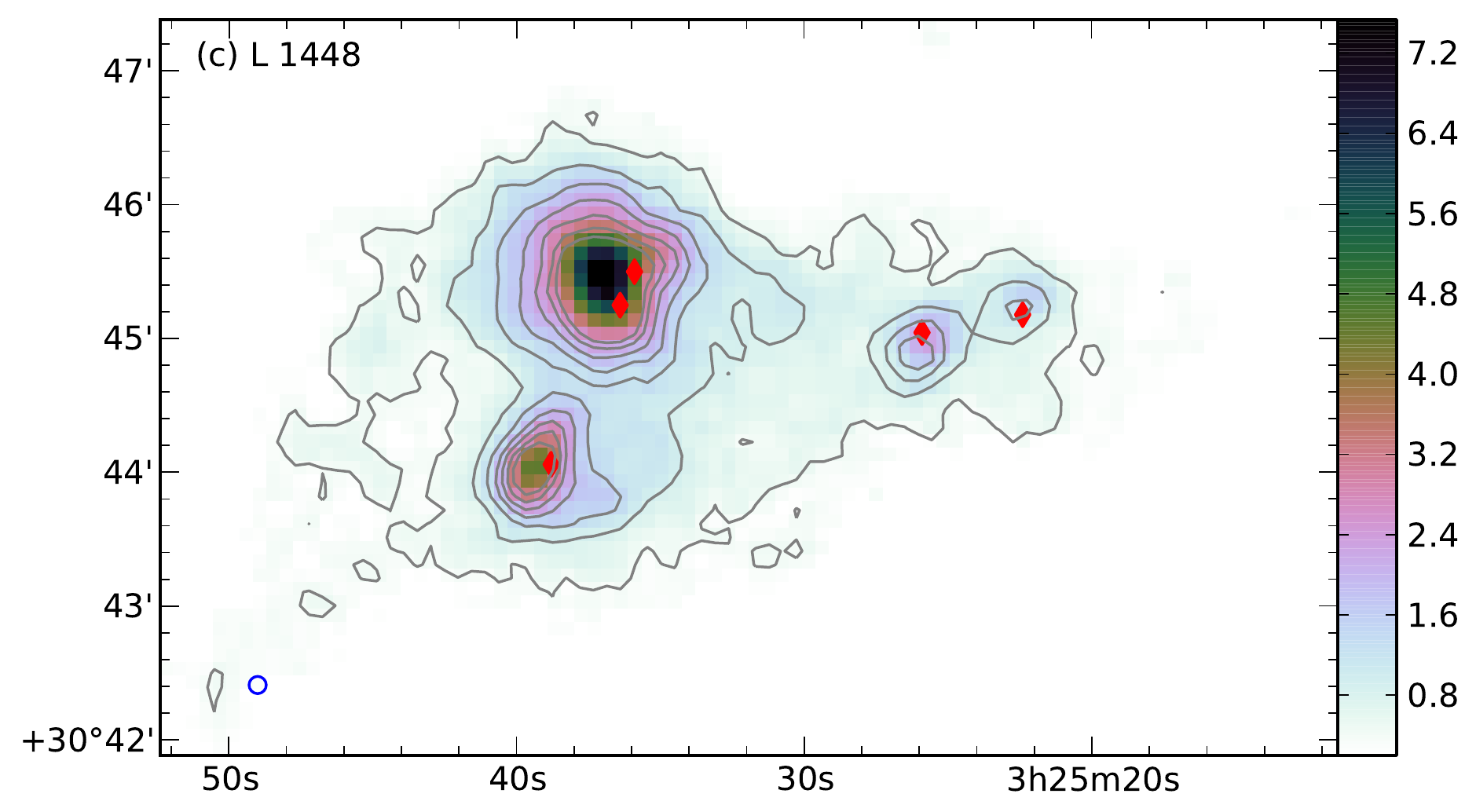}}
\smallskip
\centerline{\includegraphics[width=7cm]{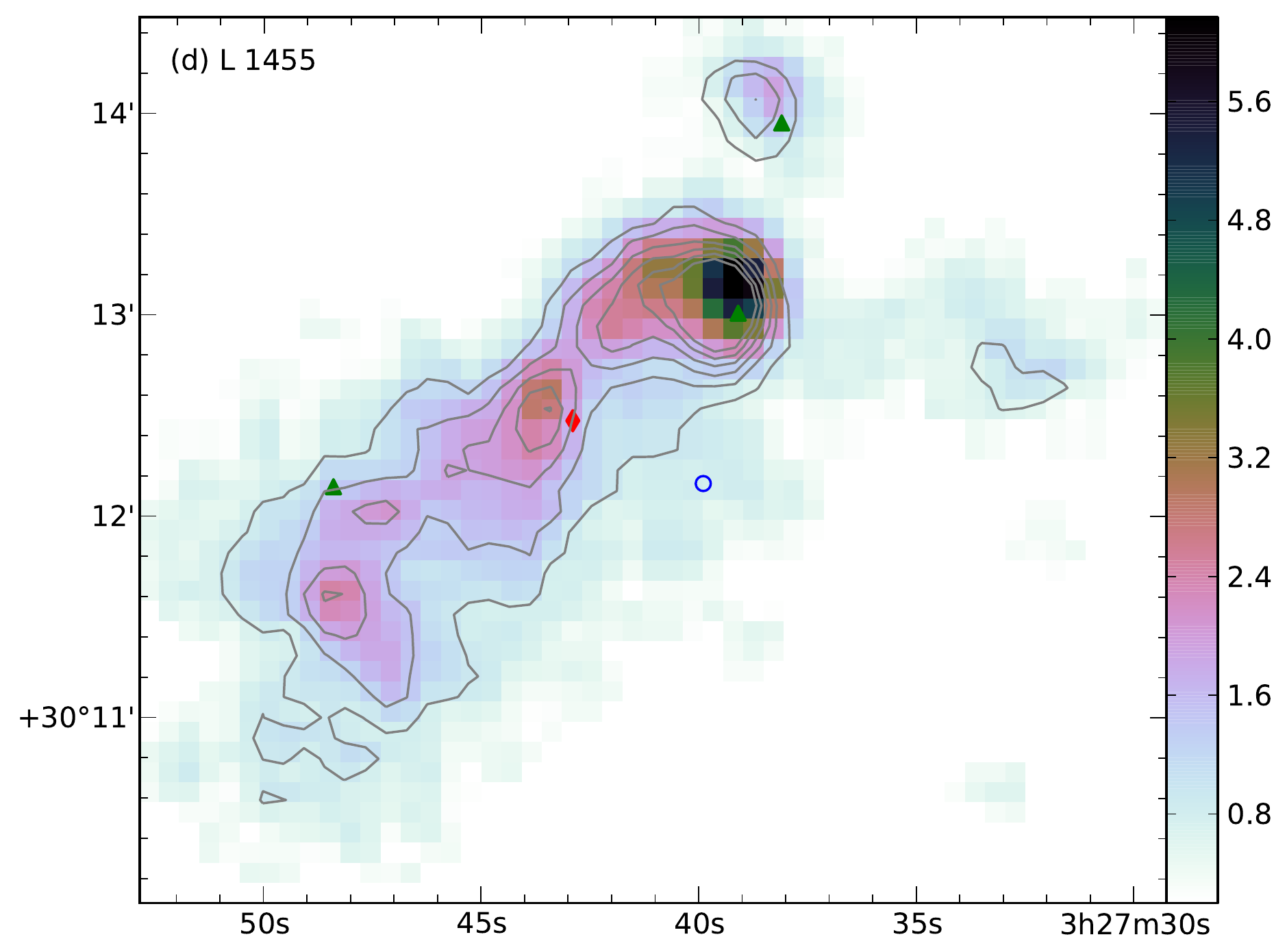}}
\caption{Integrated intensity maps of \HCOplus\ for (a) NGC~1333, (b) IC~348, (c) L~1448 and (d) L~1455. The contours for each map are as follows:  (a) from 1.0 K~\kms\ in 1.0 K~\kms\ increments; (b) from 0.5 K~\kms\ in 0.4 K~\kms\ increments; (c) from 0.4 K~\kms\ in 0.5 K~\kms\ increments; (d) from 0.9 K~\kms\ in 0.5 K~\kms\ increments. SCUBA-defined pre- and protostellar cores identified in each sub-region by H07 are overlaid on all the maps: Class 0 sources (red diamonds), Class I sources (green triangles) and starless cores (blue circles). The colour scheme is `cubehelix', as in \citet{2011BASI...39..289G}.}
\label{hcointint}
\end{figure}

\begin{figure}
\centerline{\includegraphics[width=8cm]{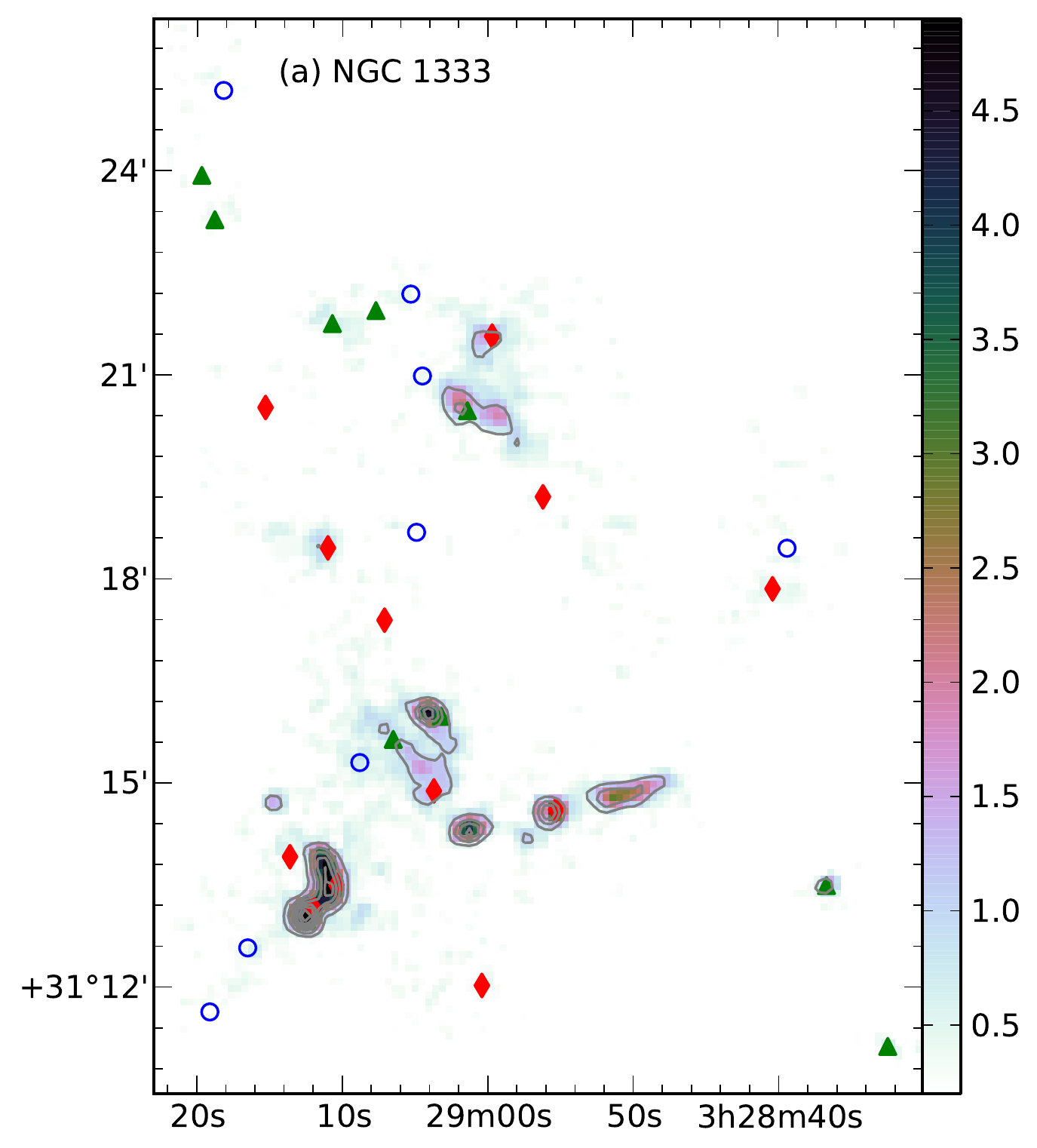}}
\smallskip
\centerline{\includegraphics[width=8cm]{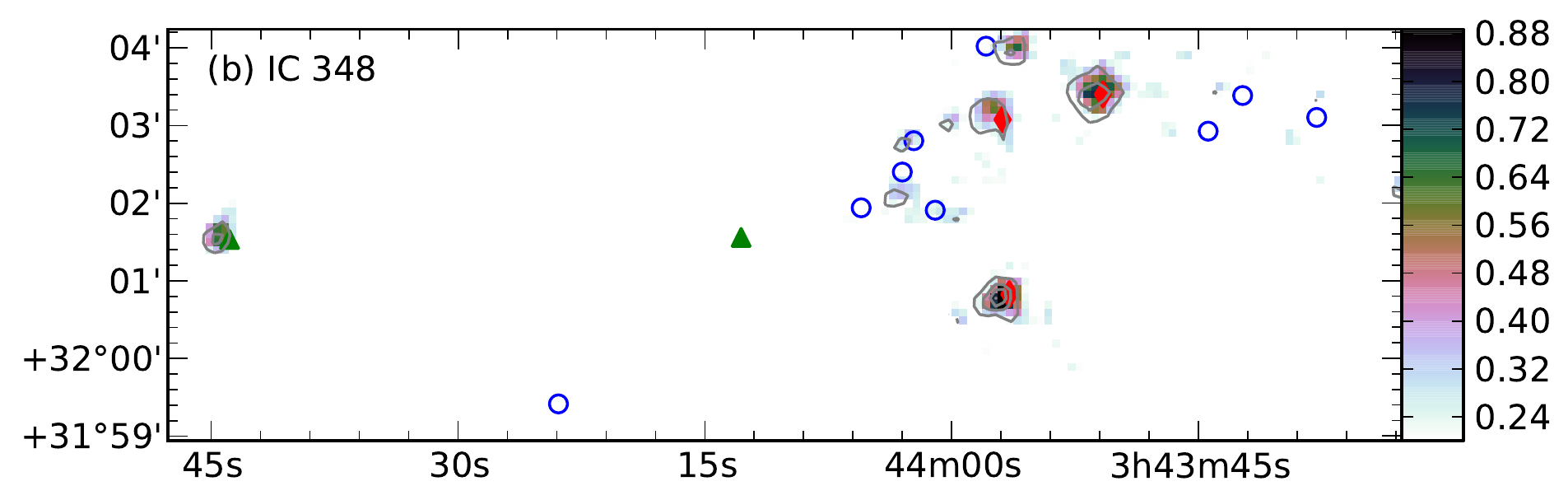}}
\smallskip
\centerline{\includegraphics[width=7cm]{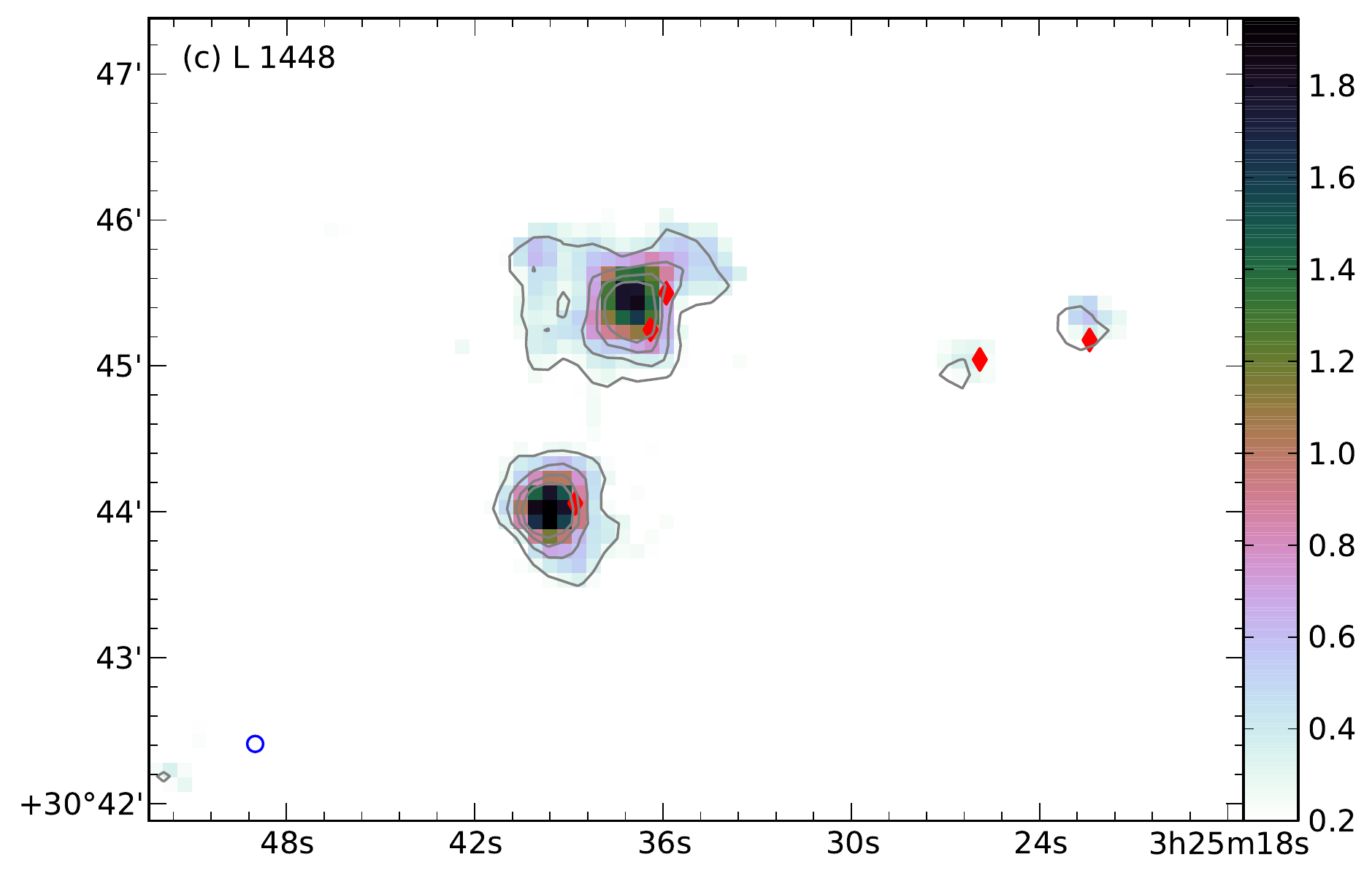}}
\smallskip
\centerline{\includegraphics[width=7cm]{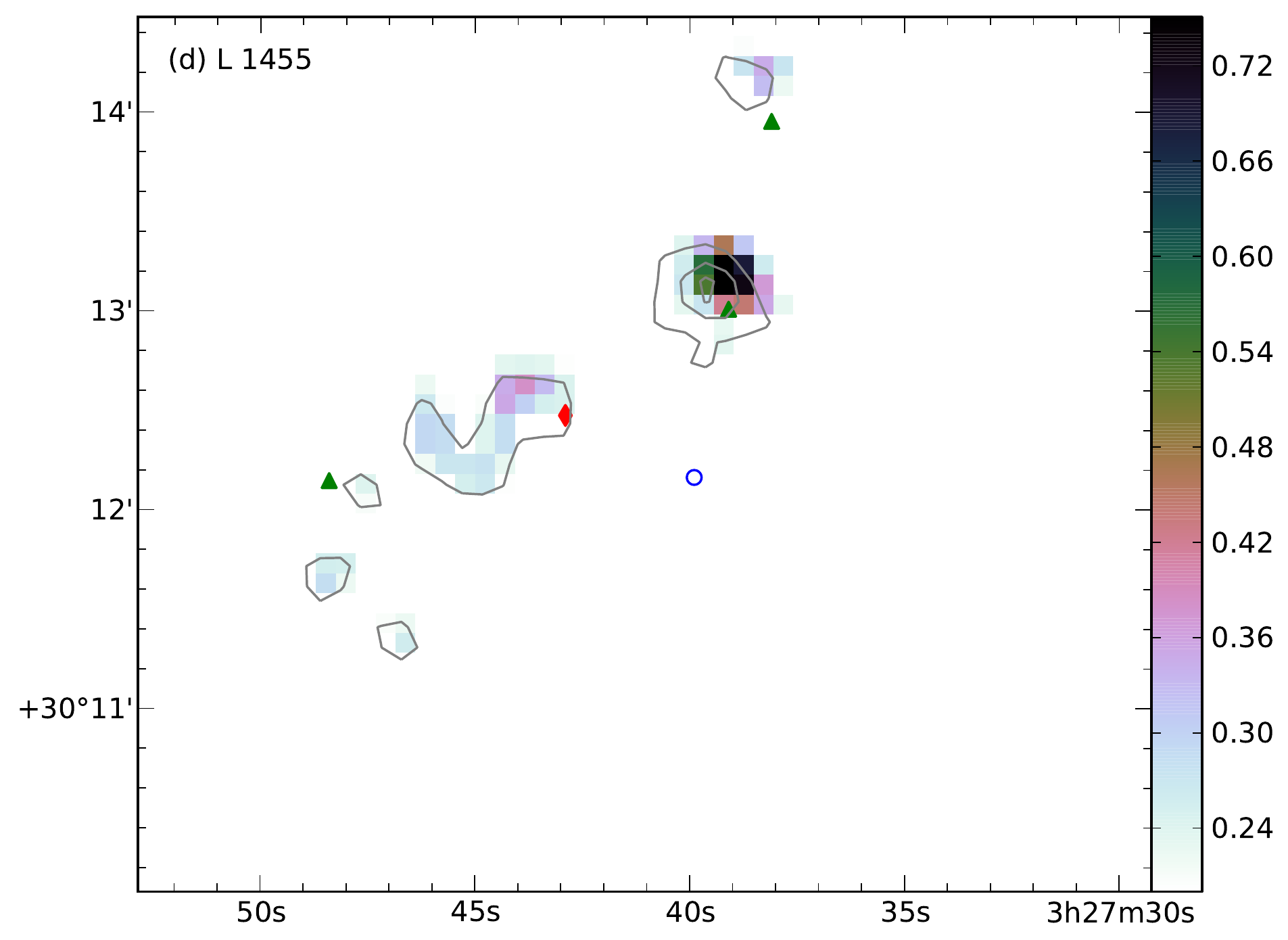}}
\caption{Integrated intensity maps of HCN for (a) NGC~1333, (b) IC~348, (c) L~1448 and (d) L~1455. The contours for each map are as follows: (a) from 1.0 K~\kms\ in 1.0 K~\kms\ increments; (b) from 0.3 K~\kms\ in 0.3 K~\kms\ increments; (c) from 0.3 K~\kms\ in 0.3 K~\kms\ increments; (d) from 0.2 K~\kms\ in 0.3 K~\kms\ increments. SCUBA-defined pre- and protostellar cores identified in each sub-region by H07 are overlaid on all the maps as in Figure \ref{hcointint}.}
\label{hcnintint}
\end{figure}

\begin{figure}
\centerline{\includegraphics[width=8cm]{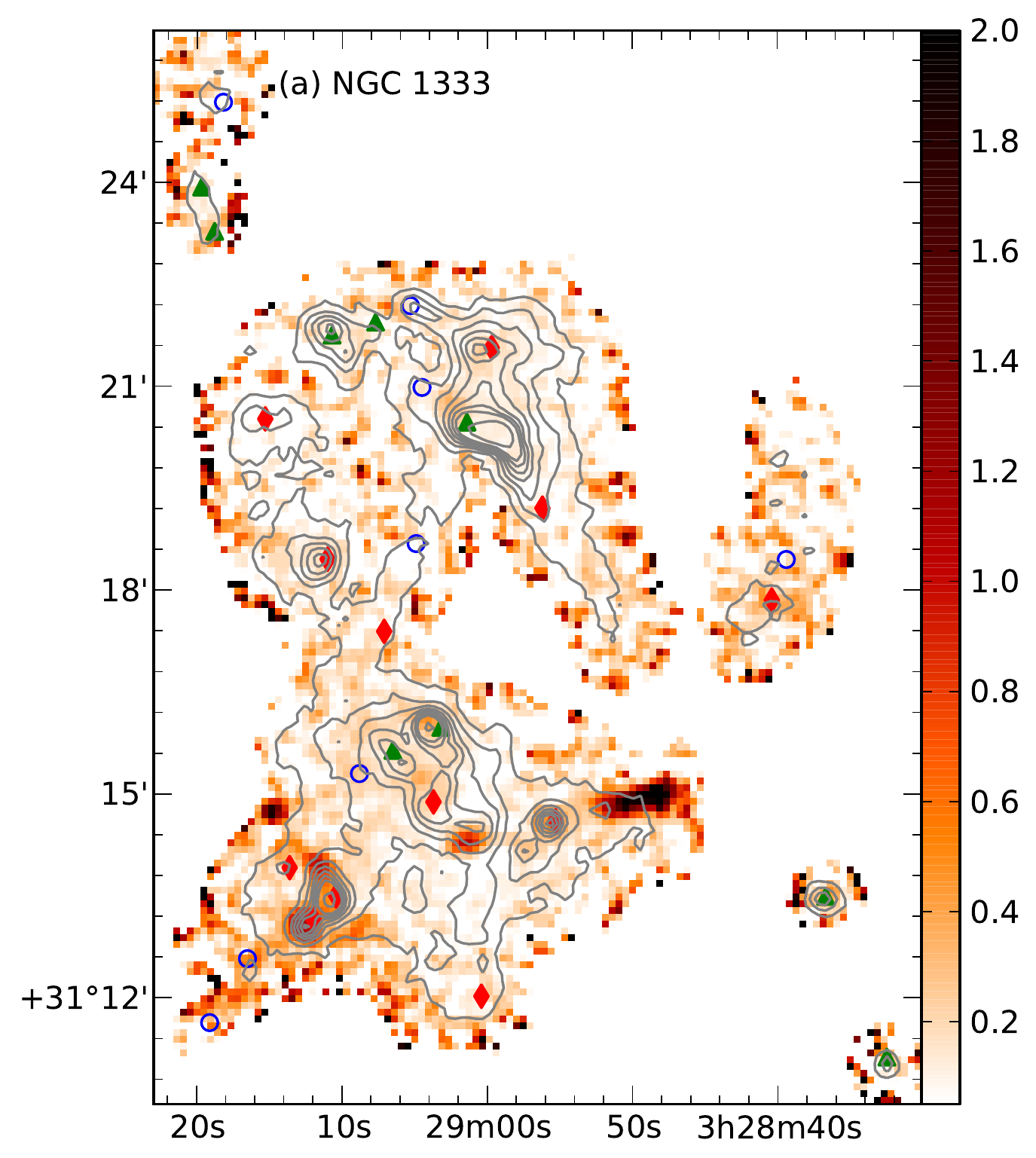}}
\smallskip
\centerline{\includegraphics[width=8.0cm]{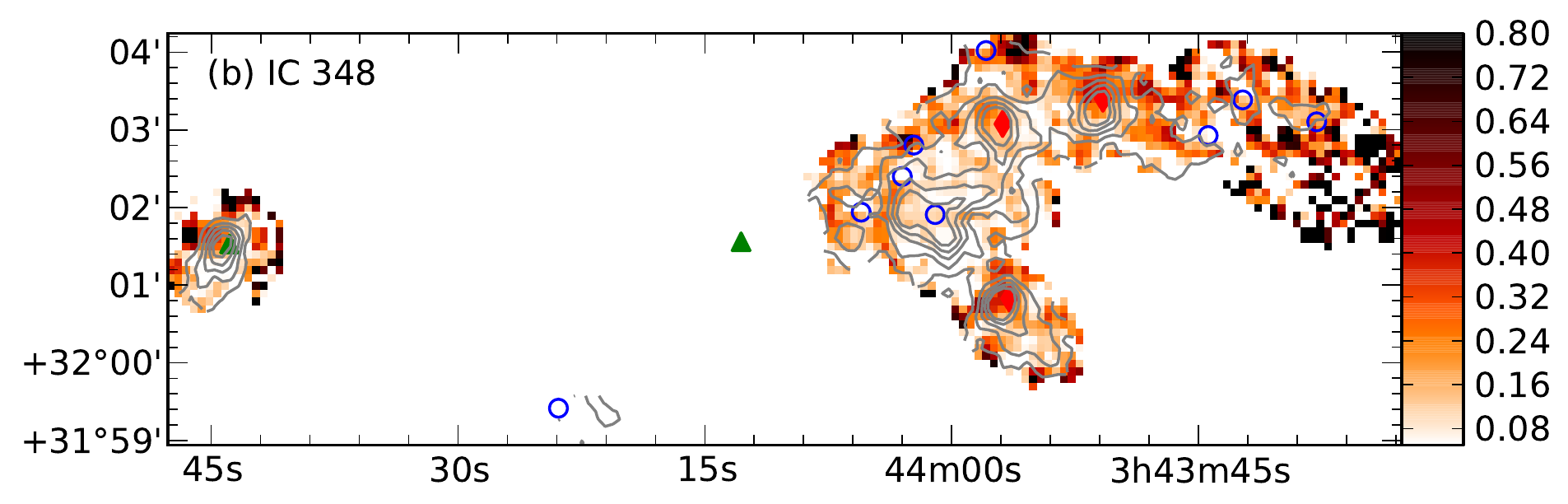}}
\smallskip
\centerline{\includegraphics[width=7.0cm]{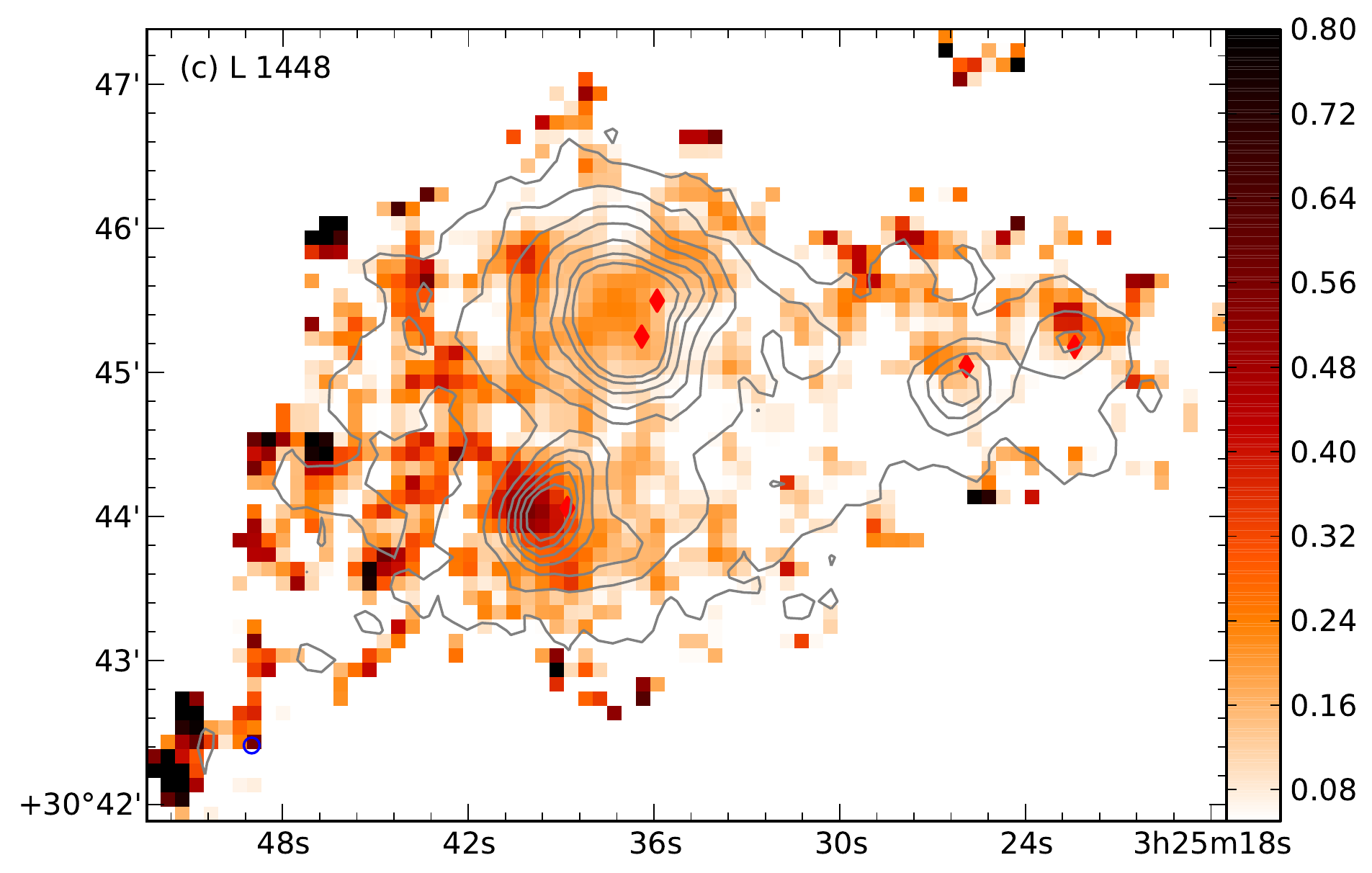}}
\smallskip
\centerline{\includegraphics[width=7cm]{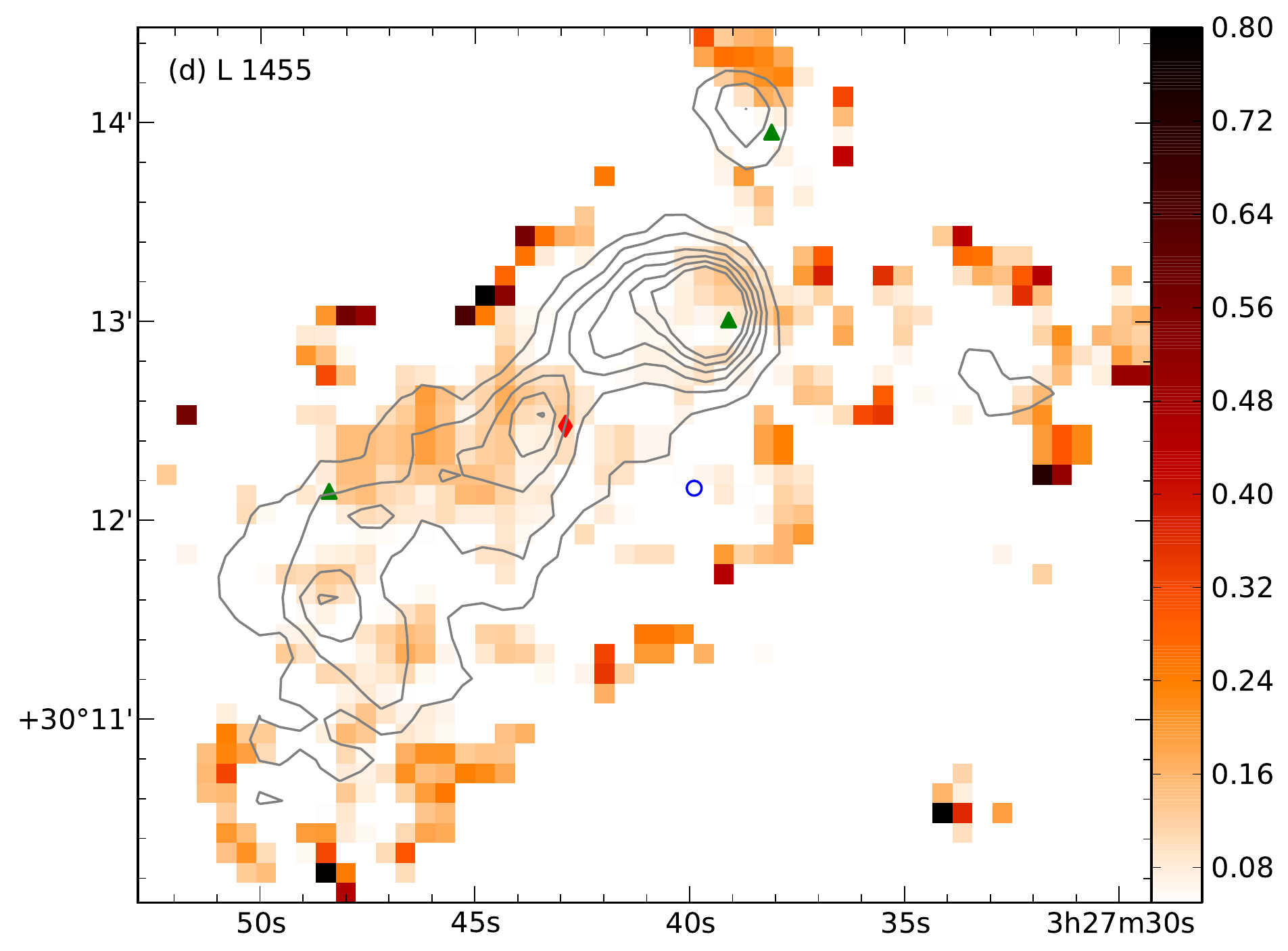}}
\caption{Ratio of the integrated intensity of HCN to that of \HCOplus\ for (a) NGC1333, (b) IC348, (c) L1448 and (d) L1455. H07 SCUBA core positions are overlaid as in Figure \ref{hcointint}. \HCOplus\ contours are included (with levels as in Figure \ref{hcointint}) for reference, and the ratio is taken over all regions where the \HCOplus\ is greater or equal to the 3 $\sigma$ noise level.}
\label{ratio}
\end{figure}

Figures \ref{hcointint} and \ref{hcnintint} show the integrated intensity maps of the \HCOplus\ and HCN emission respectively for the four Perseus sub-regions we are investigating. Hatchell et al. (2007, hereafter H07) created a SCUBA core catalogue in Perseus, and we overlay their core positions on our maps to pinpoint the positions of protostellar and starless cores. 

We observe a difference in the type of SCUBA core associated with the two molecules: only 24\% of the H07 starless cores have HCN emission at the 3$\sigma$ level and above, compared with 75\% of the protostellar H07 cores; \HCOplus\ on the other hand shows $\geq 3 \sigma$ emission for all protostellar cores, as well as 68\% of the starless cores. Therefore, we conclude that the compact HCN emission is more closely associated with protostellar objects than the \HCOplus\ emission; this is useful as it means we can use \changed{the HCN to help pinpoint the protostellar cores within the larger sample identified using the \HCOplus\ emission.}

Figure \ref{ratio} shows the ratio of the HCN integrated intensity to that of the \HCOplus\ for each of the sub-regions. One can see that on average, the HCN has a lower integrated intensity than the \HCOplus, with ratios of between 0.1 and 0.6 for the majority of the sub-regions. There are some compact clumps (particularly in NGC~1333), where the HCN/\HCOplus\ ratio is greater than one. This is most noticeable in the sub-regions around IRAS~2 and IRAS~4 (in NGC~1333), both of which are young, energetic Class 0 protostars that drive powerful outflows. In particular, at the spatial positions of the IRAS~2 outflow lobes, the HCN/\HCOplus\ ratio increases to values of 2.5. This will be discussed in greater detail in the Section 5, where we will also calculate relative abundances and enhancements of the two molecules. 

The \HCOplus\ emission is more extended than the HCN and shows more filamentary structure compared to HCN, which is concentrated in compact clumps. This is to be expected as the HCN has a slightly higher critical density than the \HCOplus, and is likely to be confined to the denser regions. We can thus tentatively constrain the gas densities in the filaments to lie between the critical densities of the two molecules (i.e. $(2 - 9) \times 10^6$ \cmcube).

\begin{figure}
\centerline{\includegraphics[width=7cm]{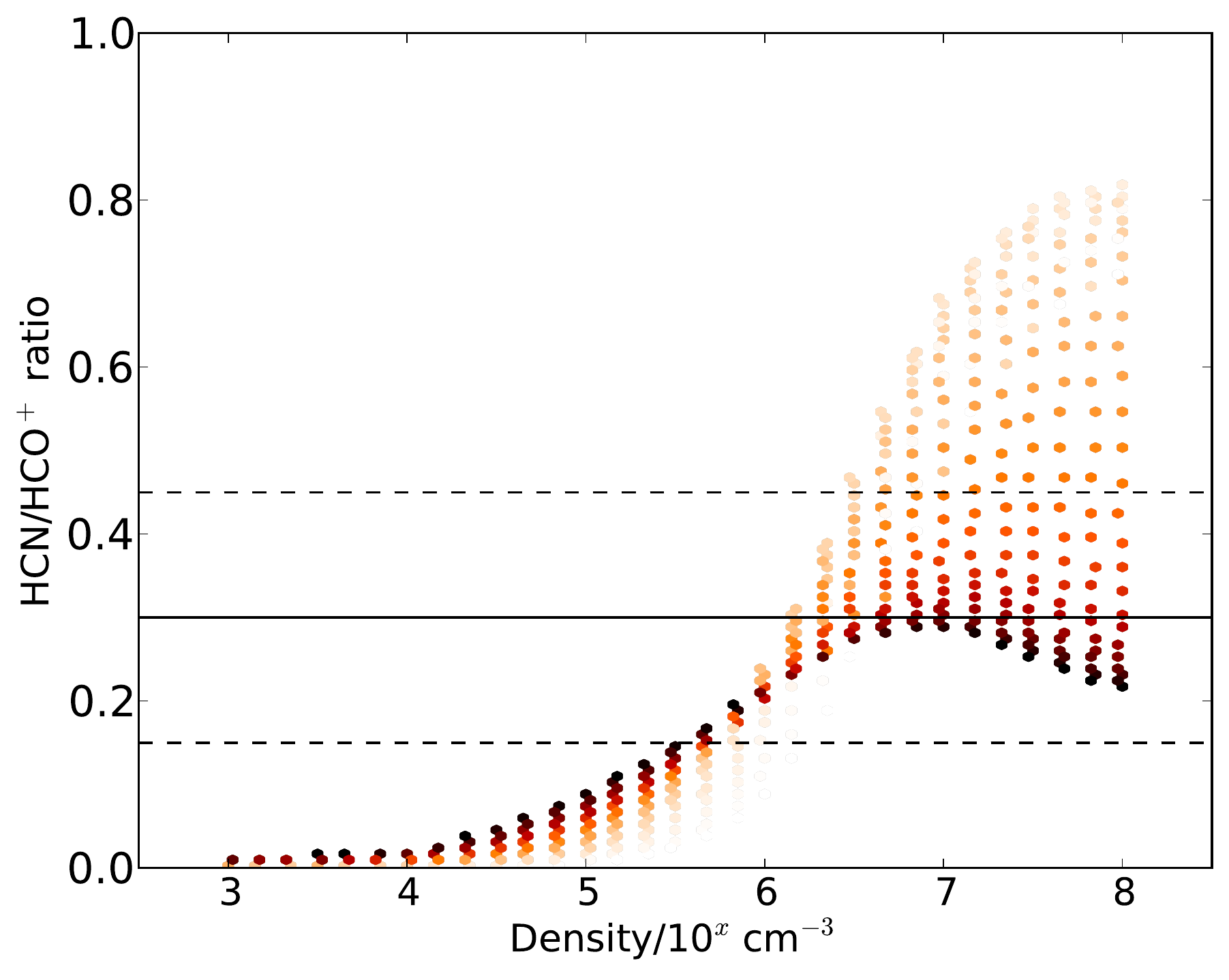}}
\caption{Ratio of the integrated intensity of HCN to that of \HCOplus\ plotted against the \Htwo\ spatial density, for kinetic temperature values (represented by dotted lines of different colours and shades) ranging between 10~K (palest) and 200~K (darkest).}
\label{rhoratio}
\end{figure}

Using RADEX \citep{2007A&A...468..627V}, we have performed simple 1-D radiative transfer modelling of the $J = 4\to3$ transitions of \HCOplus\ and HCN, over a range of temperatures ($10 - 200$~K) and densities ($10^3 - 10^8$ \cmcube). \changed{We used the most recent datafiles of the \HCOplus\ and HCN collisional rate coefficients, taken from \citet{1999MNRAS.305..651F} and \citet{2010MNRAS.406.2488D} respectively, obtained via the LAMDA database.} We assumed a constant column density for each molecule based on the column density of the filaments, as calculated from SCUBA and multiplied by the canonical abundances from Table \ref{coldenprops}. The resulting integrated intensity ratios are plotted against density in Figure \ref{rhoratio}. \changed{The mean observed ratio in the filamentary structures is $(0.30 \pm 0.15)$, corresponding to densities in the range $5 \times 10^5$ to $5 \times 10^6$~\cmcube.} This is a fairly good match for the critical density range of our two molecules, and lends support to our theory that the differences in spatial extent traced are a product of the spatial density of the filaments themselves.

%

\subsection{Velocity structure}

\begin{figure}
\centerline{\includegraphics[width=7cm]{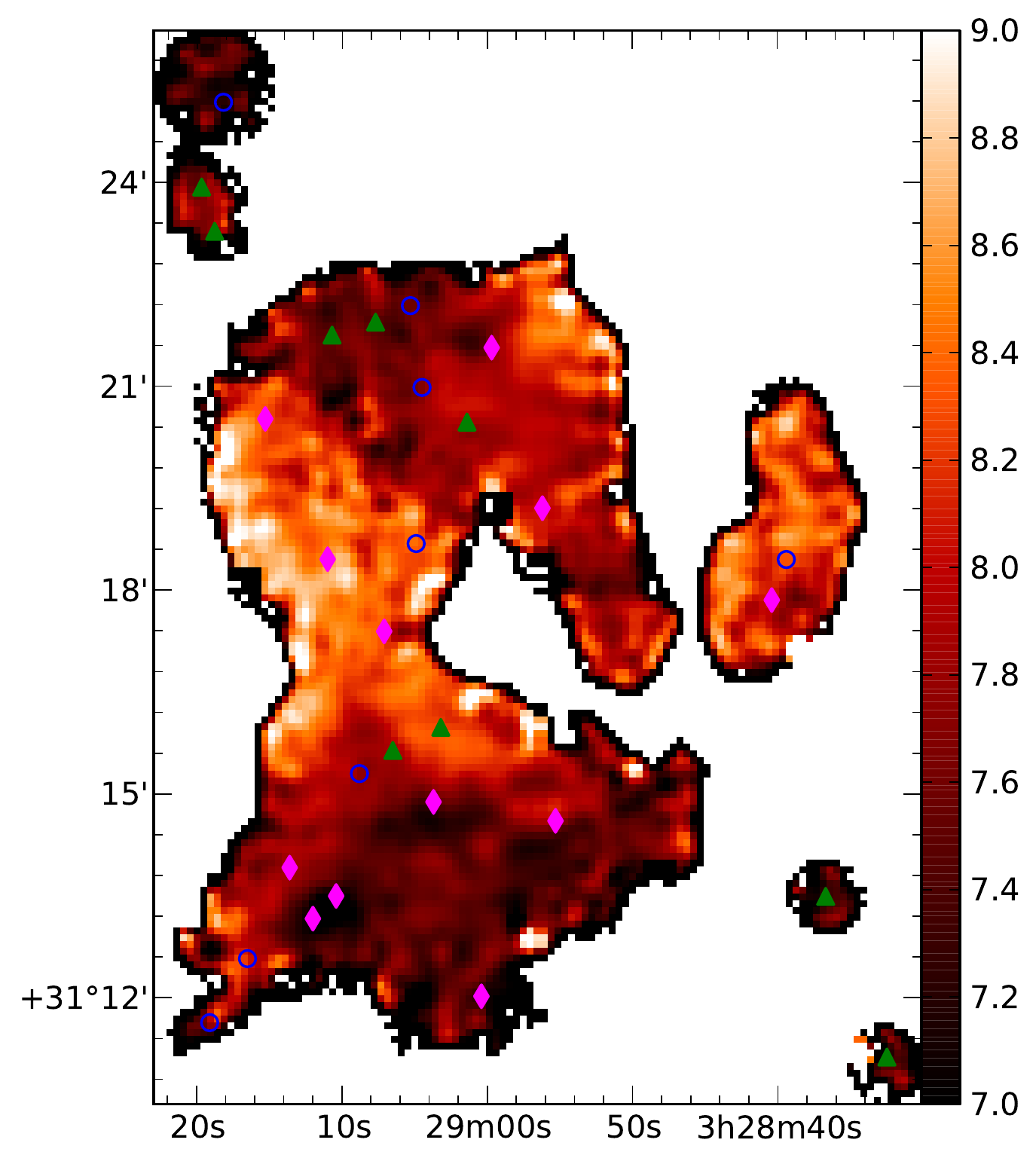}}
\smallskip
\centerline{\includegraphics[width=7cm]{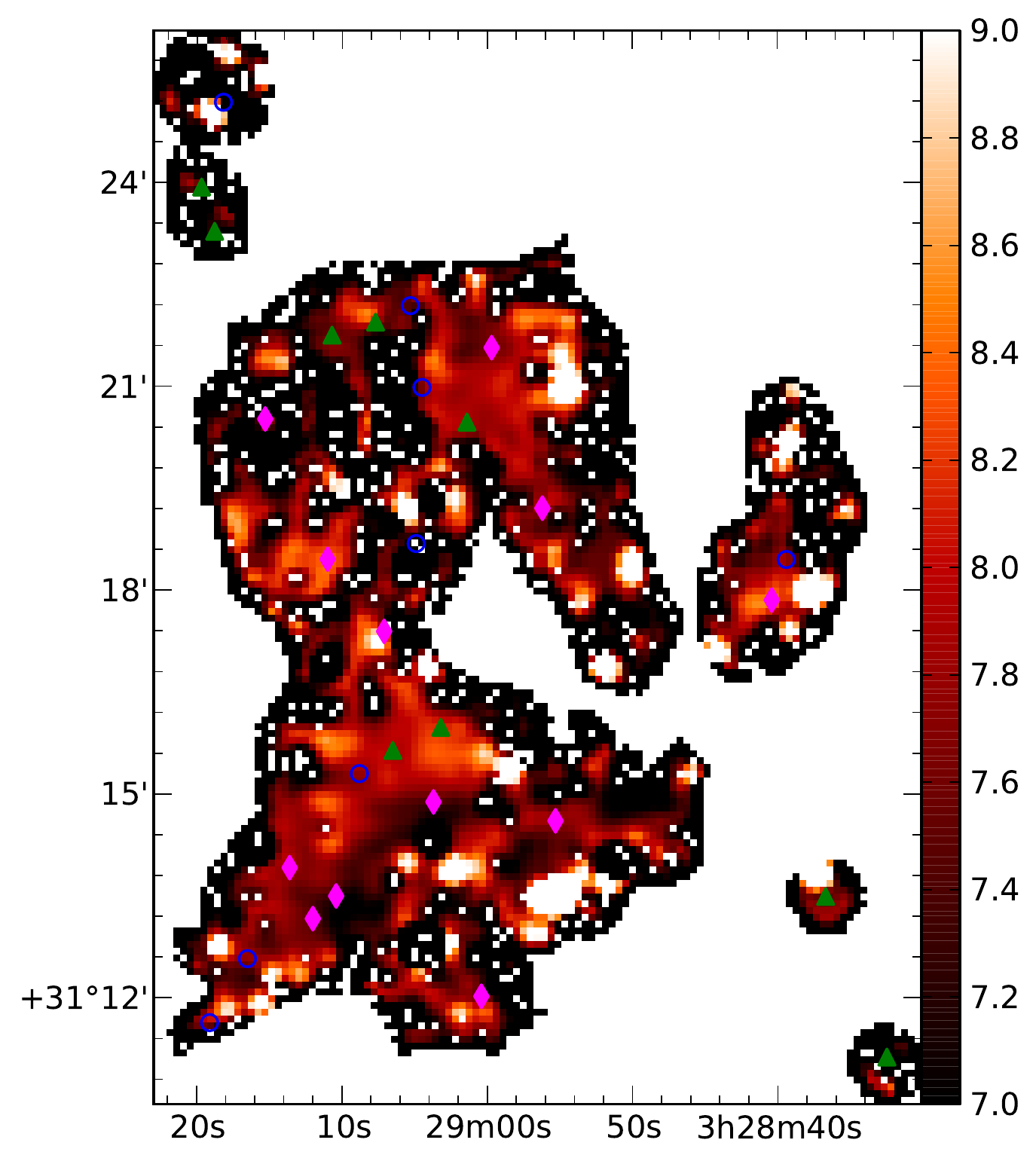}}
\caption{The velocity centroid (i.e. the line centre velocity at each spatial pixel) of \HCOplus\ (top) and HCN (bottom) for NGC~1333, with units in \kms. H07 SCUBA core positions are overlaid on all maps: Class 0 sources (magenta diamonds), Class I sources (green triangles) and starless cores (grey circles). Of the 4 sub-regions, NGC~1333 shows the greatest distinction in velocity structure between the two molecules.}
\label{ngc1333iwc}
\end{figure}

%

\begin{figure}
\centerline{\includegraphics[width=8cm]{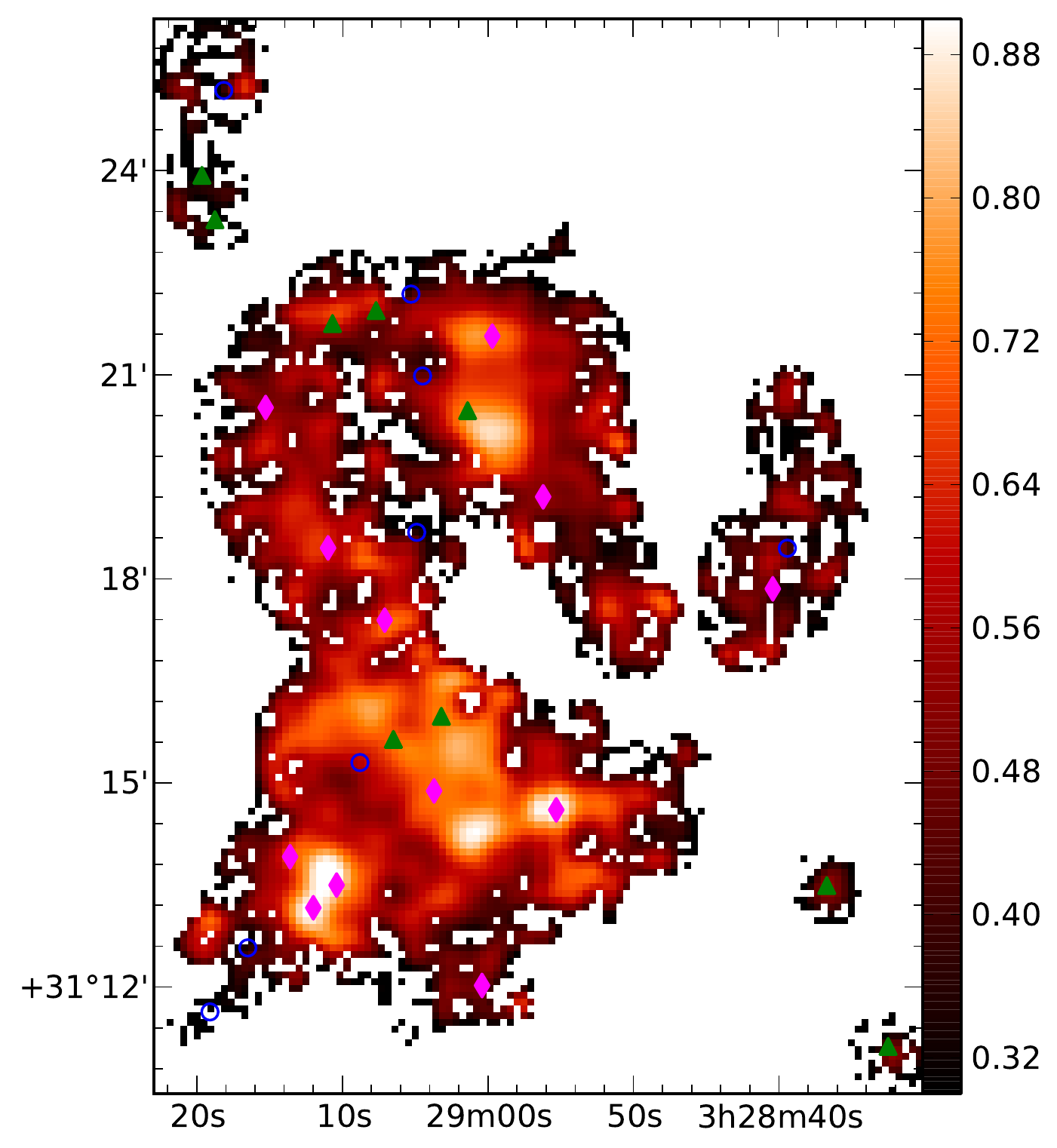}}
\caption{The velocity dispersion \changed{(i.e. the intensity-weighted dispersion or second moment, at each spatial pixel)} of \HCOplus\ for NGC~1333, with units in \kms. H07 SCUBA core positions are overlaid as in Figure \ref{ngc1333iwc}.}
\label{ngc1333iwd}
\end{figure}

\HCOplus\ shows large-scale velocity structure due to its extended coverage, particularly in NGC~1333 (see Figure \ref{ngc1333iwc}). This sub-region shows lower velocity in the north and south (especially around the Class 0 sources), and higher velocity in the central area. The velocity of HCN, by comparison, shows no clear large-scale trends or structure. 


\begin{figure}
\centerline{\includegraphics[width=5cm,angle=-90]{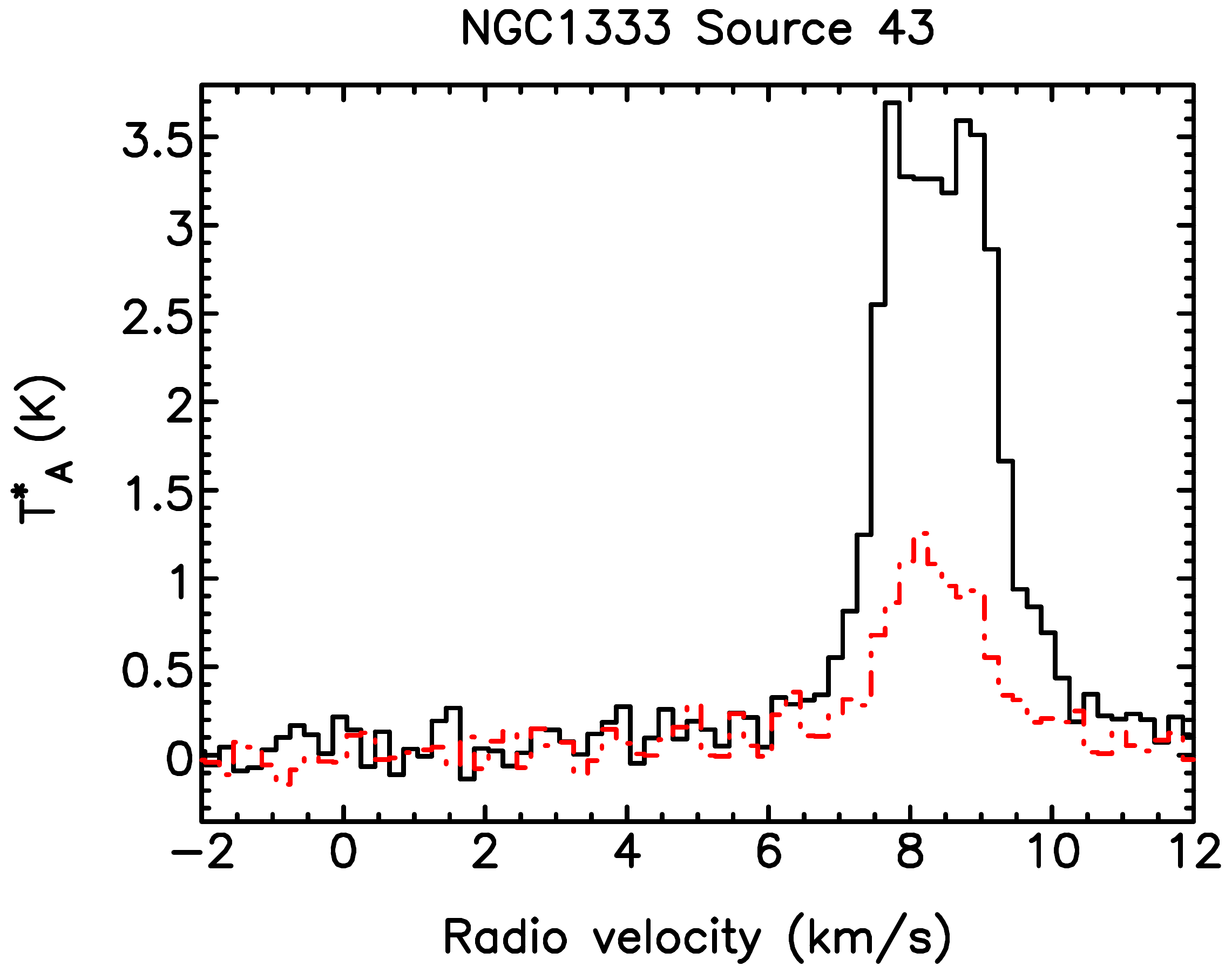}}
\smallskip
\centerline{\includegraphics[width=5cm,angle=-90]{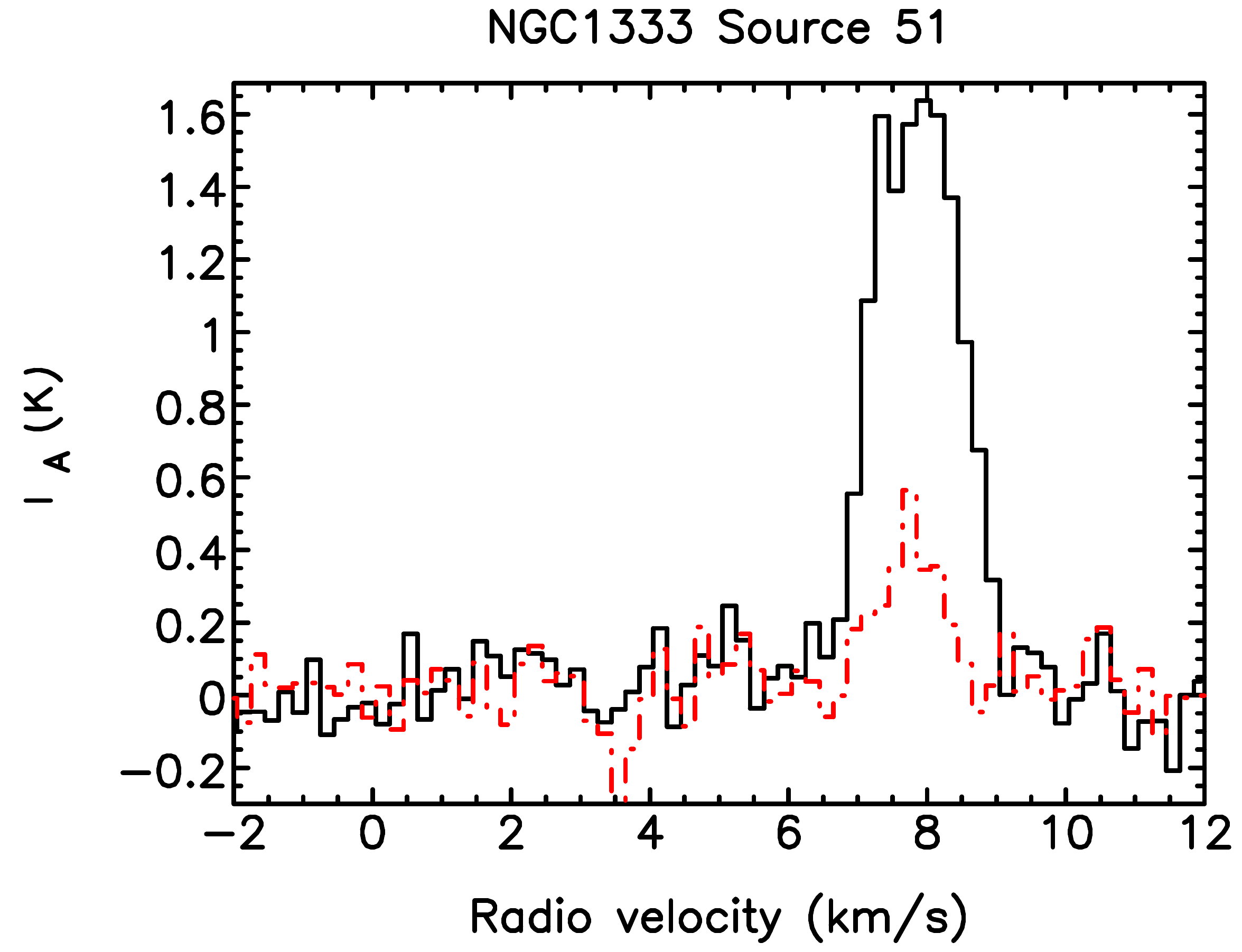}}
\caption{Representative spectra of \HCOplus\ (black) and HCN (red) for two SCUBA sources in NGC~1333: H07-labelled sources 43 (top) and 51 (bottom). The former is a Class I object, while the latter is a starless core. The velocity resolution for the two lines is 0.2~\kms.}
\label{compspec1}
\end{figure}

\changed{A comparison of the spectra of \HCOplus\ and HCN at every H07 core is presented in a supplementary (online-only) Appendix D, showing a range of different spectral profiles for the two molecules. Some sources have wide spectral lines (e.g. L~1448 Source 28), while others exhibit very narrow line profiles (e.g. NGC~1333 Source 63 and L~1455 Source 37). Some spectra exhibit signs of double-peaked blue asymmetry that have been associated with infall (e.g. NGC~1333 Source 44); others show elevated emission in the spectral linewings on either side of the main line peak, which is indicative of outflow (e.g. NGC~1333 Source 41 and 42, IC~348 Source 14).} 

Figure \ref{ngc1333iwd} shows the variation in linewidth ($\sigma$) of \HCOplus\ across the NGC~1333 sub-region, \changed{measured as the intensity-weighted dispersion ({i.e.} the square root of the second moment/variance)}: the linewidths increase from $\sim$0.5~\kms\ in the outer areas, to $\sim $1~\kms\ at protostellar SCUBA core positions, particularly those Class 0 sources showing signs of outflow activity. These younger sources and their outflows cause an increase in the local turbulence, leading to the corresponding increase in the \HCOplus\ linewidths.

\begin{figure}
\centerline{\includegraphics[width=5cm,angle=-90]{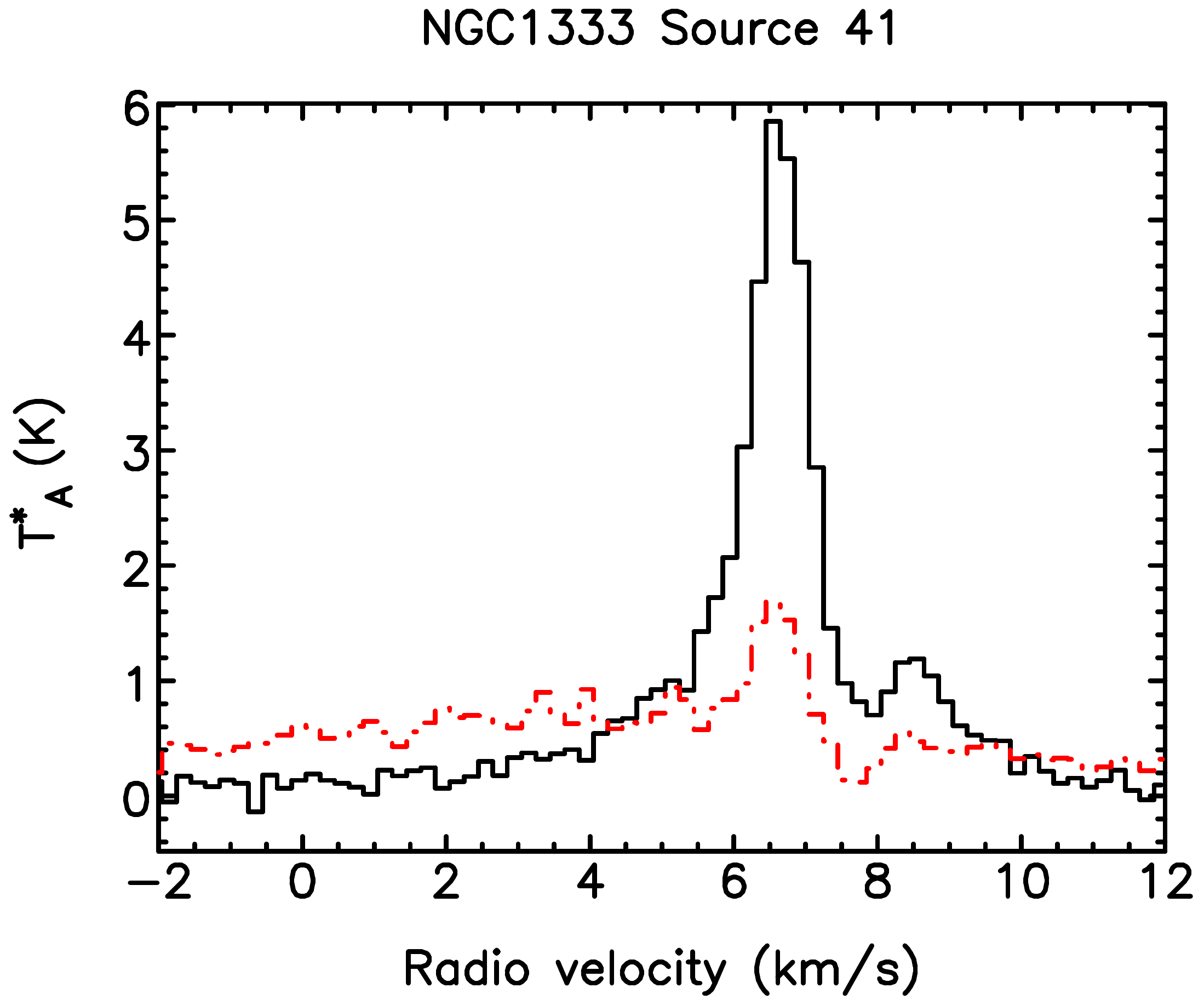}}
\smallskip
\centerline{\includegraphics[width=5cm,angle=-90]{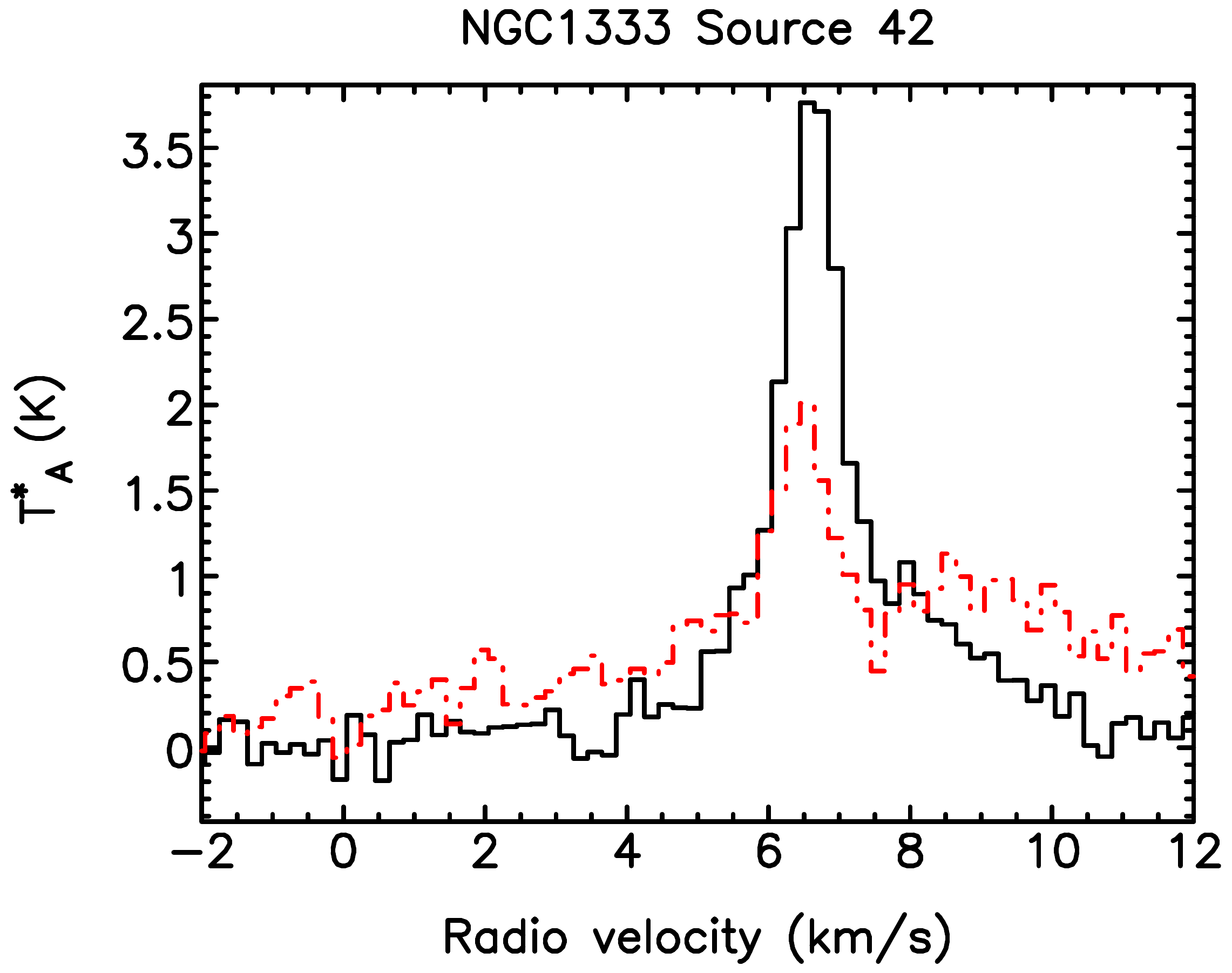}}
\caption{Representative spectra of \HCOplus\ (black) and HCN (red) for two SCUBA sources in NGC~1333 IRAS~4: H07-labelled sources 41 (IRAS~4A, top) and 42 (IRAS~4B, bottom). The velocity resolution for the two lines is 0.2~\kms.}
\label{compspec2}
\end{figure}

The \HCOplus\ lines are wider than the HCN lines ($\sigma_{\rm HCN} \sim 0.8$~$\sigma_{\rm HCO^+}$) over much of the area where they both show emission (see Figure \ref{compspec1} for example spectra). There are, however, some small areas where the HCN exhibits greater linewidths than the \HCOplus. An example of this can be seen in Figure \ref{compspec2}, which shows spectra from IRAS~4, a young Class 0 protostar. One can see that the HCN exhibits much stronger, more extended line wing emission than the \HCOplus\ causing the wider linewidth; it should be noted that even in this case, the central peak of the HCN is still narrower than that of the \HCOplus.

\section{\HCOplus\ outflow analysis}\label{sec:hcooutflows}

Multiple core catalogues of our Perseus sub-regions have been produced over the years \citep{2005A&A...440..151H,2001ApJ...546L..49S}, but for our analyses, we will be using the H07 SCUBA core catalogue. \citet{2010MNRAS.408.1516C} used the H07 catalogue in the identification of their \twelveCO\ outflow driving sources, and we will be comparing the outflow sources they have identified with those that we identify using \HCOplus. We aim to use our \HCOplus\ outflows to confirm protostellar status for those cores which have been previously defined as such by H07 using temperature and luminosity criteria. We also investigate if any of the starless cores show signs of molecular outflows, which would enable us to identify them as protostellar.

\subsection{Outflow identification}\label{sec:outflowiden}
We use three criteria to determine if a particular SCUBA H07 core is the driving source of an \HCOplus\ outflow:
\begin{enumerate}
\item We integrate over the \HCOplus\ spectral linewings, between the HWHM of the \CeighteenO\ line (as previously defined by \citealt{2010MNRAS.408.1516C}), and the furthest velocity extent of the linewing above the 1$\sigma$ noise level. This integration was done on an outflow-by-outflow basis to avoid any inaccuracies due to difference in velocity extents between outflows. We then plot these blue- and red-shifted integrated intensities as contours over an integrated intensity map of the line centre \HCOplus\ emission, and identify (by-eye) those SCUBA cores with spatially associated linewing contours.


\item We also look for evidence of linewing emission in the \HCOplus\ spectra at the SCUBA core central position for each of the H07 cores -- a core is defined as exhibiting linewing emission if the value of the \HCOplus\ emission at $\pm 1$~\kms\ from the \CeighteenO\ line centre velocity is at or above the 3$\sigma$ level. This is a similar criterion to that used by \citet{2010MNRAS.408.1516C} when identifying outflows from \twelveCO.


\item We also look for evidence of linewings (using the previous criterion) in all pixels immediately surrounding the H07 SCUBA core central position.
\end{enumerate}

We present our findings in Appendix A (Table \ref{tab:outflows}), and only consider H07 SCUBA cores as outflow driving sources if they meet at least 2 out of the 3 criteria. 24 of the 58 H07 SCUBA cores (41\%) that we investigated were determined to be outflow driving sources, based on the above criteria. This is a lower detection rate than the 69\% found by \citet{2010MNRAS.408.1516C}; this difference is to be expected, however, as the \HCOplus\ has a much higher critical density and excitation energy, and should therefore only be excited by the stronger outflows. In line with this, we find that the number of outflows driven by Class 0 sources (15/24) is almost twice that for Class I sources (9/24), which suggests that the more energetic Class 0 sources produce more favourable conditions to excite \HCOplus. 

All of the cores that we identify as likely outflow driving sources have previously been identified using \twelveCO\ \citep{2010MNRAS.408.1516C, 2007A&A...472..187H}, so we are unable to positively identify any of the previously-classified `starless' cores as protostellar. However, this \HCOplus\ outflow analysis has allowed us to confirm independently the protostellar status of many of the H07 sources.

\begin{figure}
\centerline{\includegraphics[width=8.0cm]{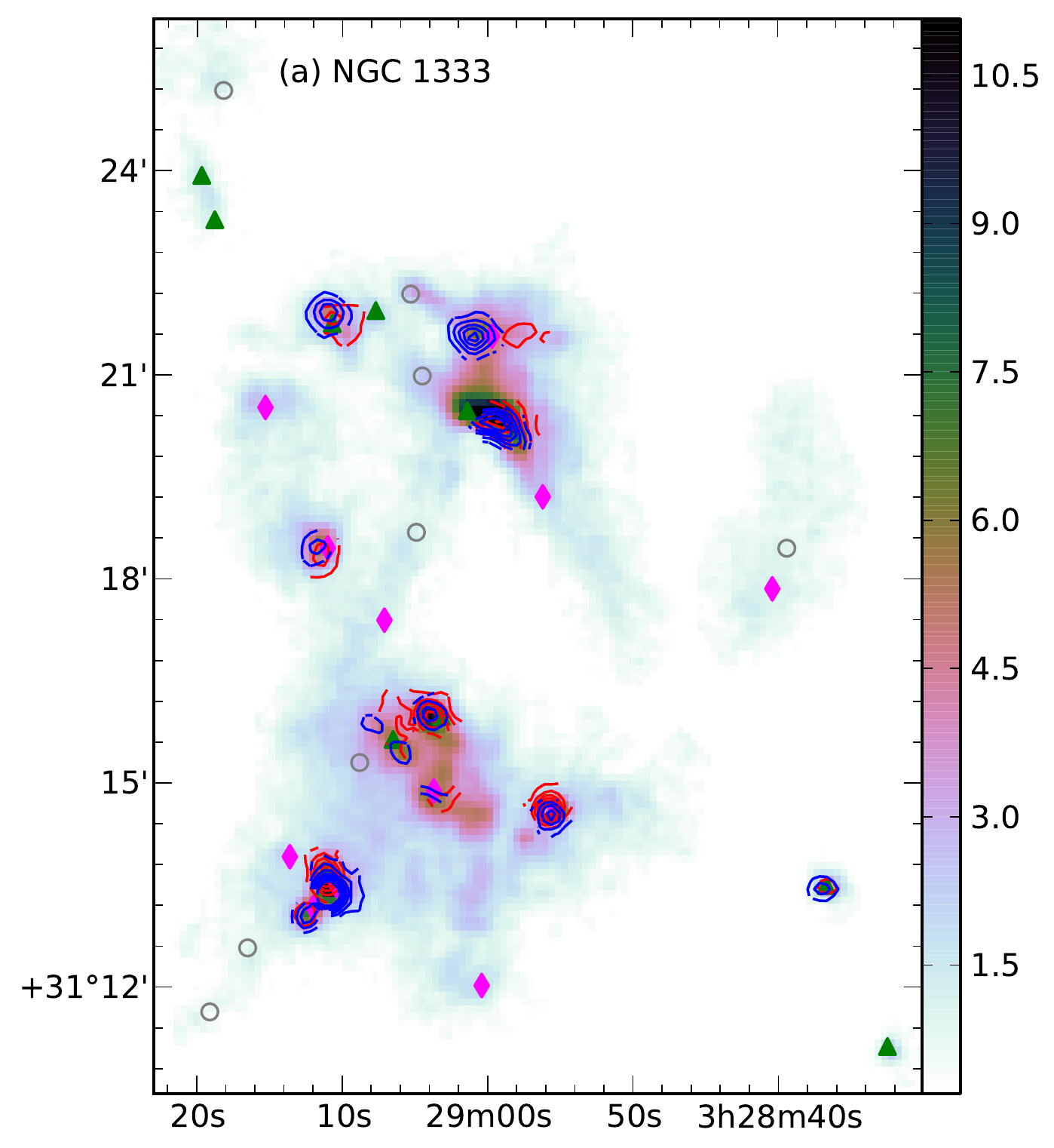}}
\smallskip
\centerline{\includegraphics[width=8.0cm]{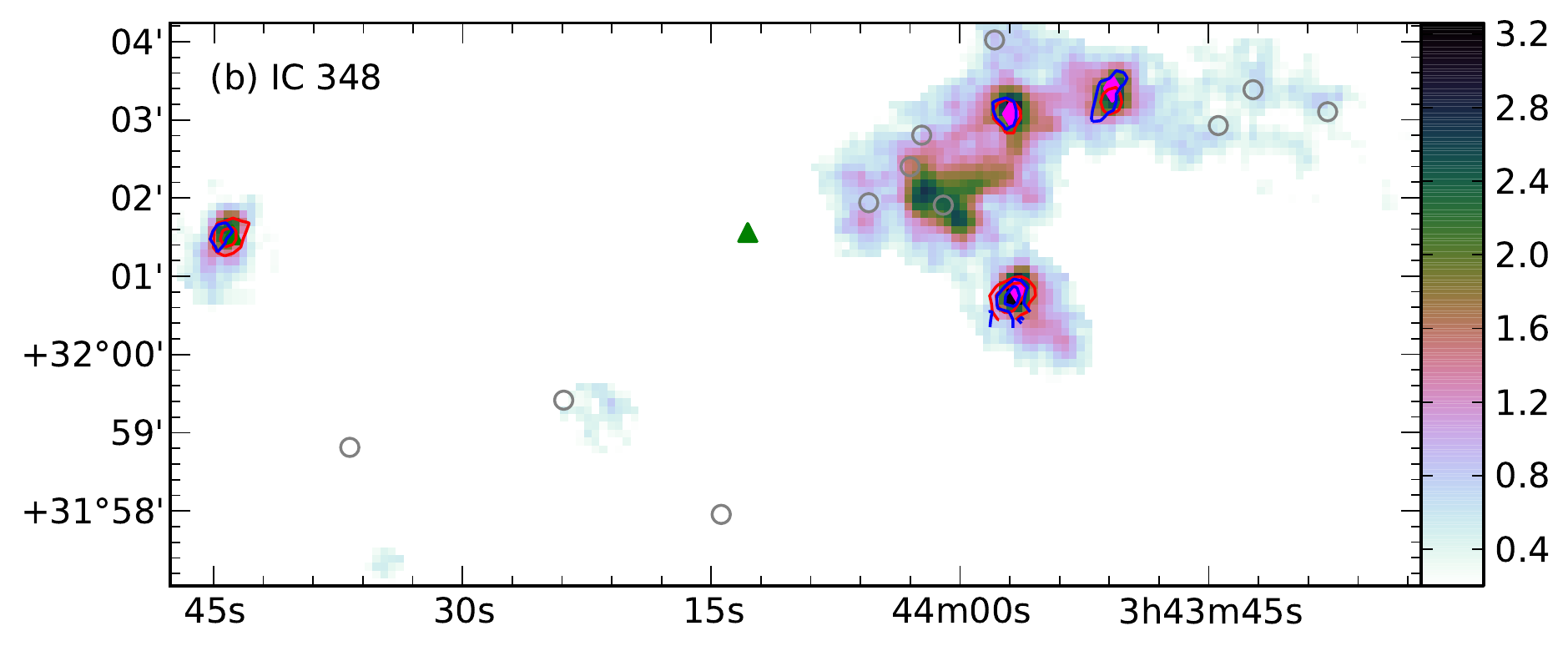}}
\smallskip
\centerline{\includegraphics[width=7.0cm]{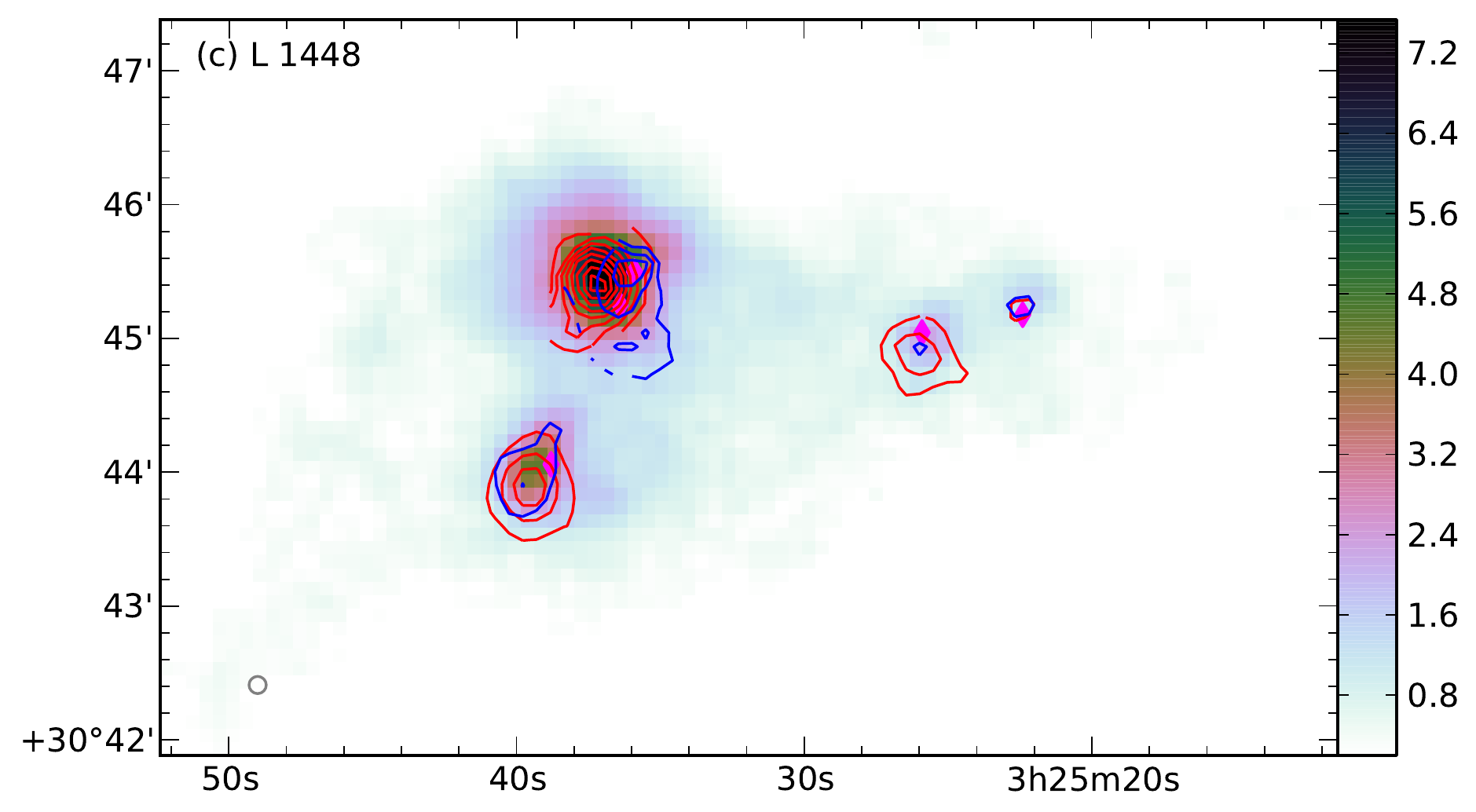}}
\smallskip
\centerline{\includegraphics[width=7.0cm]{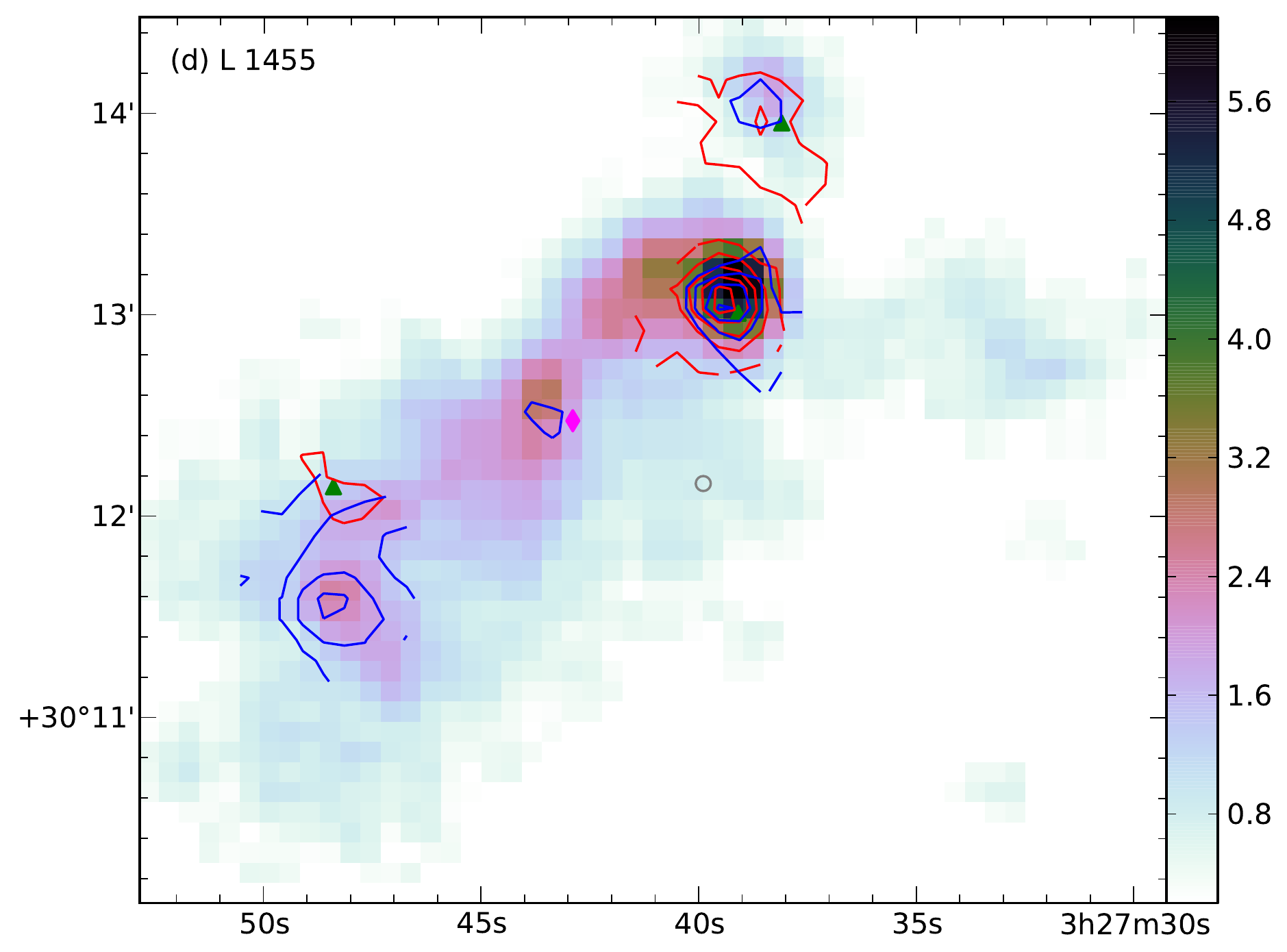}}
\caption{Integrated intensity \HCOplus\ maps for (a) NGC1333, (b) IC348, (c) L1448 and (d) L1455.  H07 SCUBA core positions are overlaid as in Figure \ref{ngc1333iwc}, outflows from each core are overlaid as red and blue contours, from 0.5 K~\kms\ in 0.5 K~\kms\ increments. The emission from each outflow is integrated over the linewings on a core-by-core basis, between the HWHM of the \CeighteenO\ line for that core and the furthest extent of the linewing above the 1$\sigma$ noise level.}
\label{outflows}
\end{figure}

We present an overview of our \HCOplus\ outflows over the 4 Perseus sub-regions in Figure \ref{outflows}, illustrating that the \HCOplus\ outflows are fairly compact and do not extend far from their driving source. A comparison can be made between \HCOplus\ and \twelveCO\ outflow spatial extents by comparing our results with Figure 1 in \citet{2010MNRAS.408.1516C}.
%


\subsection{Calculation of outflow properties}
We calculated properties for our \HCOplus\ outflows using methods first outlined in \citet{1990ApJ...348..530C}, and then expanded upon by \citet{2002A&A...383..892B}. We compare our values with those calculated from the \twelveCO\ emission by \citet{2010MNRAS.408.1516C} using the same method to quantify the differences between these two outflow tracers. 

\subsubsection{Mass}

\begin{table}
\caption{Properties of \HCOplus, HCN and \twelveCO\ used in this paper to calculate column densities and gas masses. The $\mu$, $T_0$ and $\nu_{10}$ values are obtained from the Leiden Atomic and Molecular Database -- LAMDA \citep{2005A&A...432..369S}, and the relative abundances were obtained from \citet{2004A&A...416..603J} and \citet{1982ApJ...262..590F}. \changed{It should be noted that the \HCOplus\ and HCN abundances quoted here are average protostellar envelope abundances.}}
\begin{center}
\begin{tabular}{ccccc}
\hline
Molecule & $\mu$/D & $T_0$/K & $\nu_{10}$/G~Hz & $X_{\rm gas}$\\ \hline
\HCOplus\ & 3.90  & 4.28 & 89.189 & $6.6 \times 10^{-9}$\\
HCN & 2.98 & 4.25 & 88.632 & $1.7 \times 10^{-9}$ \\
\twelveCO\ & 0.112 & 5.53 & 115.271 & $1.0 \times 10^{-4}$\\
\hline
\end{tabular}
\end{center}
\label{coldenprops}
\end{table}

We first calculate the mass of the outflow, which will be used to derive other parameters such as the momentum and kinetic energy of the outflow. The column density of \HCOplus, $N_{\rm HCO^+}$, can be calculated from the corresponding $J= 4-3$ emission using the formula below. Following \citet{2000..RW..book}, and assuming LTE, we use:
\begin{equation} \label{eq:col_density}
N_{total}=\frac{3\epsilon_0k^2T_{\rm ex}}{h\upi^2\mu_e^2\nu_{10}^2(J+1)^2}\exp\left(\frac{(J+1)(J+2)T_0}{2T_{\rm ex}}\right) \frac{\int{T_A^*\,{\rm d}v}}{\eta_{\rm MB}},
\end{equation} 
where $T_{\rm ex}$ is the excitation temperature, $\mu_e$ is the rotational electric dipole moment, and $\nu_{10}$ is the J=1-0 transition frequency for that particular molecule. $\int{T_A^*{\rm d}v}$ is the integrated intensity over the total velocity extent of the outflow. Substituting values for \HCOplus\ as in Table \ref{coldenprops}, we obtain: 

\begin{equation} \label{eq:hcocol_density}
N_{\rm HCO^+}=5.88\times 10^{13}{\rm m}^{-2} \frac{T_{\rm ex}}{\exp({-42.8 {\rm K}/T_{\rm ex}})} \int{T_A^*\,{\rm d}v},
\end{equation} 
where $T_{\rm ex}$ is taken to be 50~K, following both \citet{2010MNRAS.408.1516C} and H07. The column density is then converted into a mass using the following equation:
\begin{equation} \label{eq:mass}
M_{\rm HCO^+}= D^2(\Delta \alpha \Delta \beta)\mu_{\rm H2} m_{\rm H} \frac{N_{\rm HCO^+}}{X_{\rm HCO^+}}.
\label{mass}
\end{equation}
We take the distance to the cloud $D = 250$ pc (following \citealt{2010MNRAS.408.1516C}) and mean molecular mass per hydrogen molecule $\mu_{\rm H_2}=2.72$ (taking Helium into account), a pixel size ($\Delta \alpha$ and $\Delta \beta$) of 6\arcsecs\ and the relative abundance of \HCOplus\ to hydrogen $X_{\rm HCO^+}$ is as in Table \ref{coldenprops}. 

While it would be preferable to calculate the optical depth of the \HCOplus\ in the line wings and perform an opacity correction to the mass, we do not have data from an optically thin isotopologue (e.g. \HthirteenCOplus). We therefore assume that the \HCOplus\ is optically thin in the linewings, while keeping in mind that our results are likely to be lower limits. 

It is also possible that our assumption of LTE may be incorrect, and subthermal excitation may cause the column densities we calculate to be lower than expected.

\subsubsection{Momentum and Kinetic Energy}
The approach we use here is one of the most common methods -- the so-called ``$v_{\rm max}$" method \citep{1992A&A...261..274C}, where the total outflow momentum along the jet axis is given by:
\begin{equation}
P_{\rm out} = \left(M_{\rm r} v_{\rm max,r}\right) + \left(M_{\rm b} v_{\rm max,b}\right),
\end{equation}
where v$_{\rm max,r}$ and v$_{\rm max,b}$ are the maximal velocity extents of the red and blue outflow lobes respectively. We have taken these to be the velocities furthest away on either side from the line centre velocity (defined by \CeighteenO), where the linewing emission is still at or above the 1$\sigma$ noise level. $M_{\rm r}$ and $M_{\rm b}$ are the masses of the red and blue outflow lobes respectively, calculated using equation \ref{eq:mass}.

Similarly, the kinetic energy of the outflow is given by:
\begin{equation}
E_{\rm out} = \frac{1}{2} \left(M_{\rm r} v^2_{\rm max,r}\right) + \frac{1}{2} \left(M_{\rm b} v^2_{\rm max,b}\right).
\end{equation}
A consideration in the calculation of these values is the effect of outflow inclination -- if a lobe is inclined at an angle \emph{i} to the line of sight, then the standard correction is to divide the momentum by cos\emph{i} and the kinetic energy by  cos$^2$\emph{i}; this would have a significant effect on the values that we obtain. Some authors \citep{2002A&A...383..892B} apply a single correction to all outflows based on the assumption that the outflows are oriented in a random isotropic manner. We do not correct for inclination, and refer the reader to \citet{2013A&A...556A..76V} which contains a table of corrections for different inclination values. 


The ``$v_{\rm max}$" method will overestimate the momenta and kinetic energies of the outflows compared to the method used by \citet{2010MNRAS.408.1516C}, which involves the integral of the outflow mass over the velocity range. We have chosen this method as \citet{1992A&A...261..274C} state that the ``$v_{\rm max}$" method is more suitable for those outflows that are inclined to the line of sight, which is the case for many of our outflows. \changed{However, \citet{2013A&A...556A..76V} state that the use of different methods to calculate outflow properties will cause a scatter in the results by at most a factor of 6; they estimated this using \twelveCO\ $J = 3 \to 2$ emission, which has large velocity extents. Our \HCOplus\ outflows are generally fairly compact in velocity -- with a maximum velocity extent of $\sim 2 - 3$ \kms -- and there is} \changed{very little variation in velocity over the spatial range of the outflow. The detailed outflow kinematic structure will thus have less of an effect on the momentum and energy values for the \HCOplus\ outflows than would be the case for the \twelveCO\ outflows, which extend to much greater velocities. We therefore expect our method to overestimate the momenta and energies by at most a factor of 6.}

\subsubsection{Dynamical time}
Using the ``$v_{\rm max}$" method, the outflow dynamical time is the time taken for the bow shock travelling at the maximum velocity of the flow to travel the projected lobe length:
\begin{equation}
t_{\rm dyn} = \frac{L_{\rm lobe}}{v_{\rm max}},
\end{equation}
where the projected lobe length $L_{\rm lobe}$ is the average of the red and blue outflow lengths (as measured from the position of the central H07 driving source); and v$_{\rm max}$ is the average of the red and blue outflow maximal velocities. The dynamical time can be used as a first approximation for the age of the outflow, although it may underestimate the true age of the outflow by up to an order of magnitude in some cases \citep{1991MNRAS.252..442P}.

In addition, \citet{2004A&A...416..603J} state that while dynamical timescales are not necessarily good indicators of the true age of the protostellar driving sources, they are useful as an `order of magnitude' estimate when discussing chemical evolution in shocks within the outflows.

\subsubsection{Driving force and mechanical luminosity}

The driving force or outflow momentum flux $F_{\rm out}$ is a key input parameter for outflow models, and is very important in the investigation of the driving mechanisms of outflows. It is given by:
\begin{equation}
F_{\rm out} = \frac{P_{\rm out}}{t_{\rm dyn}},
\end{equation}
The mechanical luminosity of the outflow is given by:
\begin{equation}
L_{\rm out} = \frac{E_{\rm out}}{t_{\rm dyn}},
\end{equation}

We present a summary of our results in Table \ref{tab:hcoaverages}, which show the average properties for all outflows in each sub-region. This allows us to link the large-scale properties of the sub-regions (e.g. turbulence and source density) with differences in their outflow properties. 

\begin{table*}
\caption{Average outflow properties for all \HCOplus\ outflows in the Perseus region are presented, as well as a breakdown of these for each sub-region. The average properties for Class 0 and Class I sources in each sub-region (if more than one source is found for each protostellar class) are also presented here. Column 2: Total mass in both outflow lobes; Column 3: Total outflow momentum; Column 4: Total kinetic energy in the outflow; Column 5: Maximal velocity of the outflow (averaged over the red and blue lobes); Column 6: Dynamical time of the outflow; Column 7: Total momentum flux or driving force of the outflow; Column 8: Total mechanical luminosity of the outflow. (Errors on the last digit(s) of the average values are given in the brackets.)}
\centering
\begin{tabular}{cccccccc}
\hline
Subset  & $M_{\rm out}$  & $P_{\rm out}$ & $E_{\rm out}$ & $v_{\rm max}$ & $t_{\rm dyn}$ & $F_{\rm out}$ 			&  $L_{\rm out}$\\
	&  (\Msun)  	 & (\Msun\ \kms) &  (10$^{36}$J) & (\kms)	& ($10^3$yr)   & (10$^{-5}$ \Msun\ \kms\ yr$^{-1}$)   & (10$^{-2}$\Lsun)\\
\hline
     &                     &      &  \bf{NGC~1333}              &         &     &      \\ \hline
Average & 0.010(2) & 0.034(11) & 0.15(8) & 3.0(1.0) & 14.2(2.0) & 0.36(17) & 0.15(10) \\
Class 0 & 0.009(2) & 0.038(18) & 0.20(13) & 3.3(1.6) & 12.0(2.2) & 0.5(3) & 0.24(17) \\
Class I & 0.011(3) & 0.029(8) & 0.08(2) & 2.6(6) & 16.8(3.2) & 0.18(5) & 0.04(2) \\ \hline
     &            &             &          \bf{IC~348}              &         &     &      \\ \hline
Average & 0.0036(5) & 0.009(2) & 0.024(8) & 2.4(7)  & 15.7(2.9) & 0.08(3) & 0.018(9) \\
Class 0 & 0.0039(5) & 0.010(3) & 0.025(11) & 2.3(7) & 17.4(3.4) & 0.08(4) & 0.019(11) \\ \hline
     &            &             &          \bf{L~1448}              &         &     &      \\ \hline
Average & 0.010(2) & 0.027(8) & 0.08(3) & 2.3(5)    & 17.0(8) & 0.15(5) & 0.036(13) \\ 
(Class 0) & 	   &		   &	    &	      &		&            \\ \hline
     &            &             &          \bf{L~1455}             &         &     &      \\ \hline
Average & 0.006(2) & 0.014(4) & 0.033(10) & 2.1(4)  & 11.7(1.3) & 0.13(4) & 0.026(7) \\
Class I & 0.0076(14) & 0.018(3) & 0.043(7) & 2.13(13) & 10.1(4) & 0.17(2) & 0.034(4) \\ \hline
     &                     &      &  \bf{Perseus}              &         &     &      \\ \hline
Average & 0.007(3) 	& 0.021(11)	&  0.07(6)	&  2.6(1.0) & 14.7(2.2)	&  0.18(6)			& 0.06(6) \\
Class 0 & 0.008(3) 	& 0.025(14)  	&  0.10(9) 	& 2.7(1.2)  & 15.5(2.0)	&  0.24(3) 			&  0.10(7) \\
Class I & 0.009(2) 	& 0.024(8)   	& 0.06(3)	&  2.5(5)   & 14.0 (4.0)	&  0.18(1)		&  0.037(4) \\ \hline
\end{tabular}
\label{tab:hcoaverages}
\end{table*}

\subsection{Comparisons between protostellar classes}
We first analyse the differences in properties between outflows with Class 0 and Class I driving sources to test the following hypothesis: Class 0 sources are expected to power younger, faster, more massive and more energetic outflows than Class I sources due to a decrease in mass accretion rate \citep{1996A&A...311..858B}. We present a comparison of our results for Perseus overall, as well as for the individual sub-regions, in Table \ref{tab:hcoaverages}.

In our analyses, we perform Kolmogorov--Smirnoff (KS) tests to determine if datasets were drawn from the same underlying distribution. The KS statistic (or $p$-value) gives a value that is 1 minus the confidence level with which the null hypothesis (that the two samples originate from the same distribution) may be rejected. Generally, if the $p$-value is $< 5\%$, we can say confidently that the two samples originate from different underlying distributions.

\subsubsection{Mass, momentum and kinetic energy}
We find no statistically significant difference between the masses, momenta and kinetic energies of outflows from Class 0 and Class I sources for the Perseus region as a whole -- a KS-test gives a $p$-value of 75\% for the masses, giving a high probability that they are drawn from the same underlying distribution. 

Upon closer inspection, however, we find that the Class 0 values are down-weighted by a number of significantly less massive, less energetic outflows from IC~348, while a few massive, energetic Class I outflows in NGC~1333 increase the Perseus average overall (see Table \ref{tab:hcooutprops}). \changed{Indeed, a KS-test gives a $p$-value of 4.4\% for a comparison of the NGC~1333 and IC~348 outflow masses, which supports our conclusion that there is a statistically significant difference between the outflows from the two regions.} This reflects the environmental influence on outflows.

NGC~1333 is younger than IC~348, with a higher density of protostellar sources driving outflows; this will cause an overestimate of the mass, and hence of the momentum and energy associated with a particular outflow driving source due to confusion with surrounding sources. In addition, NGC~1333 will likely have a larger reservoir of gas surrounding the outflow sources compared to IC~348, resulting in greater availability of gas for entrainment in outflows in NGC~1333.

\subsubsection{Dynamical time, driving force and mechanical luminosity}
\changed{Although there appears to be little difference between the average dynamical times for Class 0 and Class I sources (as calculated by \HCOplus), a KS-test gives a $p$-value of 6.9\%, implying that there is a 93\% chance they originate from different underlying distributions. However, we caution that this may be a coincidence as KS-tests} \changed{show no statistically significant difference between the Class 0 and I values of both the lobe lengths ($p$-value of 39\%) and the maximal velocities ($p$-value of 56\%). This is likely due to the large degree of subjectivity (and hence significant errors) involved in estimating the lobe lengths and maximal velocities used to calculate the dynamical times.}

\changed{We have found that the dynamical time we calculate for the well-known IRAS~2A source (H07-44) -- $6.7 \times 10^3$~years -- is a match for a previous value ($5.0 \times 10^3$~years) calculated by others \citep{1998A&A...335..266B,2004A&A...416..603J} using different molecules, once the value has been corrected for the difference in distances used. However, due to the large uncertainties attached to the values we calculate, we should not attach too much significance to this result.}

\changed{The average values of outflow driving force and luminosity are greater for Class 0 than Class I sources, which appears to follow the theory of outflows driven by Class 0 sources being more powerful than those driven by Class I sources. However, KS-tests show no statistically significant difference between the Class 0 and Class I values of driving force ($p$-value of 84\%) and luminosity ($p$-value of 66\%). This result is likely linked to the lack of distinguishability of the outflow masses and velocities between Class 0 and I sources.}

\changed{In conclusion, it appears that the \HCOplus\ outflow properties do not allow us to distinguish between Class 0 and I driving sources with any degree of statistical significance.}

\subsection{\HCOplus\ -- \twelveCO\ comparison}
\begin{figure*}
\centerline{\includegraphics[width=8.0cm]{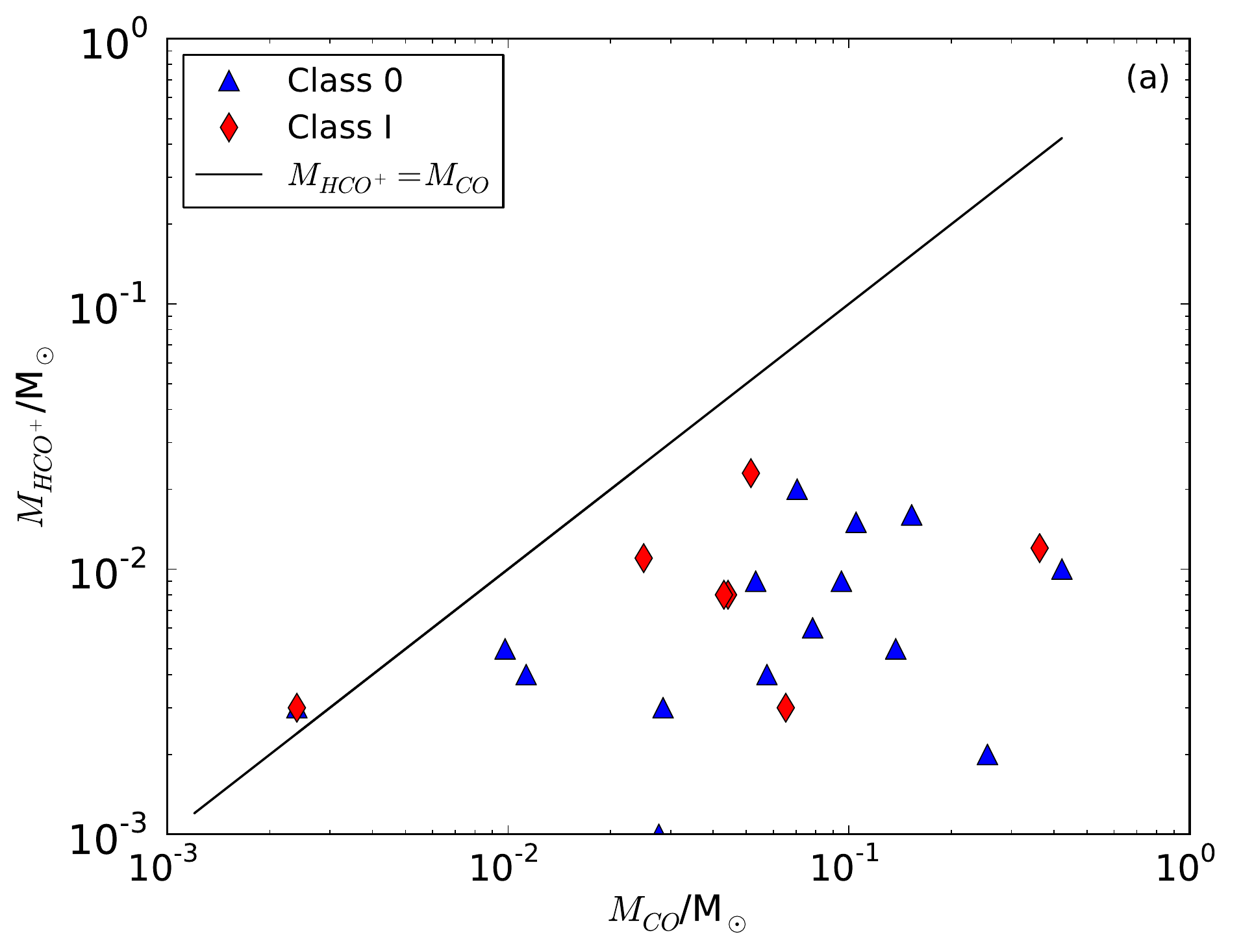}\qquad
	     \includegraphics[width=8.0cm]{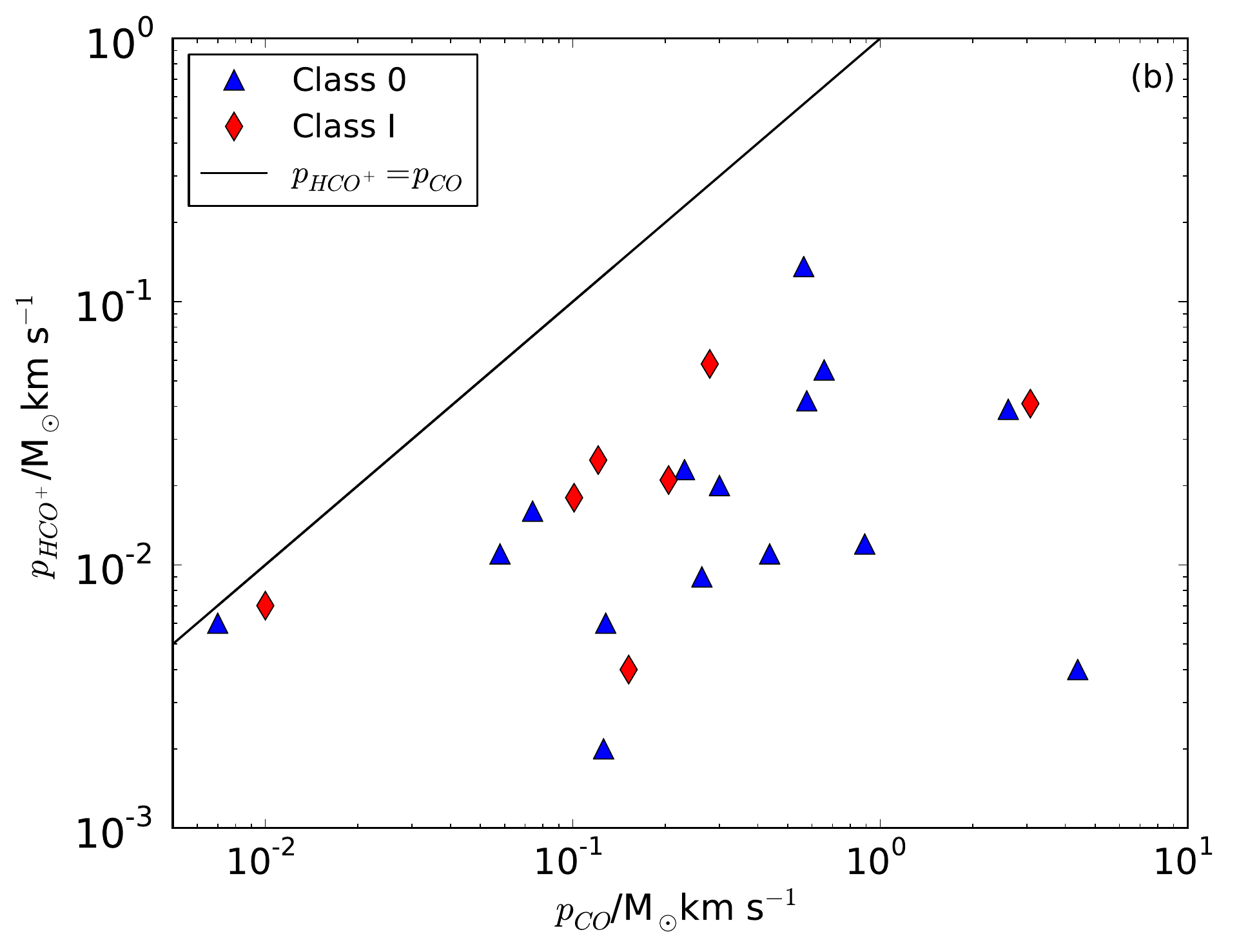}}
\smallskip
\centerline{\includegraphics[width=8.0cm]{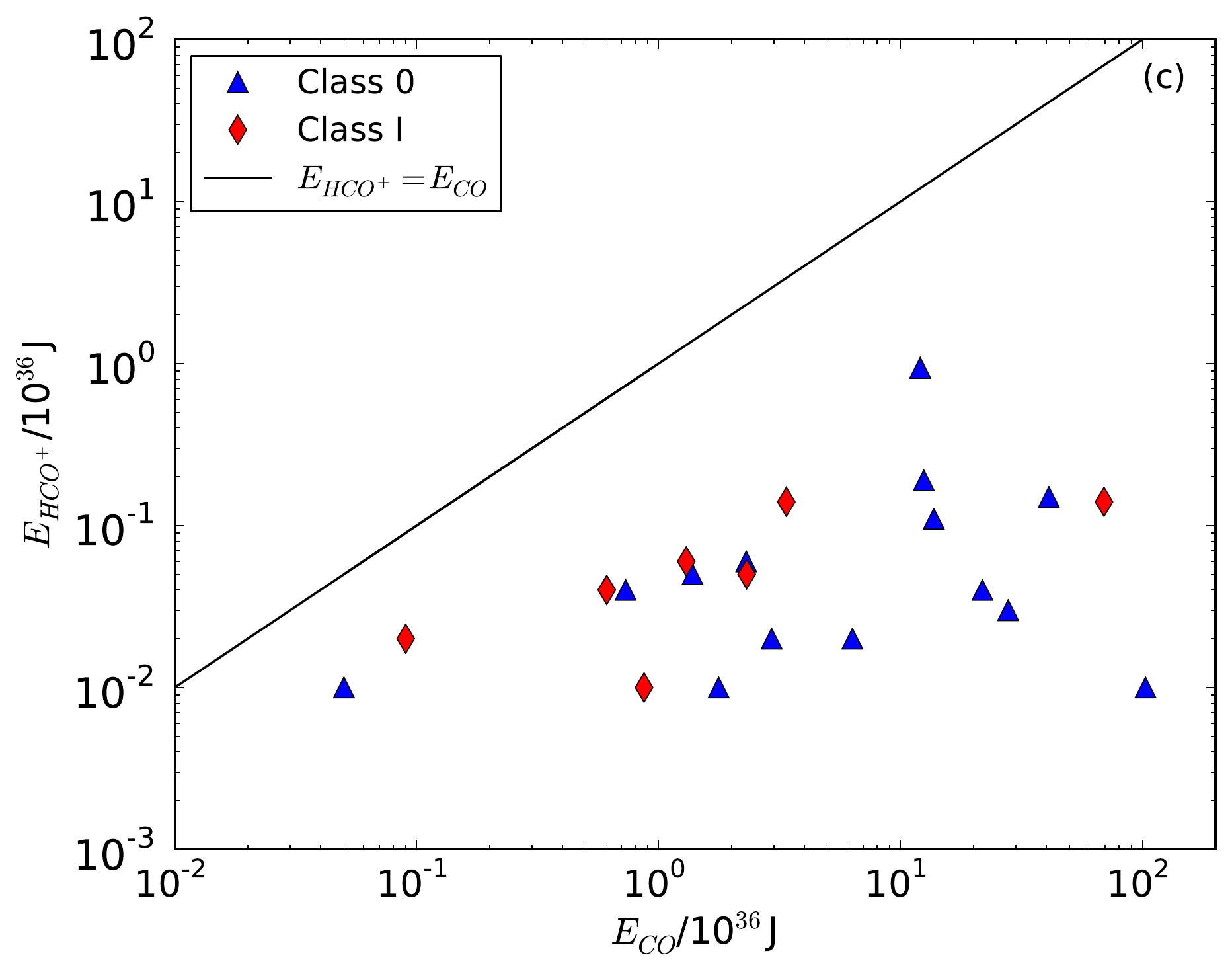}\qquad
	     \includegraphics[width=8.0cm]{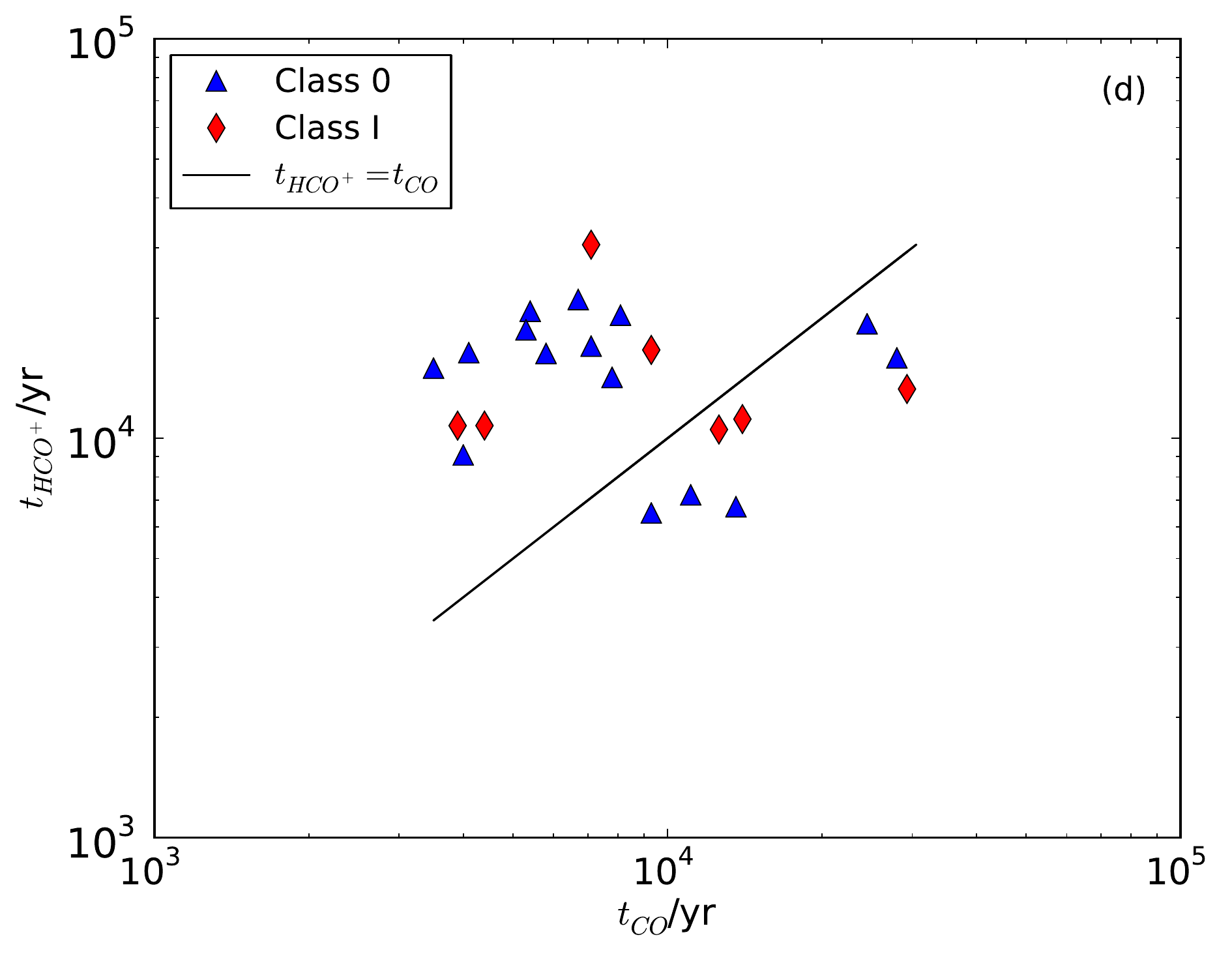}}
\caption{\HCOplus\ outflow properties are plotted against the corresponding values derived from \twelveCO: (a) Outflow mass, (b) Outflow momenta, (c) Outflow kinetic energy and (d) Outflow dynamical time. The \twelveCO\ values are obtained from \citet{2010MNRAS.408.1516C}; the points are separated into Class 0 (blue triangles) and Class I (red diamonds).}
\label{fhcoco}
\end{figure*}

We compare our \HCOplus\ outflow properties to those calculated by \citet{2010MNRAS.408.1516C} from \twelveCO\ data. These two molecules are linked chemically -- the primary routes of \HCOplus\ formation all involve CO -- and they are expected to trace broadly similar regions within the outflows \citep{2004A&A...416..603J, 2004MNRAS.351.1054R}. Our comparisons are illustrated in Figure \ref{fhcoco}.

\subsubsection{Mass, momentum and kinetic energy}
We find that the masses of the \HCOplus\ outflows are much lower than the corresponding \twelveCO\ values, \changed{with 2 exceptions; these are both low-mass sources from IC~348, and the associated errors involved in the mass calculations are proportionally higher.} We note that \citet{2010MNRAS.408.1516C} performed an opacity correction to their \twelveCO\ masses using \thirteenCO\ data, which would have increased the masses that they obtained. We were unable to perform such an opacity correction, which would result in the \HCOplus\ masses we obtain being lower than expected; this may account for some of the discrepancy between the \HCOplus\ and \twelveCO\ outflow masses. \changed{We also note that the higher critical density of \HCOplus\ will result in the emission tracing only the densest parts of the outflows, while \twelveCO\ traces the bulk of the outflow. We therefore expect the total mass traced by \HCOplus\ to be lower than that traced by \twelveCO.}

There is a significant difference in the maximal velocities $v_{\rm max}$ of \HCOplus\ and \twelveCO\ -- $v_{\rm max}$ of \HCOplus\ is on average $(2.5 \pm 0.3)$ \kms, while that of \twelveCO\ is $\sim 14$ \kms. This is a result of the \twelveCO\ having a much larger SNR in the linewings than \HCOplus, allowing it to be detected out to greater velocity extents than \HCOplus. This causes an increasing degree of discrepancy between the \HCOplus\ and \twelveCO\ outflow properties with increasing $v$-dependence; there are differences of up to 4 orders of magnitude between $E_{\rm HCO^+}$ and $E_{\rm ^{12}CO}$, as can be seen in Figure \ref{fhcoco}.

\subsubsection{Dynamical times, driving force and mechanical luminosity}
From Figure \ref{fhcoco}d we can see that a higher proportion of the sources lie to the left of the line of equality, but on average, the \HCOplus\ and \twelveCO\ dynamical times are with a factor of 2 - 3 of one another. The \HCOplus\ mainly traces the inner parts of the outflows and also the slow-moving components compared with the \twelveCO, as can be seen from the spatial extents of the outflows in Figure \ref{outflows} and the maximal velocities in Table \ref{tab:hcooutprops}. Therefore, the dynamical times calculated from the two molecules are similar because both the lengths and velocities of the outflows scale in roughly the same way with the minimum column density traced.

We find that the \HCOplus\ gives a significant underestimate of the driving forces and luminosities of the outflows when compared with \twelveCO\ due, once again, to the dependence on $v_{\rm max}^2$ and $v_{\rm max}^3$ for the driving force and luminosity respectively.

\subsection{Outflow driving force trends}
\citet{1996A&A...311..858B} found that the outflow momentum flux $F_{\rm CO}$ correlates well with the infalling envelope mass in the early stages of protostellar evolution, suggesting an underlying relationship between stellar outflow activity and protostellar evolution. \citet{2013MNRAS.431.1719M} suggest that \changed{outflow properties are determined by the accretion of mass onto the protostar, via the circumstellar disk; this process drives the outflow and is, in turn, related to the source envelope mass.} This has been confirmed by other sources \citep{2004A&A...426..503W,2010MNRAS.408.1516C}.

The bolometric luminosity $L_{\rm bol}$ of the protostar is dominated by the accretion luminosity, which is dependent on the accretion process from the circumstellar disk. There is, therefore, expected to be a correlation between $L_{\rm bol}$ and the envelope mass $M_{\rm env}$ of the source, \citep{1996A&A...311..858B,2004A&A...426..503W}. It also follows that there should be a correlation between the driving force and $L_{\rm bol}$ of the outflow driving source.

We therefore investigate the correlation of the outflow driving force $F_{\rm HCO^+}$ for our outflows, with the envelope masses $M_{\rm env}$ and bolometric luminosities $L_{\rm bol}$ (as determined by H07 from SCUBA continuum data) of the corresponding driving sources. The H07 $M_{\rm env}$ values have been corrected to account for the differences in distance estimates used for the Perseus molecular cloud -- H07 used 320~pc while we have followed \citet{2010MNRAS.408.1516C} and used 250~pc.

\begin{figure}
\centerline{\includegraphics[width=8.0cm]{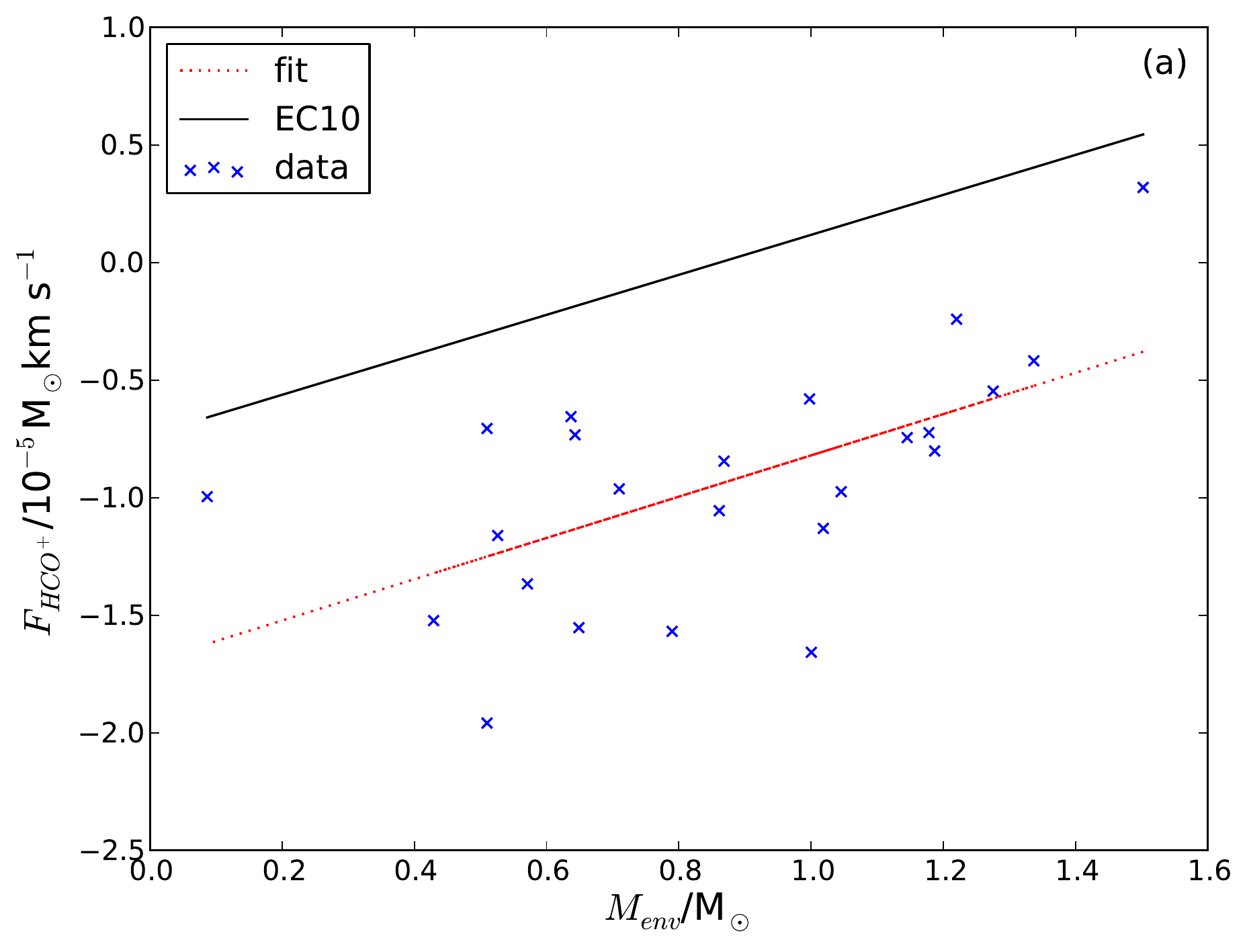}}
\smallskip
\centerline{\includegraphics[width=8.0cm]{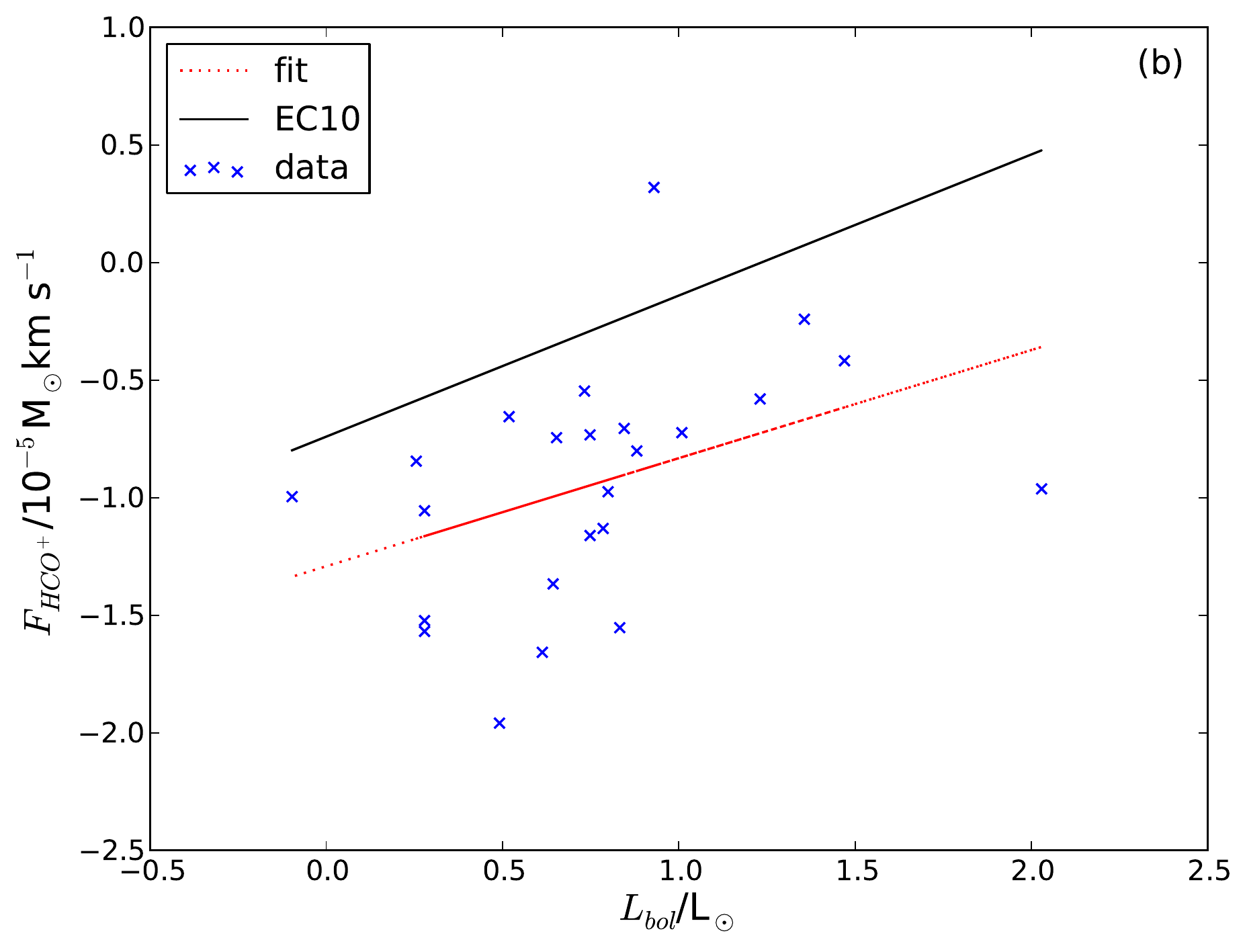}}
\caption{(a) \HCOplus\ outflow driving force $F_{\rm HCO^+}$ against envelope mass ($M_{\rm env}$)of the driving source; (b) \HCOplus\ outflow driving force $F_{\rm HCO^+}$ bolometric luminosity $L_{\rm bol}$ of the driving source. The envelope masses and bolometric luminosities were calculated from SCUBA data by \citet{2007A&A...468.1009H}. The red (dotted) lines show the fits to our data, and the black (solid) lines show the fits to \twelveCO\ data calculated by \citet{2010MNRAS.408.1516C} and are included to enable a comparison of their gradients with our data.}
\label{fhcotrends}
\end{figure}

We find the following relations: $(F_{\rm HCO^+}) \propto M_{\rm env}^{0.9 \pm 0.4}$ and $(F_{\rm HCO^+}) \propto L_{\rm bol}^{0.5 \pm 0.2}$. Both of these relations match well with the corresponding $F_{\rm CO}$ relations obtained for the same region by \citet{2010MNRAS.408.1516C}, albeit with some scatter in values, as can be seen in Figure \ref{fhcotrends}a. This similarity between \twelveCO\ and \HCOplus\ shows that the two molecules are tracing similar parts of the outflow and are indeed correlated with one another. 

Our value of the exponent in the $F_{\rm HCO^+}$--$M_{\rm env}$ relation also matches that found by \citet{2013A&A...556A..76V} for outflow sources in Ophiuchus of ($0.86 \pm 0.19$), which has qualitatively similar levels of turbulence and clustering to Perseus. This similarity in the correlation relation between the two completely separate regions points to similar outflow driving mechanisms.

\section{HCN Outflow analysis}\label{sec:hcnoutflows}

\subsection{Outflow identification}
The outflows in HCN were identified using the same basic criteria as for \HCOplus. However, since the HCN linewidths are generally narrower and the signal is lower than the \HCOplus, we relaxed criteria (ii) and (iii) slightly: a $2\sigma$ detection (instead of $3\sigma$) at $\pm 1$ \kms\ from the line centre was deemed sufficient for a positive outflow linewing identification.

We detected a total of 10 outflows: 6 in NGC~1333, 3 in L~1448 and 1 in L~1455. We did not detect any outflows in IC~348, a further indication that it is much less active than the other three sub-regions investigated. Of these 10 outflows, 7 are associated with Class 0 sources and the remaining 3 with Class I sources, showing that HCN is excited in the youngest, most energetic outflows.

\subsection{HCN outflow properties}
With so few outflows, separating them by evolutionary stage of their driving source will not yield statistically significant results. We therefore only consider the average values for Perseus as a whole when comparing the HCN outflow properties with those calculated in the previous section for \HCOplus.

We calculated the HCN outflow masses using Equation \ref{eq:col_density}, substituting values for HCN from Table \ref{coldenprops}. The other outflow properties were calculated using the same methods as in the previous section and are presented in Appendix C (Table \ref{tab:hcnoutprops}). 

\begin{figure*}
\centerline{\includegraphics[width=8.0cm]{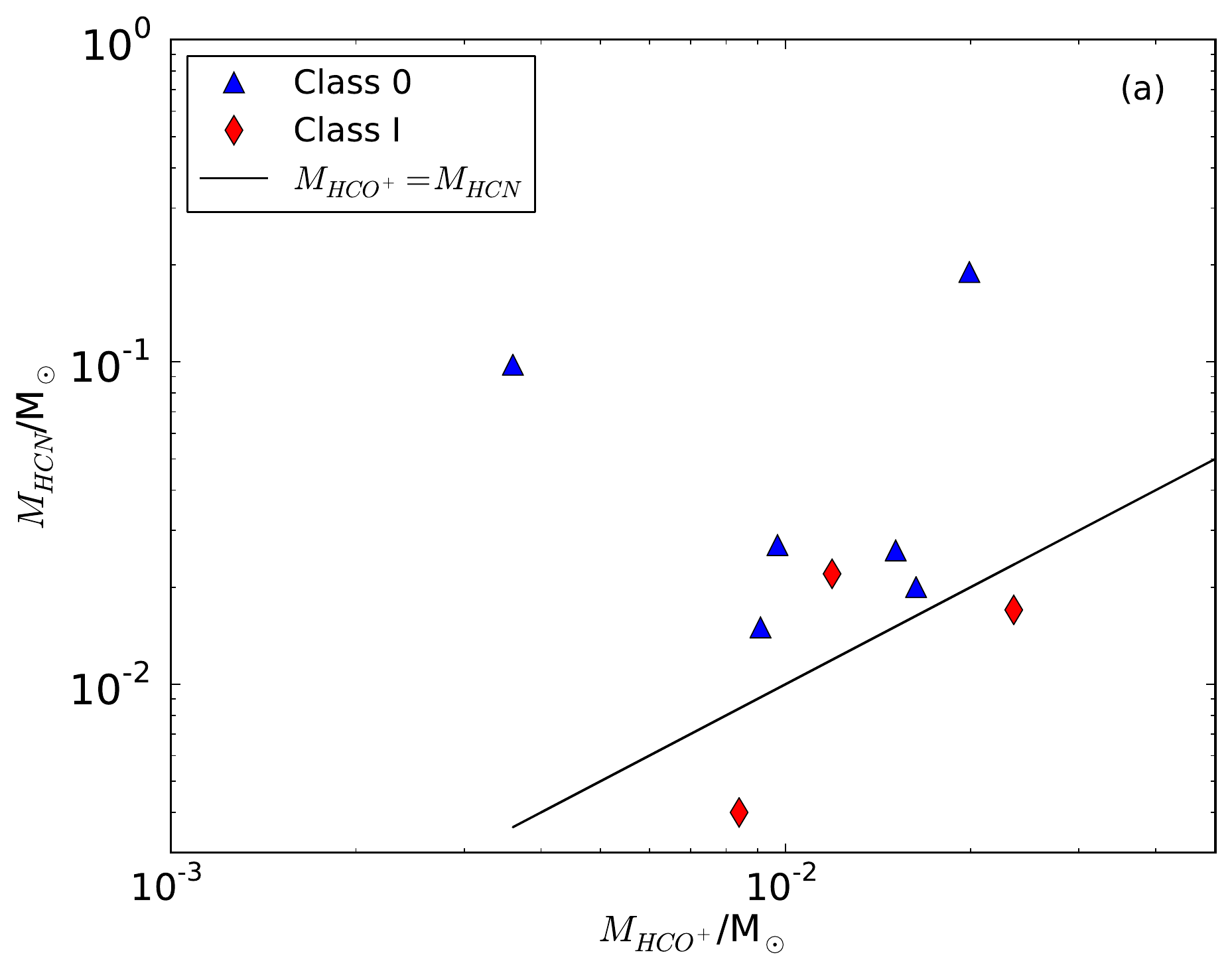}\qquad
	     \includegraphics[width=8.0cm]{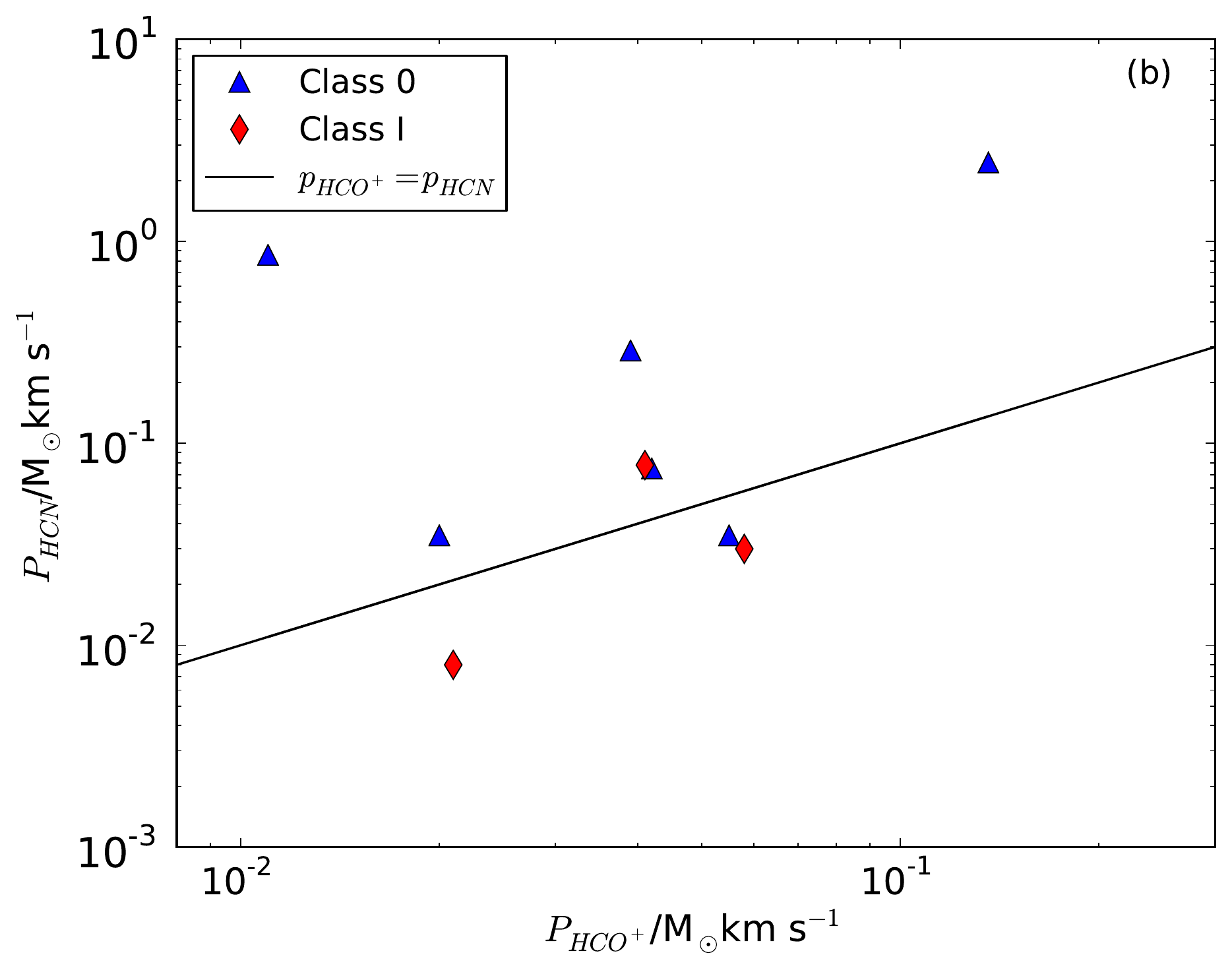}}
\smallskip
\centerline{\includegraphics[width=8.0cm]{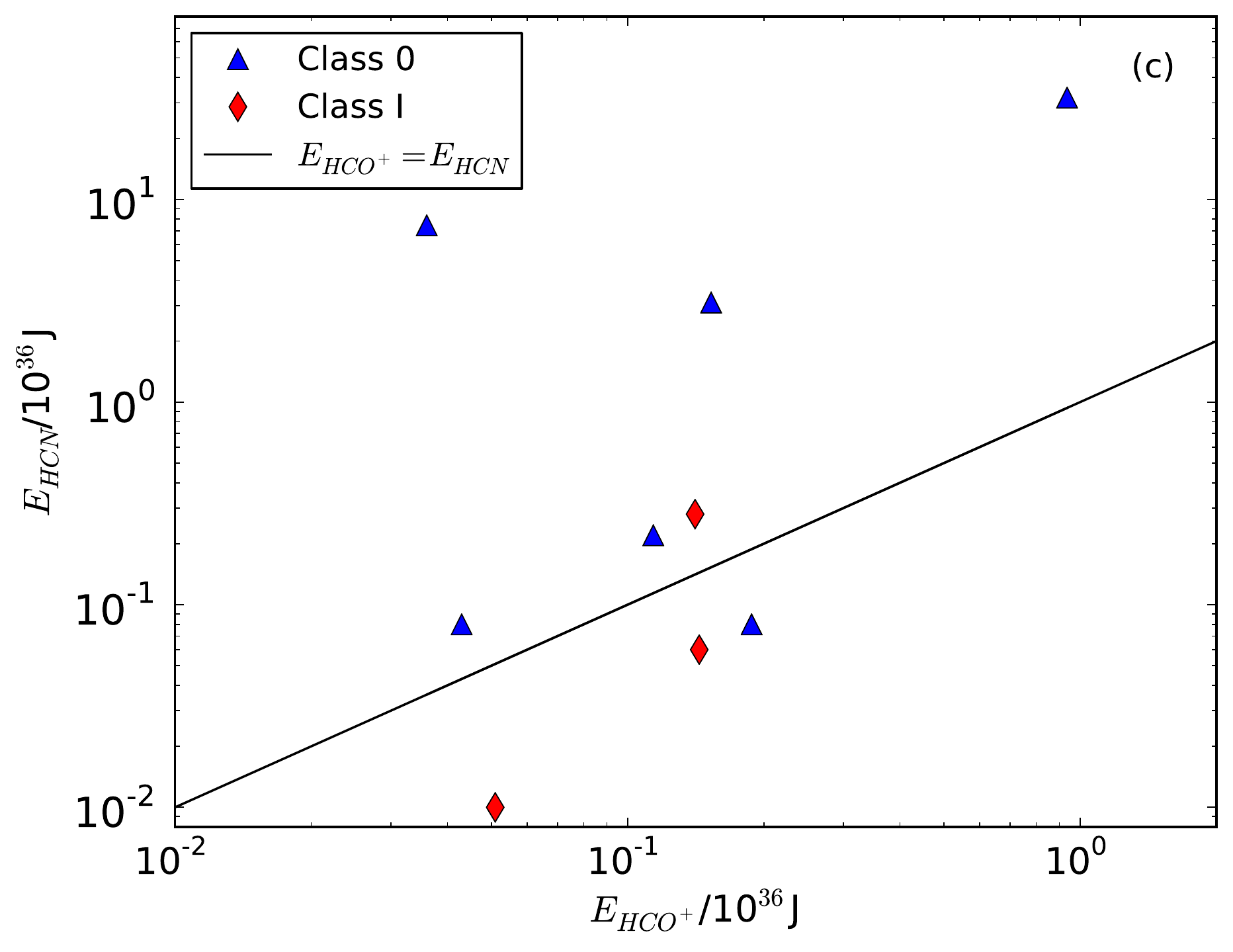}\qquad
	     \includegraphics[width=8.0cm]{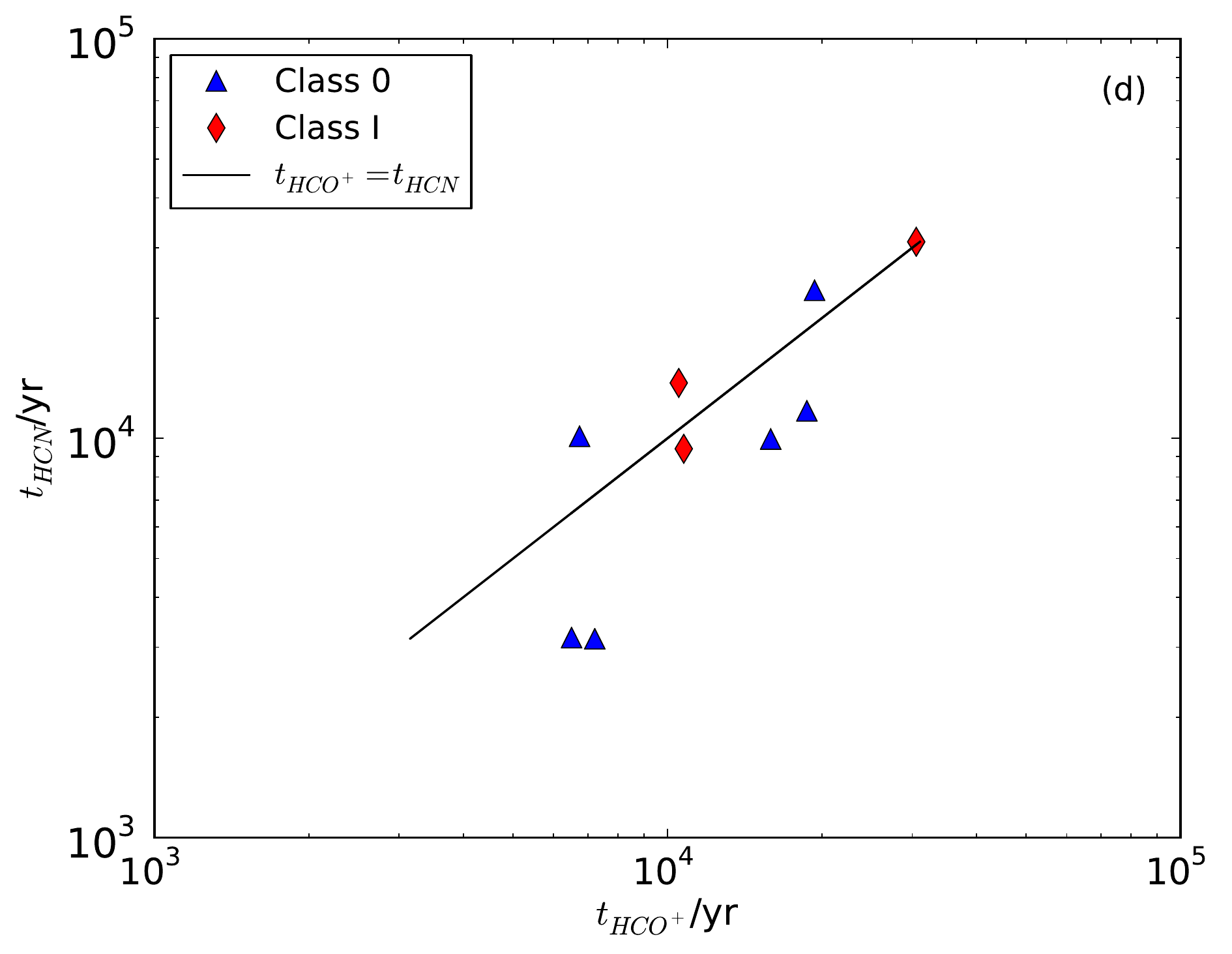}}
\smallskip
\centerline{\includegraphics[width=8.0cm]{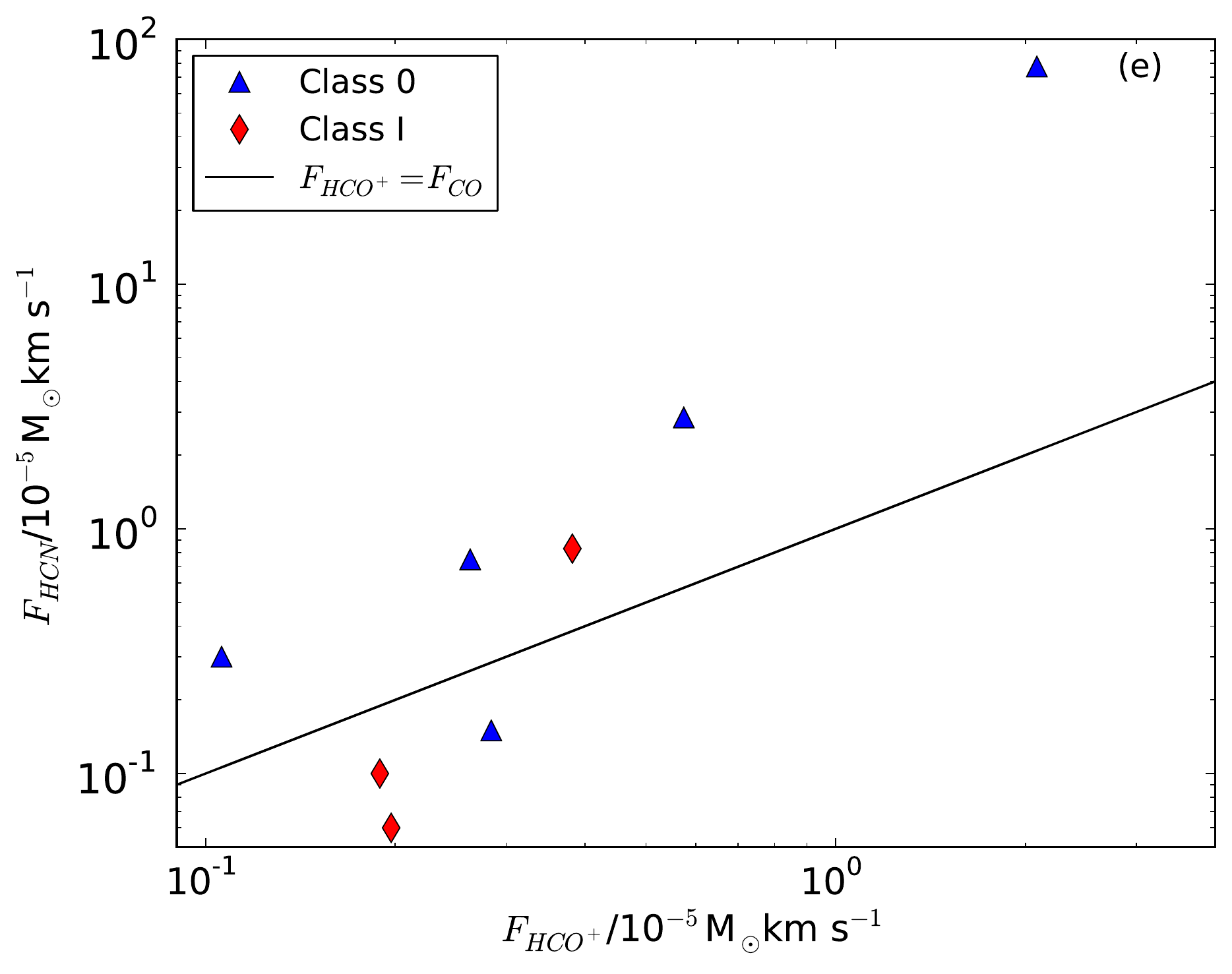}\qquad
	    \includegraphics[width=8.0cm]{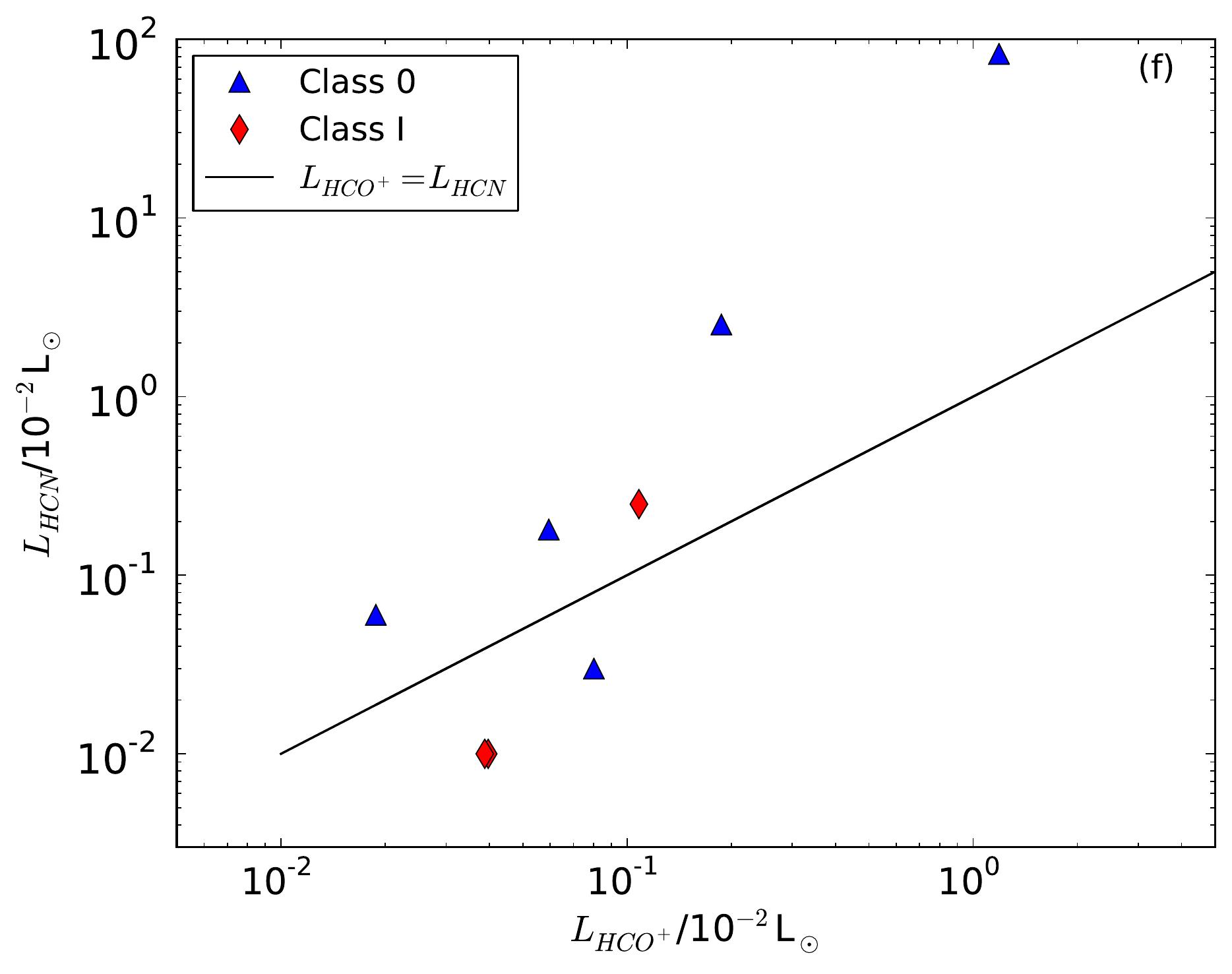}}
\caption{HCN outflow properties are plotted against the values derived from \HCOplus\ in Section \ref{sec:hcooutflows}: (a) Outflow mass, (b) Outflow momenta, (c) Outflow kinetic energy, (d) Outflow dynamical time, (e) Outflow momentum flux or driving force and (f) Outflow mechanical luminosity. The \twelveCO\ values are obtained from \citet{2010MNRAS.408.1516C}; the points are separated into Class 0 (blue triangles) and Class I (red diamonds).}
\label{fhcncomp}
\end{figure*}

One can see immediately that the 3 outflows powered by IRAS~4A, 4B and 2A have values about an order of magnitude higher than the others, as can also be seen from Figure \ref{fhcncomp}. We present both the average values for \emph{all} HCN outflows as well as the average values for the 7 `normal' outflows in Table \ref{tab:hcnhcocomp} to illustrate the effect of these more powerful outflows on the average outflow properties.

\begin{table*}
\caption{The first row gives the average properties of all HCN outflows identified, the second row gives the average values for the `normal' HCN outflows, excluding those from IRAS~2 and IRAS~4. Column 2: Total mass in both outflow lobes; Column 3: Total outflow momentum; Column 4: Total kinetic energy in the outflow; Column 5: Dynamical time of the outflow; Column 6: Total momentum flux or driving force of the outflow; Column 7: Total mechanical luminosity of the outflow. (Errors on the last digit(s) of the average values are given in the brackets.)}
\centering
\begin{tabular}{ccccccc}
\hline
Subset  & $M_{\rm out}$  & $P_{\rm out}$ & $E_{\rm out}$ & t$_{\rm dyn}$ & $F_{\rm out}$ 			&  $L_{\rm out}$\\
	&  (\Msun)  	 & (\Msun\ \kms) &  (10$^{36}$J) &  ($10^3$yr)   & (10$^{-5}$ \Msun\ \kms\ yr$^{-1}$)   & (10$^{-2}$\Lsun)\\
\hline
HCN average (all) &  0.048(18)	& 0.40(24)	& 4.4(3.1)	& 13(3)		& 11(8)		&  11(8) \\
HCN average (normal)  &  0.019(3)	& 0.08(4)	& 0.5(4)	& 16(3)		& 0.7(4)	& 0.4(4) \\ \hline 
\HCOplus\ average & 0.007(3) 	& 0.021(11)	&  0.07(6)	& 15(2)		&  0.18(6)	& 0.06(6) \\ \hline
\end{tabular}
\label{tab:hcnhcocomp}
\end{table*}

\subsection{\HCOplus\ -- HCN comparison: the `normal' outflows}
As the excitation temperatures and critical densities of \HCOplus\ and HCN are fairly similar (see Table \ref{mol_trans}), we might expect them to trace similar regions within the outflows. We therefore expect that barring any significant differences in chemical enhancement of the two molecules, the outflow properties that we calculate using the two molecules should be similar.

Discounting the 3 most massive and energetic outflows, we do indeed find that all the average HCN outflow properties are within a factor of 2 to 3 of the average \HCOplus\ values, as can be seen in Table \ref{tab:hcnhcocomp}. In addition, we find that the 7 `normal' HCN and \HCOplus\ outflows all have similar maximal velocities -- an average $v_{\rm max}$ of 2.5 $\pm$ 0.5 \kms\ for HCN outflows compared to 2.4 $\pm$ 0.3 \kms\ for \HCOplus\ outflows.

Given the large errors in the calculation of outflow properties \citep{2004A&A...426..503W}, we conclude that for the majority of relatively less-active `normal' outflows, \HCOplus\ and HCN outflow properties are generally in good agreement with one another. We therefore conclude that the majority of the outflows exhibit similar levels of activity and excite the molecular emission through similar processes.

\subsection{\HCOplus\ -- HCN comparison: the IRAS~2 and 4 outflows}
The HCN outflow properties calculated for the three most massive outflows driven by IRAS~2A, 4A and 4B are all at least an order of magnitude higher than their corresponding \HCOplus\ values. These outflows also have much higher maximal velocity extents than any of the other outflows, extending out to 13~\kms\ for IRAS~4. Even though this is only about half that of the \twelveCO\ outflow velocity (26~\kms), it is still much larger than the \HCOplus\ values, as can be seen in Figure \ref{compspec2}. 

The extremely high HCN values for these outflows indicate that there are significantly different processes occurring in these outflows compared to the other `normal' outflows. As these outflows are all driven by young, powerful Class 0 sources, it is reasonable to assume that there will be a large amount of shock activity occurring within the outflow lobes. 

HCN is known to be enhanced in shocked regions \citep{1990MNRAS.244..668P}, and the large HCN masses and velocities indicate that there is a great deal of shock-induced production of HCN occurring in these outflows compared with the others. Whether this enhanced production of HCN is a result of temperature or density differences caused by the shocks, or simply due to higher shock speeds, or a combination of all three is difficult to determine without detailed modelling. It is, nevertheless, apparent that the enhancement of HCN is very dependent on the physical properties within the outflows.

\section{\HCOplus\ -- HCN abundance comparison}\label{sec:abundances}
In the previous section, we found that the HCN appears to be enhanced considerably in some outflows compared to \HCOplus. An analysis of HCN and \HCOplus\ abundances and enhancements relative to \twelveCO\ is required to quantify this enhancement. 

We have chosen to perform our initial analysis on one particular outflow -- that associated with IRAS~2 -- as the outflow lobes are inclined at such an angle as to be well-separated spatially from the protostar, and should enable a good comparison of `quiescent' versus `shocked' gas.

\subsection{IRAS~2}
IRAS~2 (also known as IRAS 03258+3104) was first discovered by \citet{1987MNRAS.226..461J}, and submillimetre continuum imaging suggests that it consists of 3 different objects: young stellar sources 2A and 2B (both of which are well-isolated and detected at mid-IR wavelengths); and starless condensation 2C. IRAS~2A has been shown to be a true Class 0 protostar \citep{2009A&A...502..199B}, and is known to drive 2 strong outflows along axes that are almost perpendicular to one another:
\begin{itemize}
\item outflow A: a narrow, collimated, jet-like EW outflow, which is thought to a prototype for an extremely young Class 0-driven outflow. 
\item outflow B: a wider, shell-like outflow that lies roughly NS. 
\end{itemize}
The orientation and morphology of the two flows suggests that IRAS~2A may be a close binary, with the two stars at different evolutionary stages. \changed{This has recently been confirmed by \citet{2014arXiv1401.6672C}, who have resolved IRAS~2A into 2 distinct sources, MM~1 and MM~2. They find that MM~1 (a bright continuum source) likely drives the main NS jet (outflow B), while MM~2 (a weaker source) likely drives an EW jet (outflow A).}

\begin{figure}
\centerline{\includegraphics[width=8.3cm]{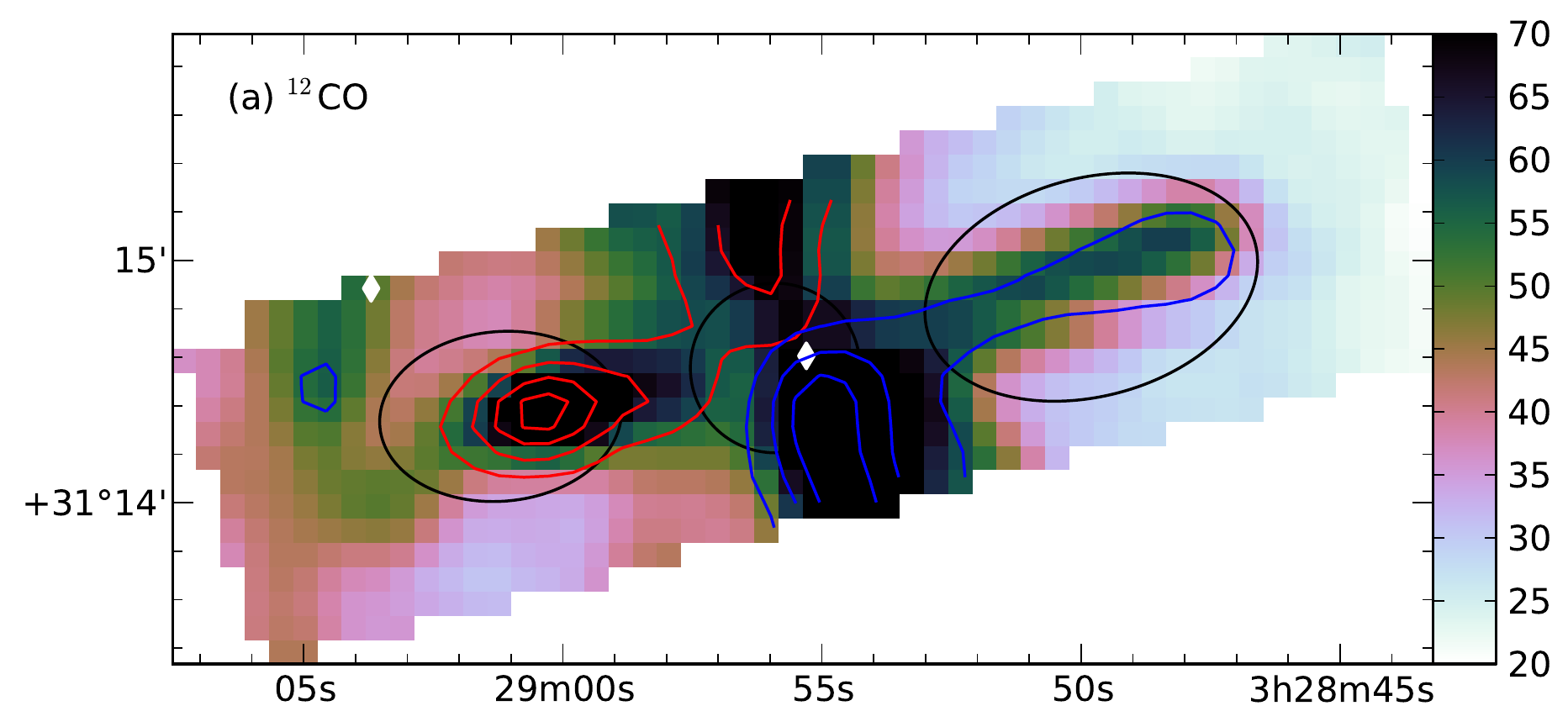}}
\smallskip
\centerline{\includegraphics[width=8.3cm]{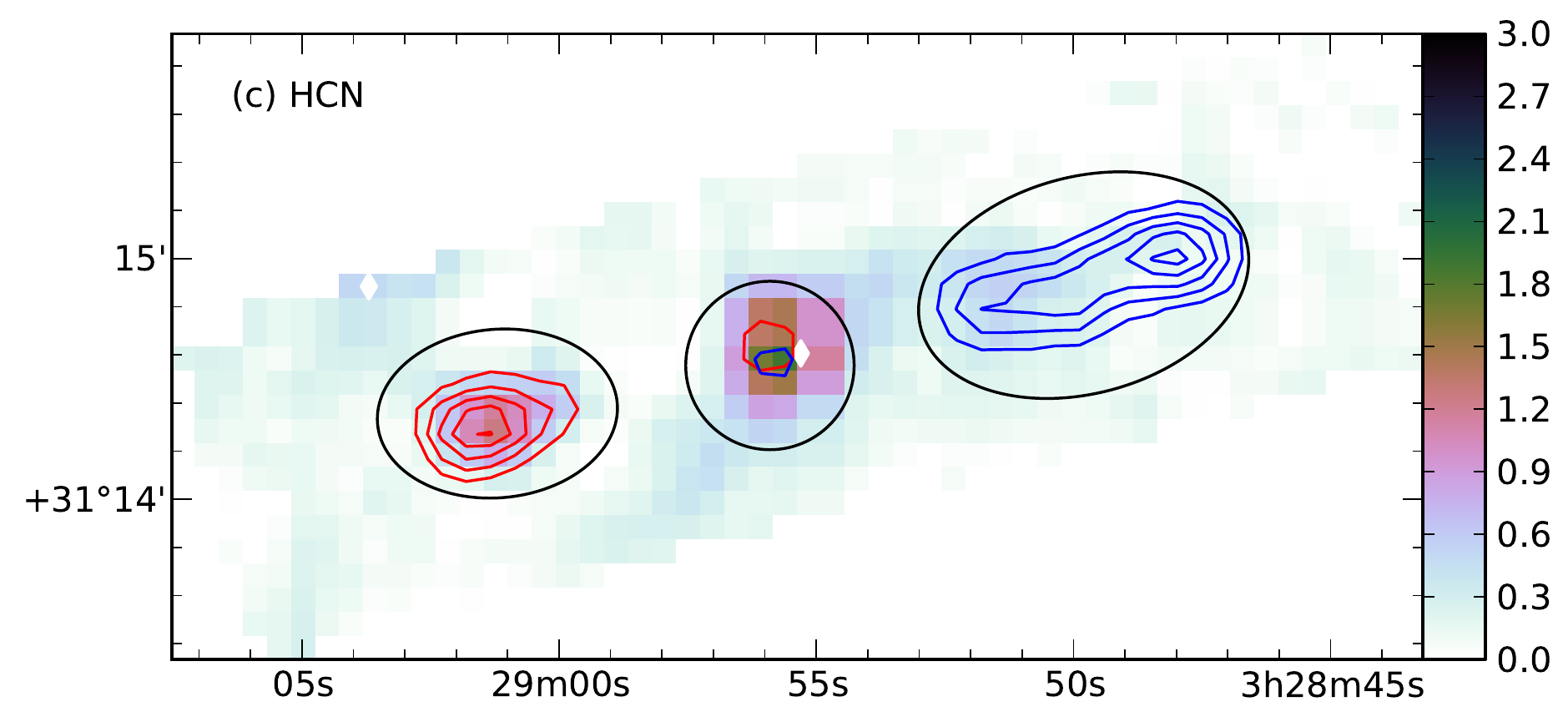}}
\smallskip
\centerline{\includegraphics[width=8.3cm]{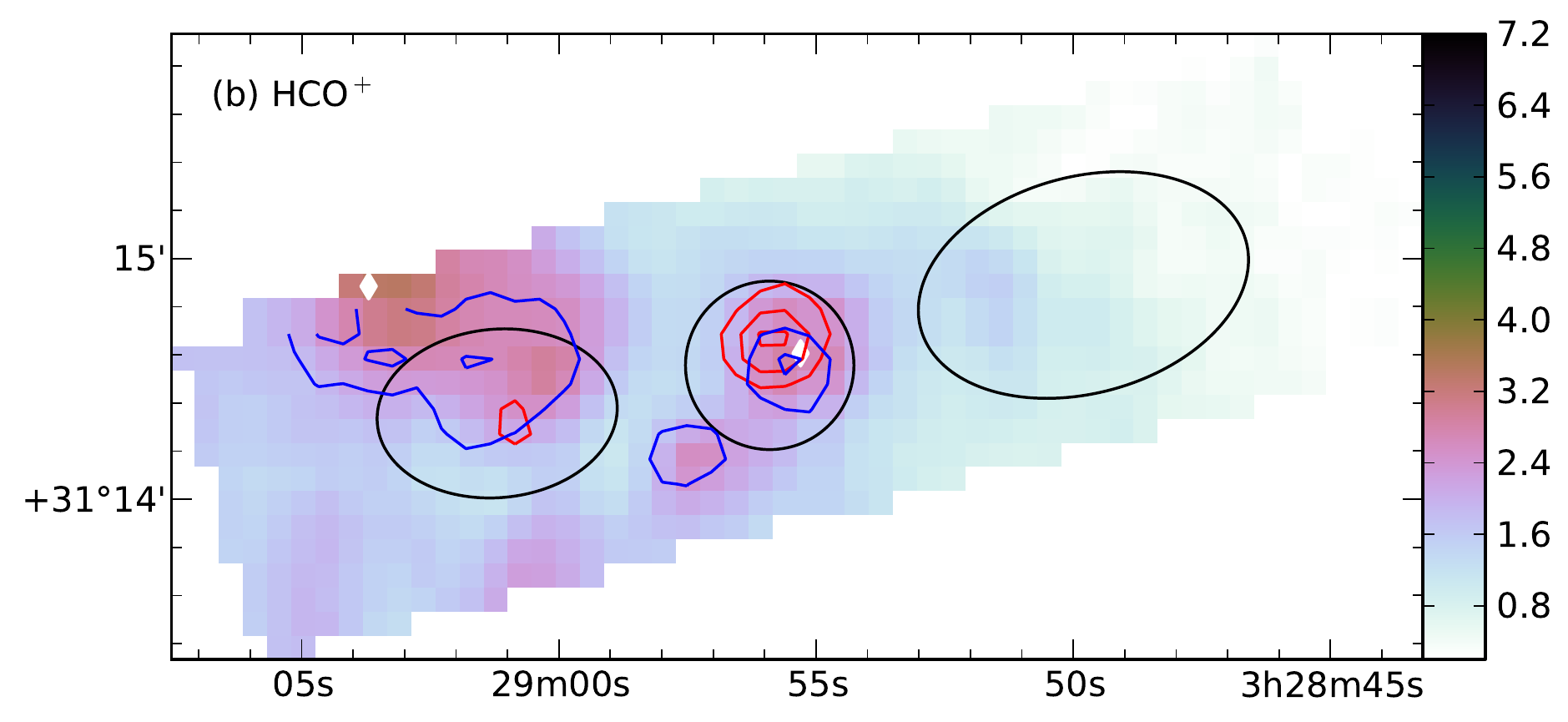}}
\caption{The integrated intensities (in K~\kms) of: (a) \twelveCO, (b) HCN and (c) \HCOplus\ for NGC~1333 IRAS~2A, integrated over the line centre (6.3 to 9.3 \kms). Overlaid are the red (9.3 to 25.6 \kms) and blue ($-10$ to 6.3 \kms) integrated intensity linewings. Black ellipses are overlaid to mark the three spatial regions -- east, west and source centre -- that were considered in this analysis; and white diamonds mark the positions of H07 SCUBA cores.}
\label{compall}
\end{figure}

\subsection{Overall analysis of IRAS~2A outflows}
We present an initial comparison of IRAS~2A and its outflows as traced by the \twelveCO, \HCOplus\ and HCN in Figure \ref{compall}. One can see immediately that the \twelveCO\ (Figure \ref{compall}a) traces both outflows well, although outflow A is narrower and more elongated than outflow B, most likely due to the jet-like nature of the latter. The peaks of the red and blue contours for outflow A are also clearly separated from the central driving source due to its angle of inclination. There is little differentiation between the central source and the outflows in the \twelveCO\ line centre integrated intensity map, most likely due to the high optical depth at the line centre. 

The HCN map (Figure \ref{compall}b) also shows a clear spatial separation of the lobes for outflow A, even more so than that demonstrated by \twelveCO. This is in line with the results of \citet{2004A&A...415.1021J}, who found that the HCN extended the furthest spatially in their data. 

Within outflow A itself, there is a difference in the spatial distribution of HCN: the emission in the red lobe is much more compact while that in the blue lobe is more extended. This could result from the inclination of the outflow, but \citet{2004A&A...415.1021J} have suggested a scenario where the highly-collimated protostellar EW outflow is progressing into a region with a steep density gradient. Our results support this as a higher density of material in the direction of the blue lobe would result in more extended HCN emission compared to the lower density material in the red lobe.

HCN shows little/no evidence for the presence of outflow B. The presence of HCN in one of the outflows but not the other points to a definite difference in the chemical activity and densities/temperatures within the lobes of the two outflows. \changed{This is a reflection of the fact that the two outflows have different driving sources, with intrinsically different properties;} it also reflects the properties of the regions that the two outflows are progressing into.

The \HCOplus\ map (Figure \ref{compall}c) is almost the exact opposite of the HCN: there is little indication for the presence of outflow A, while concentrated, overlapping linewings are seen at the base of outflow B. This also matches the findings of \citet{2004A&A...415.1021J}, who state that the \HCOplus\ emission they observe does not extend far from the central protostar. Our \HCOplus\ emission also shows a higher line centre integrated intensity at the position of the central driving source -- an indication of the young, protostellar nature (Class 0) of the core.


\subsection{Analysis of IRAS~2A Outflow A (EW)}
\subsubsection{Calculation of abundances and enhancement factors}
In our analysis, we consider three distinct spatial regions for each of the molecules: namely, the position of the central driving source, and the red and blue outflow lobes which lie to the east and west respectively. For each of these spatial regions in turn, we consider two distinct velocity regimes: the blue outflow wing (between -10 to 6.3 \kms) and the red outflow wing (between 9.3 and 25.6 \kms). 

\citet{2010A&A...522A..91T} calculated the relative abundances and enhancements of multiple molecules for the two Class 0 outflows, L~1448-mm and IRAS~04166+2706. We follow their methods and compare our values to theirs to determine how typical the IRAS~2A outflow is of Class 0 sources; we also compare our values to those calculated by \citet{2004A&A...415.1021J} who investigated a large number of molecules in outflow A of IRAS~2A.

We first calculate the column densities of the molecules in the 2 velocity regimes, for each spatial region. We use Equation \ref{eq:col_density}, substituting the corresponding values for each molecule from Table \ref{coldenprops}. We assume a temperature of 50~K for the outflow lobes (as we have done in our earlier outflow analyses), and 12~K for the central driving source -- a lower temperature would be expected for the cold, central Class 0 protostar. These values match those obtained through modelling of IRAS~2A and its outflow by \citet{2004A&A...415.1021J}.

The column densities for \HCOplus\ and HCN are then normalised by the corresponding \twelveCO\ column density to produce CO-normalised abundances (which we will refer to simply as \emph{abundances}). We also calculate the enhancement factor $F_{\rm enh}$ for each molecule using the following relation:
\begin{equation}
F_{\rm enh} = \frac{N_{\rm mol}}{N_{\rm ^{12}CO}} \times \frac{X_{\rm ^{12}CO}}{X_{\rm mol}},
\end{equation}
where the values of $X_{\rm mol}$ are given in Table \ref{coldenprops}. \changed{It should be noted that the \HCOplus\ and HCN abundances used in this case are average protostellar envelope abundances, and may be higher in regions with $n < 7 \times 10^4$~\cmcube, where the species has not frozen out onto the dust grains \citep{2004A&A...416..603J}.} These values are referred to as the \emph{canonical} abundances, to differentiate them from the CO-normalised abundances mentioned above. Our results and the comparisons of the abundances and enhancement factors for the four molecules are presented in Table \ref{abunprops}. 

\subsubsection{Comparisons between molecules}
\begin{table}
\caption{Comparison of the enhancement factors for \HCOplus\ and HCN in the two lobes of outflow A and in the driving source (IRAS~2A). The enhancement factors are calculated by comparing the relative abundances of each molecule calculated in this paper to other generally accepted abundance values for each molecule.}
\centering
\begin{tabular}{ccccc}
\hline
Molecule   &  Blue lobe & \multicolumn{2}{c} {Source centre}  & Red lobe \\ \hline
	   &   Blue wing	  & Blue wing   & Red wing		&  Red wing \\ \hline
\HCOplus\  &    0.1  		  & 0.3  	  &   	0.4   		& 0.2 \\
HCN  	   &   6.3  		  & 1.1  	  &  2.1  		& 5.5 \\
\hline
\end{tabular}
\label{abunprops}
\end{table}

We find that the HCN is more enhanced than the \HCOplus\ in all regimes by over an order of magnitude. The difference in enhancement between the two molecules has also been observed in this outflow by \citet{2004A&A...415.1021J}, and in two other outflows by \citet{2010A&A...522A..91T}, who all attributed it to shock chemistry. 

\citet{1990MNRAS.244..668P} show by modelling shock waves that the $n$(HNC)/$n$(HCN) ratio is greatly decreased in shocks, which could explain the increase in HCN enhancement; for example, the reaction $ \rm HNC + H \rightarrow HCN + H$ is facilitated by the high temperatures in outflow shocks. \citet{2004A&A...415.1021J} cite direct release of HCN from grain mantles as a further contributing factor to the enhancement of HCN abundance.

The stronger the shock, the higher the temperatures and the greater the enhancement. An H$_2$ object \citep{2008MNRAS.387..954D} has been associated with the spatial position of the red lobe, which is an indication that the shock in the red lobe is strong. The aforementioned increase in density in the blue lobe could result in a more focused shock and cause the increased HCN enhancement factor in the blue lobe.

The very low abundances of \HCOplus\ that we observe could result from the destruction of \HCOplus\ in the passage of the outflow shock by reactions with water \citep{1998ApJ...499..777B}. This matches our observations of the spatial extent of \HCOplus\ -- it is only present at the base of outflow B, and traces material only in the aftermath of shocks.

\subsubsection{Comparisons between outflows}
We find that the levels of HCN and \HCOplus\ enhancement in this outflow are about a factor of 100 less than that found for the L~1157 outflow \citep{2010A&A...522A..91T, 1997IAUS..182..153B}, implying that the L~1157 outflow likely has much stronger shocks than the IRAS~2A outflow. Despite outflow A appearing extremely active compared to other outflows in our study of Perseus, it does not stand out in comparison with truly `chemically-active' sources -- those which exhibit great enhancements of several orders of magnitude in many tens of molecules. 

One consideration is that as the peak of chemical activity is transient due to the freeze-out of enhanced molecules \citep{2011IAUS..280...88T}, our outflow may have have passed its peak. However, as the timescales of chemical activity are on the order of the length of a Class 0/Class I stage, our outflow is still young enough ($\sim 5 \times 10^3$ years) that this is likely not the cause of the `lack' of activity compared to L~1157. 

\changed{The molecular richness of the cloud that our outflow jet is propagating into could be a cause of the apparent lack of activity in IRAS~2A. If there is a lack of material of sufficient complexity and diversity in the outflow region, there will be less visible signs of} \changed{chemical activity from a wide range of more complex molecules. This lack of richness may be caused by either insufficient density of the molecular cloud (less likely as the \HCOplus\ and HCN are still present), or by the shock not being of sufficient strength.}

\subsubsection{Discussion of errors}
Of course, there are significant uncertainties present in our calculations. The abundances we assume for our molecules are subject to large uncertainties and can vary by a factor of a few. In particular, our assumption of a constant \twelveCO\ abundance of $X_{^{12} \rm CO} = 10^{-4}$ may be incorrect. There is likely to be a large degree of CO freeze-out in the cold dense protostellar core of IRAS~2A, making a lower \twelveCO\ abundance more likely. However, the \changed{high levels of shock activity occurring will likely sputter the CO off the dust grains, releasing it into the gas phase}; we therefore feel that the use of the canonical value of $X_{^{12}\rm CO} = 10^{-4}$ to calculate abundances in the outflow lobes is acceptable.

From comparisons between \twelveCO\ and \CeighteenO, we inferred the mean optical depth of \twelveCO\ in the outflow line wings $\tau_{^{12} \rm CO}$ to lie between 5 and 10, meaning that \twelveCO\ is fairly optically thick. The optical depth of the \twelveCO\ is therefore a potential source of error, as a decrease in column density due to a higher optical depth of \twelveCO\ could be misinterpreted as a sign of abundance enhancement for \HCOplus\ and HCN. It is reassuring to note however that the values of our molecular abundances and enhancement factors are within a factor of a few of those calculated for the same molecular transitions in the same outflow by \citet{2004A&A...415.1021J}.

Ultimately, while the \emph{absolute} abundances may not be accurate, this will have little effect on the relative values of enhancement factors, and comparisons between molecules can still be made. Therefore, our result that HCN is over an order of magnitude more abundant than the \HCOplus\ is still valid. 

\section{Conclusions}\label{sec:concl}

We have carried out an analysis and comparison of the \HCOplus\ and HCN $J = 4\to3$ emission in several sub-regions of the Perseus molecular cloud and we have found the following results:
\begin{enumerate}
\item The \HCOplus\ shows much more extended structure than the HCN which is largely confined to compact clumps. We can constrain densities of the filamentary structures that are solely traced by \HCOplus\ to lie between the critical densities of the \HCOplus\ and HCN transitions.
\item HCN is predominantly excited around protostellar objects, unlike the \HCOplus\ emission, which is also associated with a high proportion of starless SCUBA dust cores -- we can use the combination of a $3 \sigma$ detection for both molecules to pinpoint the truly protostellar cores.
\item \HCOplus\ shows large-scale velocity structure across the individual sub-regions, particularly in NGC~1333; HCN on the other hand shows very little change in velocity across a particular sub-region, except around those protostars that power the most energetic outflows.
\item \HCOplus\ emission is a good tracer of outflow activity -- it identifies over 50\% of the outflow sources that have been previously identified by \twelveCO, and those it misses are generally significantly weaker. 
\item \HCOplus\ outflow driving forces exhibit similar trends with $M_{\rm env}$ and $L_{\rm bol}$ when compared with \twelveCO, which indicates that they are excited similarly within the outflow. It also exhibits a similar trend with $M_{\rm env}$ to that calculated in Ophiuchus, indicating similar outflow driving mechanisms in the two separate regions.
\item The outflow properties calculated for HCN are on average within a factor of 2 of those calculated for the same outflows using \HCOplus. This is an indication that the \HCOplus\ and HCN are excited similarly in the majority of the outflows that do not exhibit strong shock activity.
\item HCN traces the most energetic outflows, especially in certain cases (e.g. IRAS~4) where the outflow wings extend further in velocity than in \HCOplus. The presence of large enhancements of HCN in particular outflows is thought to be an indication of shock activity within the outflow, and illustrates the youth and energy of the driving source. 
%
%
\item The increased enhancement of HCN in the blue lobe of outflow A in IRAS~2A lends support to the theory of a density gradient in the EW direction; its increased enhancement in the red lobe reflects the increased strength of the shocks in that lobe.
%
\item \HCOplus\ is the least enhanced molecule and is generally confined to the base of the outflow, and the central driving source. This illustrates the effects of shocks on the chemistry in outflows (as traced by HCN), compared with that in protostellar cores and their envelopes (as traced by \HCOplus).
\end{enumerate}

A natural extension to this work would be to investigate all the other outflows in Perseus that are traced by both HCN and \HCOplus, to determine how typical IRAS~2A is as an outflow source, and whether other outflows in the region exhibit similar molecular enhancements. This is intended as the subject of a future paper. 

In conclusion, we find that disentangling the effects of the physical environments (density, temperature) of outflows, from the actual chemical processes occurring within the outflows is very difficult. A next step in the analysis would be the modelling of these protostellar sources and their outflows using 3D radiative transfer code, such as ARTIST \citep{2011IAUS..270..451P}, to better constrain the physical conditions giving rise to the emission we observe.

\section*{Acknowledgments}
S. Walker-Smith and J. Hatchell are funded by the Science and Technology Facilities Council of the UK. The James Clerk Maxwell Telescope is operated by the Joint Astronomy Centre on
behalf of the Science and Technology Facilities Council of the United Kingdom,
the National Research Council of Canada, and (until 31 March 2013) the
Netherlands Organisation for Scientific Research.

\appendix

\section{Results of identification of \HCOplus\ outflows}
We present a table showing our criteria for a positive outflow identification.

\clearpage
\onecolumn 

\makeatletter
\let\@makecaption=\SFB@maketablecaption
\makeatother

\tablecaption{Results for outflow testing in NGC 1333. Column 1 gives the source designation as defined by H07; Columns 2,3 and 4 give the results of the outflow determination using the 3 criteria defined in Section \ref{sec:outflowiden}. For these columns, `R' and `B' indicate the presence of only red or blue linewings respectively. A `--' entry indicates that the source does not have \HCOplus\ at the $5\sigma$ level. A question mark indicates that the linewings appear to be present, but are slightly obscured by the double-peaked/asymmetric nature of the \HCOplus\ line. Column 5 states the presence of \twelveCO\ outflows as defined by \citet{2010MNRAS.408.1516C}. Column 6 indicates the protostellar status of each core, and Column 7 shows any other designations that the source is associated with. Column 8 indicates any interesting features in the shape of the profile -- DP (double-peaked), RA and BA (Red and blue asymmetry respectively); the final column indicates the \HCOplus\ outflow driving status of each core.\label{tab:outflows}}
\tablehead{\hline H07  & \HCOplus\  & HCO+ wings  & HCO+ wings & CO  & Protostellar  & Other & Remarks & \HCOplus \\	
sources & lobes & (core peak) & (surrounding pixels) & lobes & Class & Designation &  & Outflow \\ \hline}
\tabletail{\hline \multicolumn{9}{r}{\emph{Continued on the next page}}\\}
\tablelasttail{\hline}
\begin{supertabular*}{\textwidth}{ccccccccc}
     &            &             &          &      \bf{NGC~1333}          &      &         &     &      \\ \hline
41 & 2 & 2	 & 	2	 & 	2	 & 	0	 & IRAS4B &  & D\\
42 & 2 & 2	 & 	2	 & 	2	 & 	0	 & IRAS4A &  & D\\
43 & 2 & 2?	 & 	2	 & 	2	 & 	I	 & SVS13,HH7-MMS1 &  & P\\
44 & 2 & 2	 & 	2?	 & 	2x2	 & 	0	 & IRAS2A & DP (BA) & D\\
45 & 2 & B	 & 	B	 & 	2	 & 	I	 & SVS12 &   & D\\
46 & R & B	 & 	N	 & 	2	 & 	0	 & IRAS7 &   & P\\
47 & 2 & 2	 & 	B	 & 	2	 & 	0	 & ----  & RA  & D\\
48 & N & B?	 & 	N?	 & 	?	 & 	0	 & IRAS4C & DP (BA) & M\\
49 & 2 & 2	 & 	N?	 & 	2	 & 	I	 & IRAS03255+3103 & DP (BA) & P\\
50 & 2 & B	 & 	B	 & 	?	 & 	I	 & HH7-11,MMS4 &  & D\\
51 & N & N	 & 	N	 & 	N	 & 	S	 & ----  &  & M\\
52 & 2 & 2	 & 	N	 & 	2	 & 	0	 & HH7-11,MMS6 &  BA  & P\\
53 & N & N	 & 	N	 & 	N	 & 	S	 & ----  &  & N\\
54 & B & 2?	 & 	B?	 & 	R	 & 	I	 & ----  &  & P\\
55 & N & N	 & 	N	 & 	N	 & 	0	 & ----  &  & N\\
56 & N & R?	 & 	N	 & 	2	 & 	I	 & ----  &  & M\\
57 & N & N	 & 	N	 & 	N	 & 	S	 & ----  &  & N\\
58 & --	& --	 & 	--	 & 	N	 & 	0	 & ----  &  & N\\
59 & --	& N	 & 	--	 & 	N	 & 	S	 & ----  &  & N\\
60 & N	& N	 & 	--	 & 	N	 & 	S	 & ----  &  & N\\
61 & --	& --	 & 	--	 & 	N	 & 	0	 & ----  &  & N\\
62 & N & B?	 & 	N	 & 	2	 & 	0	 & ----  &  & M\\
63 & N & N	 & 	N	 & 	2	 & 	I	 & ----  &  & N\\
65 & N & N	 & 	N	 & 	2	 & 	0	 & IRAS4B1 &   & N\\
66 & N & N	 & 	N	 & 	N	 & 	S	 & ----  &  & N\\
67 & N & N	 & 	N	 & 	B	 & 	I	 & ----  &  & N\\
68 & N & N	 & 	N	 & 	B	 & 	0	 & ----  &  & N\\
69 & -- & --	 & 	--	 & 	B	 & 	I	 & ----  &  & N\\
70 & N & N	 & 	R?	 & 	N	 & 	0	 & ----  &BA  & M\\
71 & -- & --	 & 	--	 & 	2	 & 	0	 & ----  &  & N\\
72 & -- & N	 & 	--	 & 	N	 & 	S	 & ----  &  & N\\
74 & N & N & 	N	 & 	N	 & 	I	 & ----  &  & N\\
Bolo44 & N & 	N	 & 	N	 & 	N	 & 	S	 & ----  &  & N\\ \hline
     &            &             &          &      \bf{IC~348}          &      &         &    &       \\ \hline
12 & 2 & 2	 & 	2	 & 	2	 & 	0	 & HH211 &  & D\\
13 & 2 & 2	 & 	2	 & 	2	 & 	0	 & IC348 MMS &  & D\\
14 & 2 & 2	 & 	2	 & 	2	 & 	I	 & ---- &  & D\\
15 & 2? & 2	 & 	2	 & 	2	 & 	0	 & ---- &  & P\\
16 & N & N	 & 	N	 & 	N	 & 	S	 & ---- &  & N\\
17 & N & N	 & 	N	 & 	N	 & 	S	 & ---- &  & N\\
18 & R? & N	 & 	N	 & 	N	 & 	S	 & ---- &  & M\\
19 & N & N	 & 	N	 & 	N	 & 	S	 & ---- &  & N\\
20 & N & N	 & 	N	 & 	N	 & 	S	 & ---- &  & N\\
21 & N & N	 & 	N	 & 	N	 & 	S	 & ---- &  & N\\
23 & N & N	 & 	N	 & 	N	 & 	S	 & ---- &  & N\\
24 & N & N	 & 	N	 & 	N	 & 	S	 & ---- &  & N\\
25 & N & N	 & 	N	 & 	N	 & 	S	 & ---- &  & N\\
26 & N & N	 & 	N	 & 	N	 & 	S	 & ---- &  & N\\
101 & N & N	 & 	N	 & 	R	 & 	I	 & IRAS03410+3152 &  & N\\
Bolo111 & N & N	 & 	N	 & 	N	 & 	S	 & ---- &  & N\\
Bolo113 & N & N	 & 	N	 & 	N	 & 	S	 & ---- &  & N\\ \hline
     &            &             &          &      \bf{L~1448}          &      &         &     &      \\ \hline
27 & 2 & 2	 & 	2	 & 	2	 & 	0	 & L1448NW &  & D\\
28 & 2 & 2	 & 	2	 & 	2	 & 	0	 & L1448N A/B &  & D\\
29 & 2 (B?) & 2	 & 	2	 & 	2	 & 	0	 & L1448C &  & D\\
30 & 2? & 2	 & 	2	 & 	2	 & 	0	 & L1448 IRS2 &  & P\\
31 & R & 2	 & 	2	 & 	2	 & 	0	 & ---- &  & P\\
32 & N & N	 & 	N	 & 	N	 & 	S	 & ---- &  & N\\
Bolo11 & N & N	 & 	N	 & 	N	 & 	S	 & ---- &  & N\\ \hline
     &            &             &          &      \bf{L~1455}          &      &         &     &      \\ \hline
35 & 2 & 2	 & 	2	 & 	2	 & 	I	 & L1455 FIR4 &  & D\\
36 & N & 2	 & 	2	 & 	2	 & 	0	 & ---- &  & P\\
37 & N & 2	 & 	2	 & 	2?	 & 	I	 & L1455 PP9 &  & P\\
39 & 2? & 2	 & 	2?	 & 	2	 & 	I	 & L1455 FIR1/2 &  & D\\
40 & N & N	 & 	N	 & 	N	 & 	S	 & ---- &  & N\\ \hline
\end{supertabular*}

\section{Properties of \HCOplus-identified outflows}
We present the properties calculated for all the sources that exhibit \HCOplus-linewing emission and have been identified as outflows.

\begin{table*}
\caption{Outflow properties for those sources that display outflow signatures in \HCOplus, separated by sub-region. Column 3: Total mass in both outflow lobes; Column 4: Total outflow momentum; Column 5: Total kinetic energy in the outflow; Column 6: Average length of outflow lobe; Column 7: Maximum velocity extent reached (as an average of the red and blue maximal velocities); Column 8: Dynamical time of the outflow; Column 9: Total momentum flux or driving force of the outflow; Column 10: Total mechanical luminosity of the outflow. The average properties for all outflows in each sub-region are also given, as are the average properties for each protostellar class in each sub-region. (Errors on the last digit(s) of the average values are given in the brackets.)}
\centering
\begin{tabular}{cccccccccc}
\hline
H07 &  Class & $M_{\rm out}$  & $P_{\rm out}$ & $E_{\rm out}$ & $L_{\rm lobe}$ & v$_{\rm max}$ & t$_{\rm dyn}$ & $F_{\rm out}$ &  $L_{\rm out}$\\
ID  &  	    & (\Msun)        & (\Msun\ \kms) & (10$^{36}$J) & (pc)            &  (\kms)       &  ($10^3$yr) & (10$^{-5}$ \Msun\ \kms\ yr$^{-1}$) & (10$^{-2}$\Lsun)\\
\hline
     &            &             &          &      &  \bf{NGC~1333}              &         &     &      \\ \hline
41 & 0 & 0.0199 & 0.136 & 0.936 & 0.043 & 6.50 & 6.5 & 2.09 & 1.19 \\
42 & 0 & 0.0036 & 0.011 & 0.036 & 0.022 & 3.00 & 7.2 & 0.158 & 0.0413 \\
43 & I & 0.0119 & 0.041 & 0.141 & 0.039 & 3.50 & 10.7 & 0.382 & 0.1082 \\
44 & 0 & 0.0097 & 0.039 & 0.153 & 0.028 & 4.00 & 6.7 & 0.574 & 0.1873 \\
45 & I & 0.0235 & 0.058 & 0.144 & 0.078 & 2.50 & 30.5 & 0.189 & 0.0388 \\
46 & 0 & 0.0057 & 0.011 & 0.022 & 0.031 & 2.00 & 15.0 & 0.074 & 0.0122 \\
47 & 0 & 0.0087 & 0.023 & 0.063 & 0.042 & 2.50 & 16.0 & 0.143 & 0.0320 \\
49 & I & 0.0025 & 0.004 & 0.005 & 0.020 & 1.50 & 13.3 & 0.028 & 0.0033 \\
50 & I & 0.0094 & 0.024 & 0.062 & 0.037 & 2.75 & 13.0 & 0.185 & 0.0395 \\
52 & 0 & 0.0043 & 0.009 & 0.017 & 0.042 & 2.00 & 20.3 & 0.043 & 0.0071 \\
54 & I & 0.0080 & 0.018 & 0.042 & 0.038 & 2.25 & 16.6 & 0.109 & 0.0206 \\ \hline
\multicolumn{2}{c}{Average} & 0.010(2) & 0.034(11) & 0.15(8) & 0.038(4) & 3.0(4) & 14.2(2.0) & 0.36(17) & 0.15(10) \\
\multicolumn{2}{c}{Class 0} & 0.009(2) & 0.038(18) & 0.20(13) & 0.034(3) & 3.3(6) & 12.0(2.2) & 0.5(3) & 0.24(17) \\
\multicolumn{2}{c}{Class I} & 0.011(3) & 0.029(8) & 0.08(2) & 0.042(9) & 2.5(3) & 16.8(3.2) & 0.18(5) & 0.04(2) \\ \hline
     &            &             &          &      &  \bf{IC~348}              &         &     &      \\ \hline
12 & 0 & 0.0051 & 0.016 & 0.052 & 0.030 & 3.25 & 9.1 & 0.180 & 0.0476 \\
13 & 0 & 0.0035 & 0.006 & 0.010 & 0.040 & 1.75 & 22.3 & 0.027 & 0.0039 \\
14 & I & 0.0026 & 0.007 & 0.022 & 0.033 & 3.00 & 10.8 & 0.069 & 0.0165 \\
15 & 0 & 0.0033 & 0.006 & 0.012 & 0.037 & 1.75 & 20.8 & 0.030 & 0.0047 \\ \hline
\multicolumn{2}{c}{Average} & 0.0036(5) & 0.009(2) & 0.024(8) & 0.035(2) & 2.4(3) & 15.7(2.9) & 0.08(3) & 0.018(9) \\
\multicolumn{2}{c}{Class 0} & 0.0039(5) & 0.010(3) & 0.025(11) & 0.036(2) & 2.2(4) & 17.4(3.4) & 0.08(4) & 0.019(11) \\ \hline
     &            &             &          &      &  \bf{L~1448}              &         &     &      \\ \hline
27 & 0 & 0.0163 & 0.055 & 0.188 & 0.057 & 2.90 & 19.4 & 0.284 & 0.0802 \\
28 & 0 & 0.0151 & 0.042 & 0.114 & 0.045 & 2.75 & 15.9 & 0.263 & 0.0594 \\
29 & 0 & 0.0091 & 0.020 & 0.043 & 0.041 & 2.15 & 18.7 & 0.106 & 0.0188 \\
30 & 0 & 0.0023 & 0.004 & 0.006 & 0.029 & 1.65 & 17.0 & 0.022 & 0.0029 \\
31 & 0 & 0.0055 & 0.012 & 0.033 & 0.031 & 2.10 & 14.2 & 0.088 & 0.0189 \\ \hline
\multicolumn{2}{c}{Average} & 0.010(2) & 0.027(8) & 0.08(3) & 0.040(5) & 2.3(2) & 17.0(8) & 0.15(5) & 0.036(13) \\ \hline
     &            &             &          &     &    \bf{L~1455}             &         &     &      \\ \hline
35 & 0 & 0.0084 & 0.021 & 0.051 & 0.027 & 2.50 & 10.5 & 0.197 & 0.0397 \\
36 & I & 0.0012 & 0.002 & 0.003 & 0.025 & 1.50 & 16.4 & 0.011 & 0.0013 \\
37 & I & 0.0110 & 0.025 & 0.055 & 0.023 & 2.00 & 11.2 & 0.221 & 0.0410 \\
39 & I & 0.0035 & 0.009 & 0.022 & 0.020 & 2.25 & 8.6 & 0.101 & 0.0207 \\ \hline
\multicolumn{2}{c}{Average} & 0.006(2) & 0.014(4) & 0.033(10) & 0.024(1) & 2.1(2) & 11.7(1.3) & 0.13(4) & 0.026(7) \\
\multicolumn{2}{c}{Class I} & 0.0076(14) & 0.018(3) & 0.043(7) & 0.023(1) & 2.2(1) & 10.1(4) & 0.17(2) & 0.034(4) \\ \hline
\end{tabular}
\label{tab:hcooutprops}
\end{table*}

\section{Properties of HCN-identified outflows}
We present the properties calculated for all the sources that exhibit HCN-linewing emission and have been identified as outflows.

\begin{table*}
\caption{Outflow properties for the individual sources that display outflow signatures in HCN. Column 3: Total mass in both outflow lobes; Column 4: Total outflow momentum; Column 5: Total kinetic energy in the outflow; Column 6: Average length of outflow lobe; Column 7: Maximum velocity extent reached (as an average of the red and blue maximal velocities); Column 8: Dynamical time of the outflow; Column 9: Total momentum flux or driving force of the outflow; Column 10: Total mechanical luminosity of the outflow. The final three rows give the average properties for all the HCN outflows, as well as the average properties when the outflows are separated by protostellar class. The errors on the last digit of these averages are given in brackets next to each number.}
\centering
\begin{tabular}{cccccccccc}
\hline
Source & Class & $M_{\rm out}$  & $P_{\rm out}$ & $E_{\rm out}$ & $L_{\rm lobe}$ & v$_{\rm max}$ & t$_{\rm dyn}$ & $F_{\rm out}$ &  $L_{\rm out}$\\
Name  &        & (\Msun) & (\Msun\ \kms) & (10$^{36}$J) & (pc) &  (\kms)   &  (10$^3$yr) & (10$^{-5}$ \Msun\ \kms\ yr$^{-1}$) & (10$^{-2}$\Lsun)\\
\hline
NGC~1333 41 & 0 & 0.190 & 2.456 & 31.84 & 0.042 & 13.00 & 3.2 & 77.45 & 82.90 \\
NGC~1333 42 & 0 & 0.098 & 0.857 & 7.45 & 0.028 & 8.75 & 3.1 & 27.20 & 19.53 \\
NGC~1333 43 & I & 0.022 & 0.078 & 0.28 & 0.034 & 3.50 & 9.4 & 0.83 & 0.25 \\
NGC~1333 44A & 0 & 0.027 & 0.288 & 3.10 & 0.106 & 10.25 & 10.1 & 2.85 & 2.53 \\
NGC~1333 44B & 0 & 0.061 & 0.169 & 0.47 & 0.024 & 2.75 & 8.4 & 2.02 & 0.46 \\
NGC~1333 45 & I & 0.017 & 0.030 & 0.06 & 0.054 & 1.70 & 31.1 & 0.10 & 0.01 \\
L~1448 27 & 0 & 0.020 & 0.035 & 0.08 & 0.054 & 2.25 & 23.5 & 0.15 & 0.03 \\
L~1448 28 & 0 & 0.026 & 0.075 & 0.22 & 0.028 & 2.75 & 10.9 & 0.75 & 0.18 \\
L~1448 29 & 0 & 0.015 & 0.035 & 0.08 & 0.030 & 2.50 & 11.7 & 0.30 & 0.06 \\
L~1455 35 & I & 0.004 & 0.008 & 0.01 & 0.025 & 1.75 & 13.8 & 0.06 & 0.015 \\ \hline
\multicolumn{2}{c}{Average} & 0.048(18) & 0.40(24) & 4.4(3.0) & 0.042(8) & 4.9(1.3) & 12.4(2.7) & 11.2(7.8) & 10.6(8.3) \\
\multicolumn{2}{c}{Class 0} & 0.062(24) & 0.6(3) & 6.2(4.4) & 0.045(11) & 6.0(1.7) & 10.0(2.6) & 16(11) & 15(11) \\
\multicolumn{2}{c}{Class I} & 0.015(4) & 0.04(2) & 0.12(7) & 0.037(7) & 2.3(5) & 18.1(5.4) & 0.3(2) & 0.09(6) \\ \hline
\end{tabular}
\label{tab:hcnoutprops}
\end{table*}

\clearpage

\section{Sample \HCOplus\ and HCN spectra}
We present the \HCOplus\ and HCN spectra at the position of every H07 SCUBA core identified, where there is emission from both molecules, and the \HCOplus\ is at least at the $3\sigma$ level. The \HCOplus\ spectra are plotted in black (solid lines), and the HCN spectra are plotted in red (dot-dashed lines).

\spectra{n1333source41}{n1333source42}{n1333source43}
\spectra{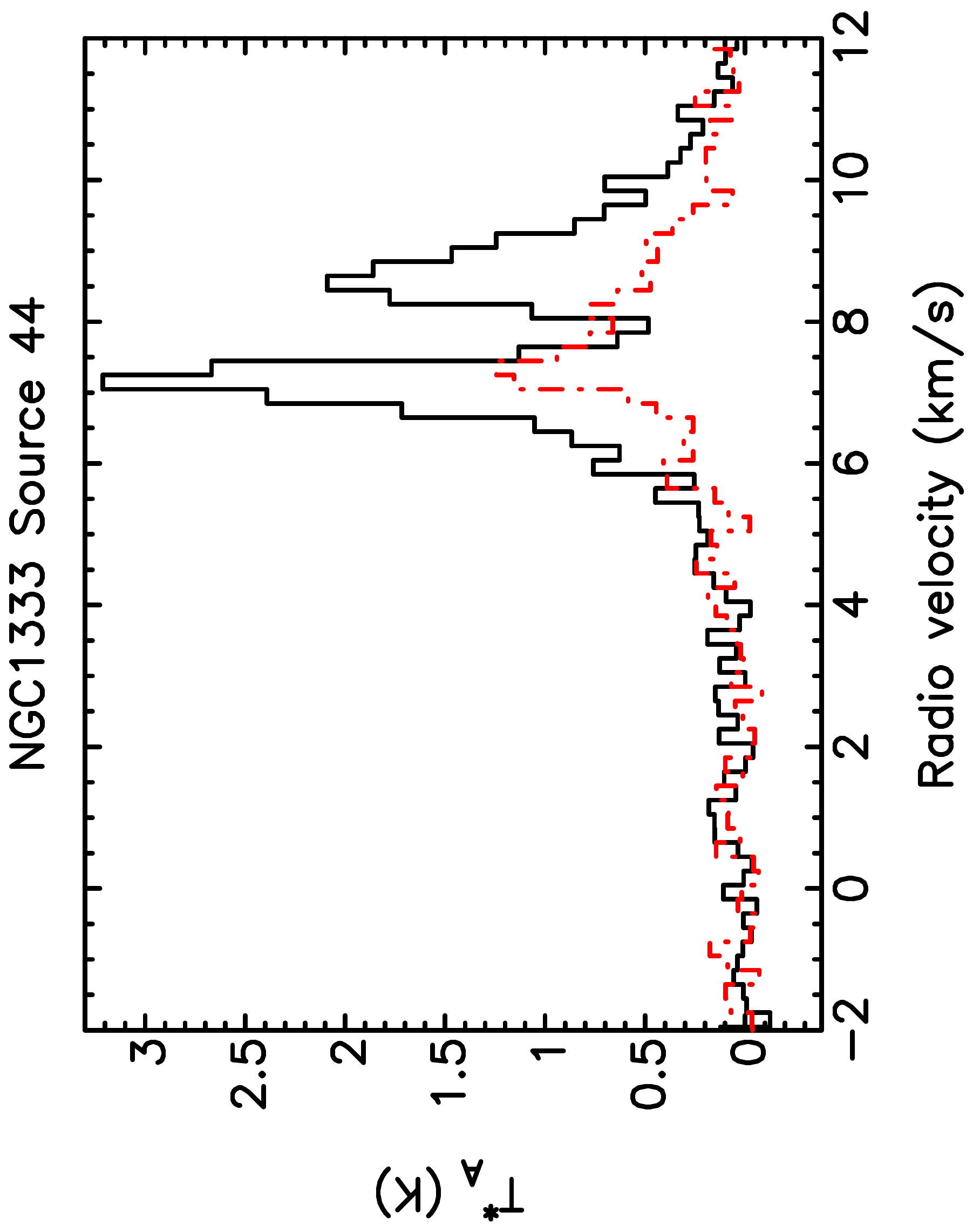}{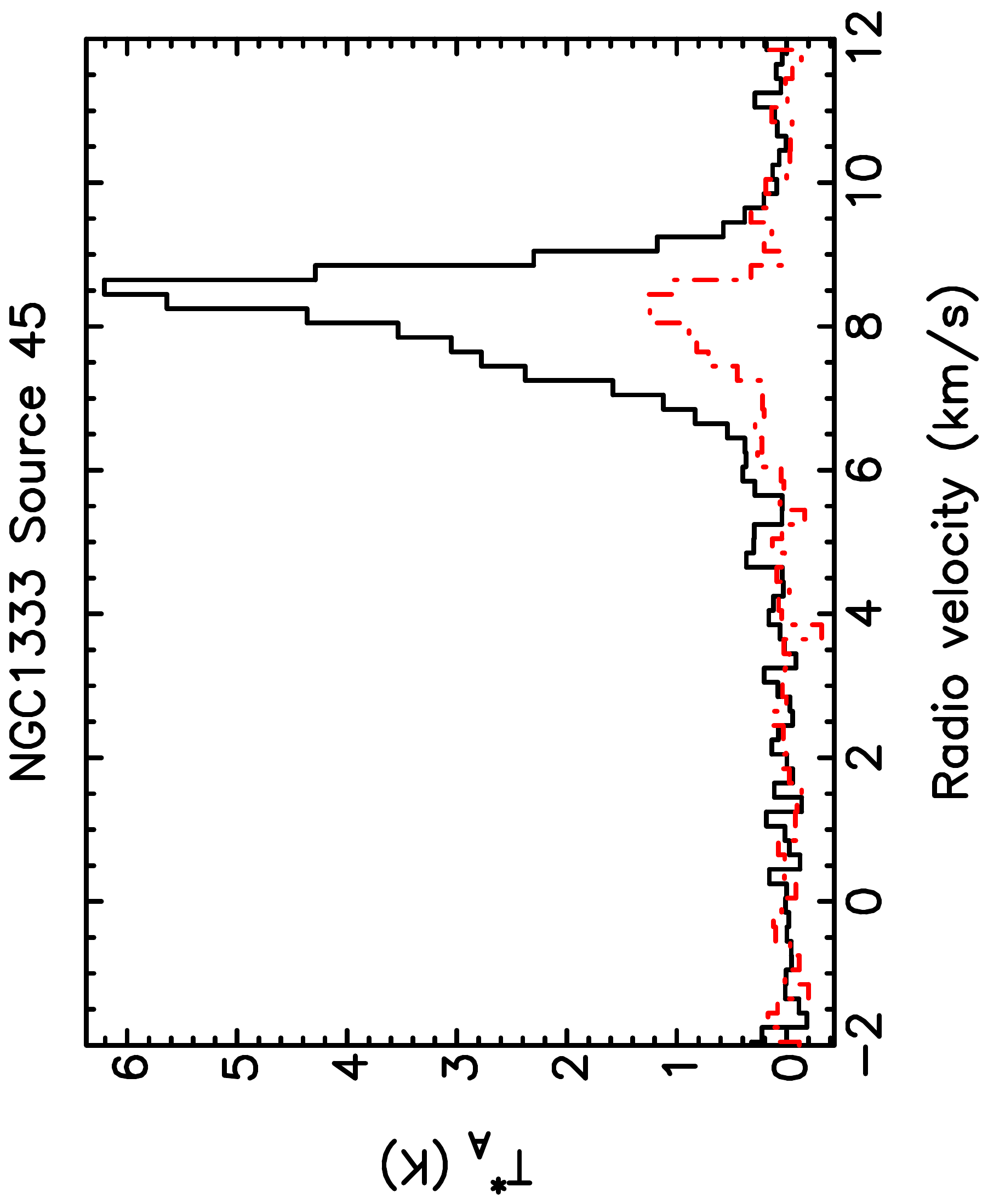}{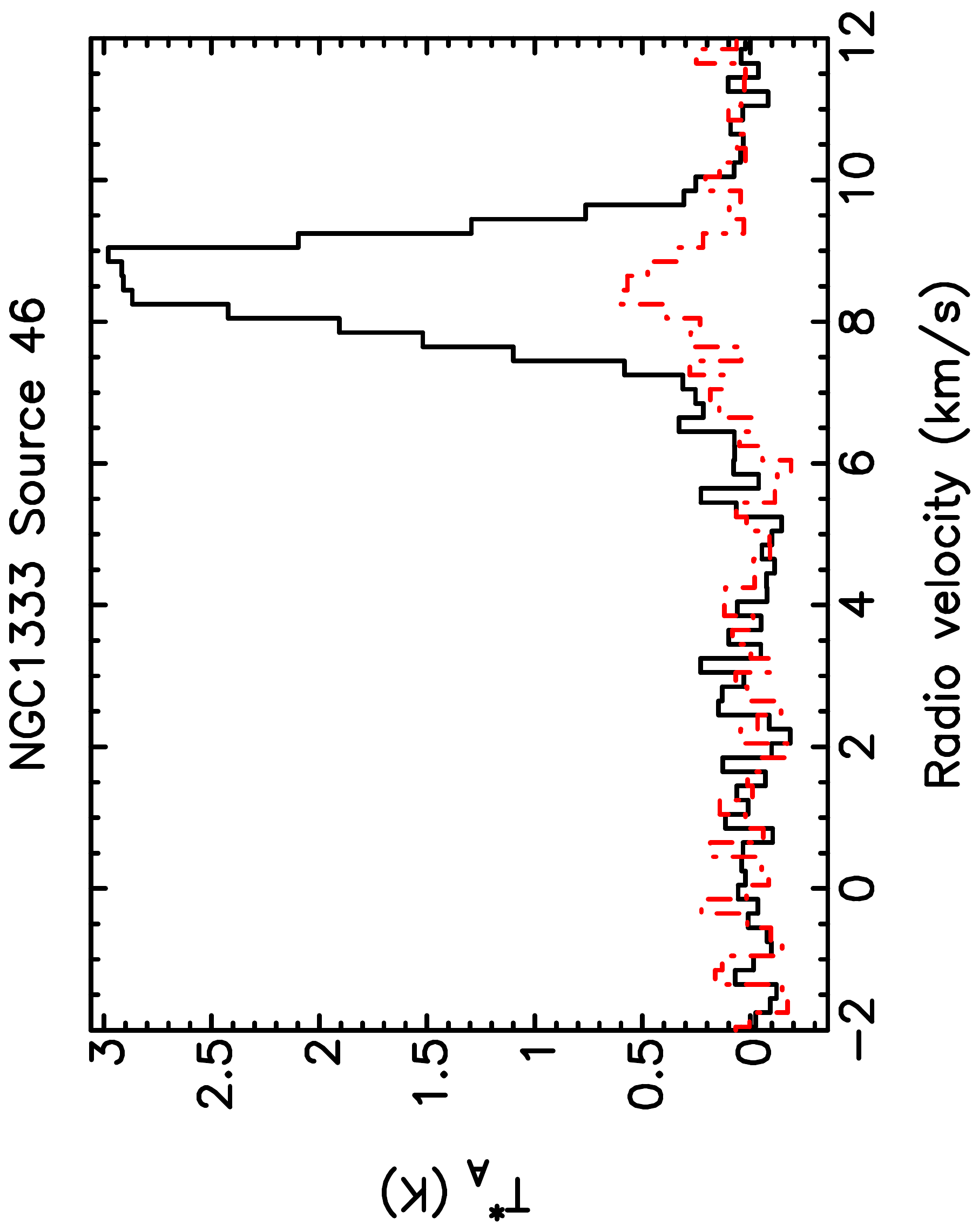}
\spectra{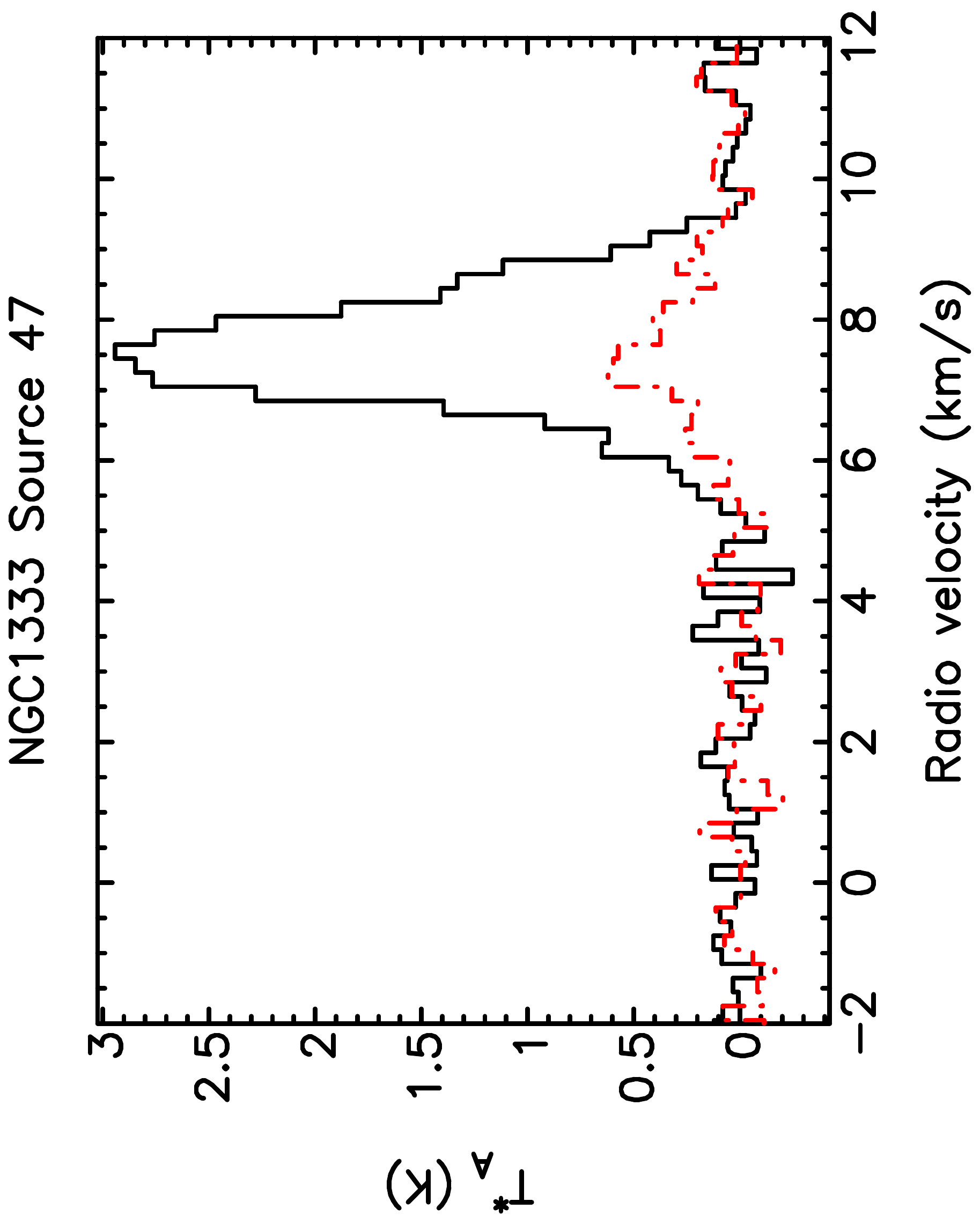}{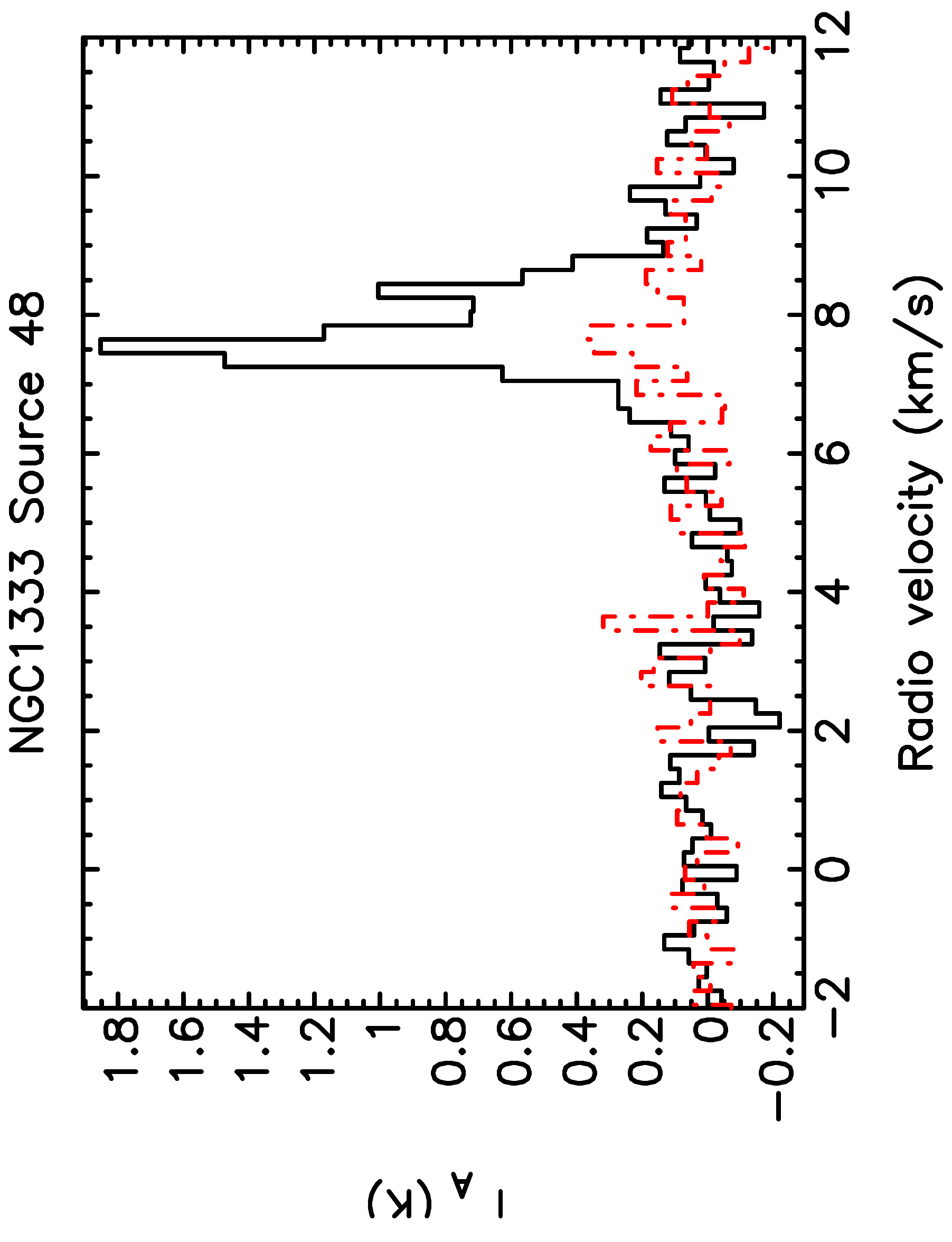}{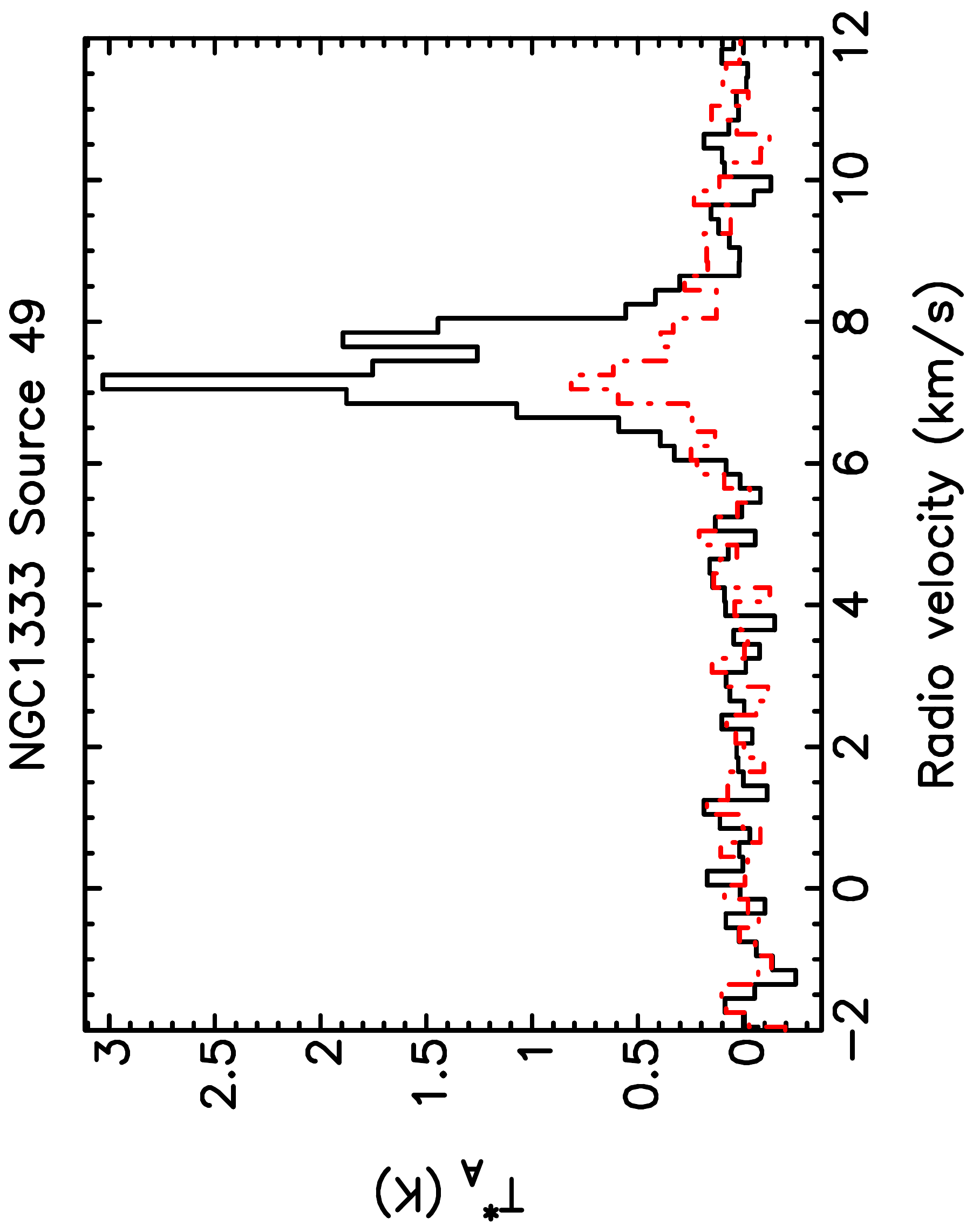}
\spectra{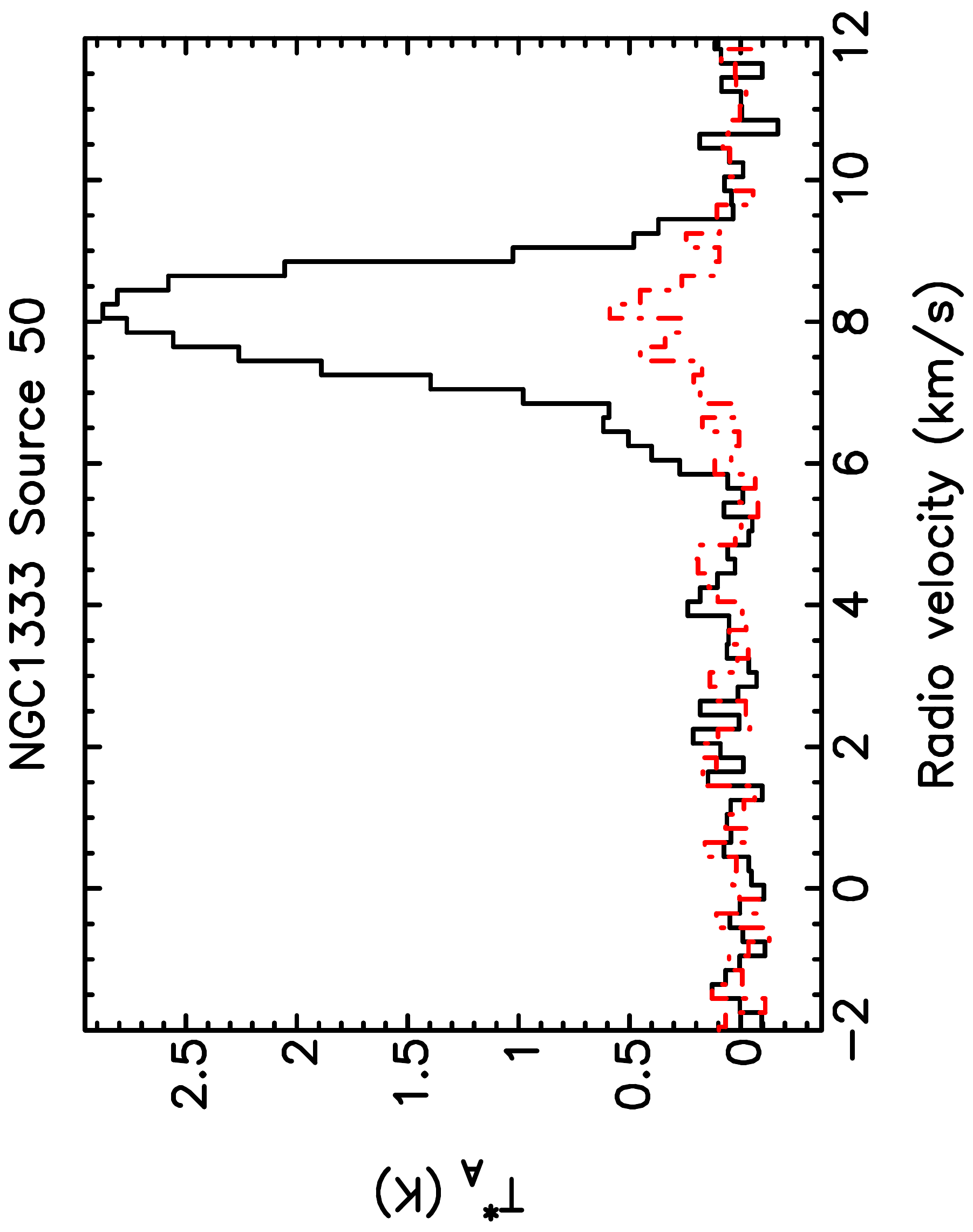}{n1333source51}{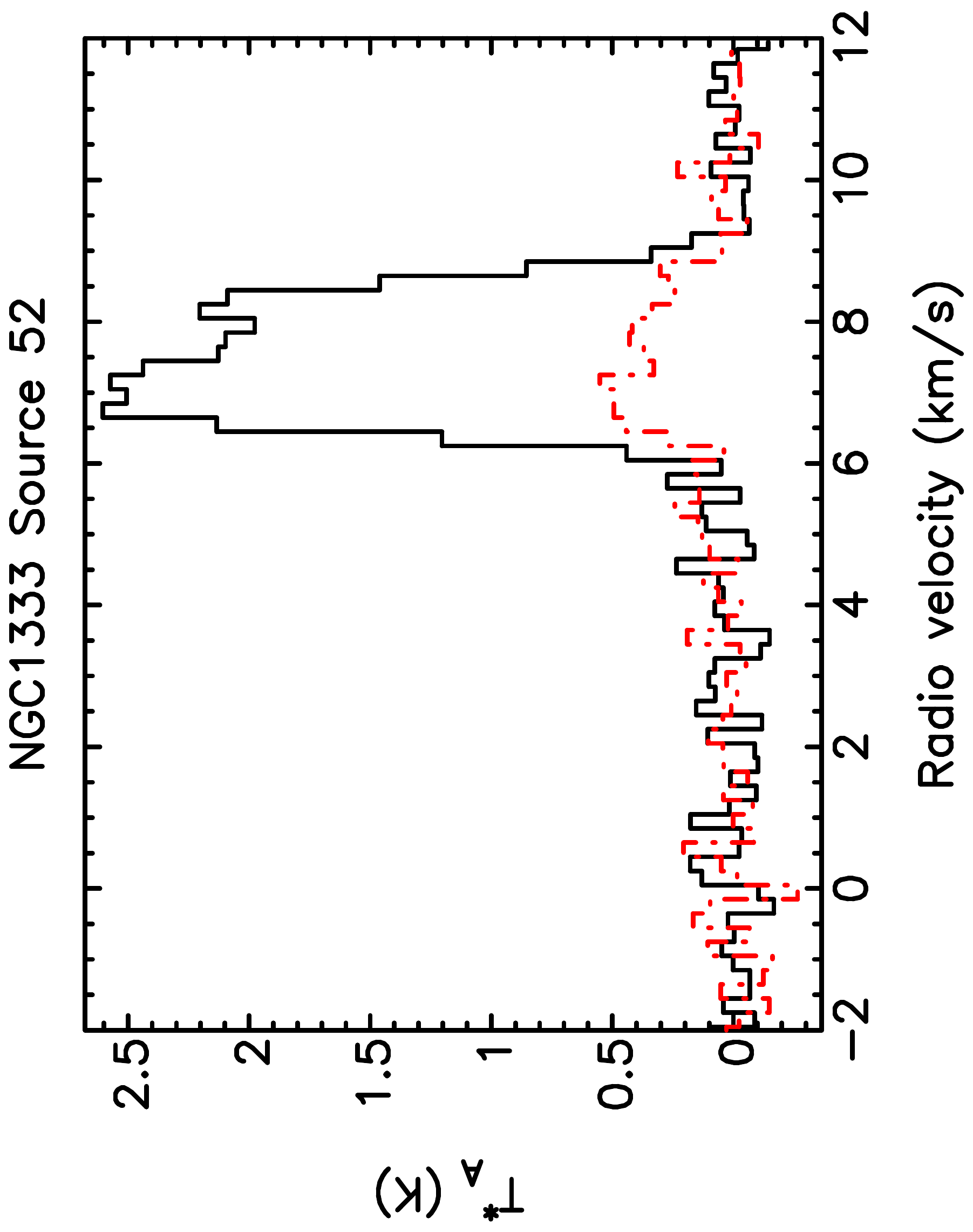}
\spectra{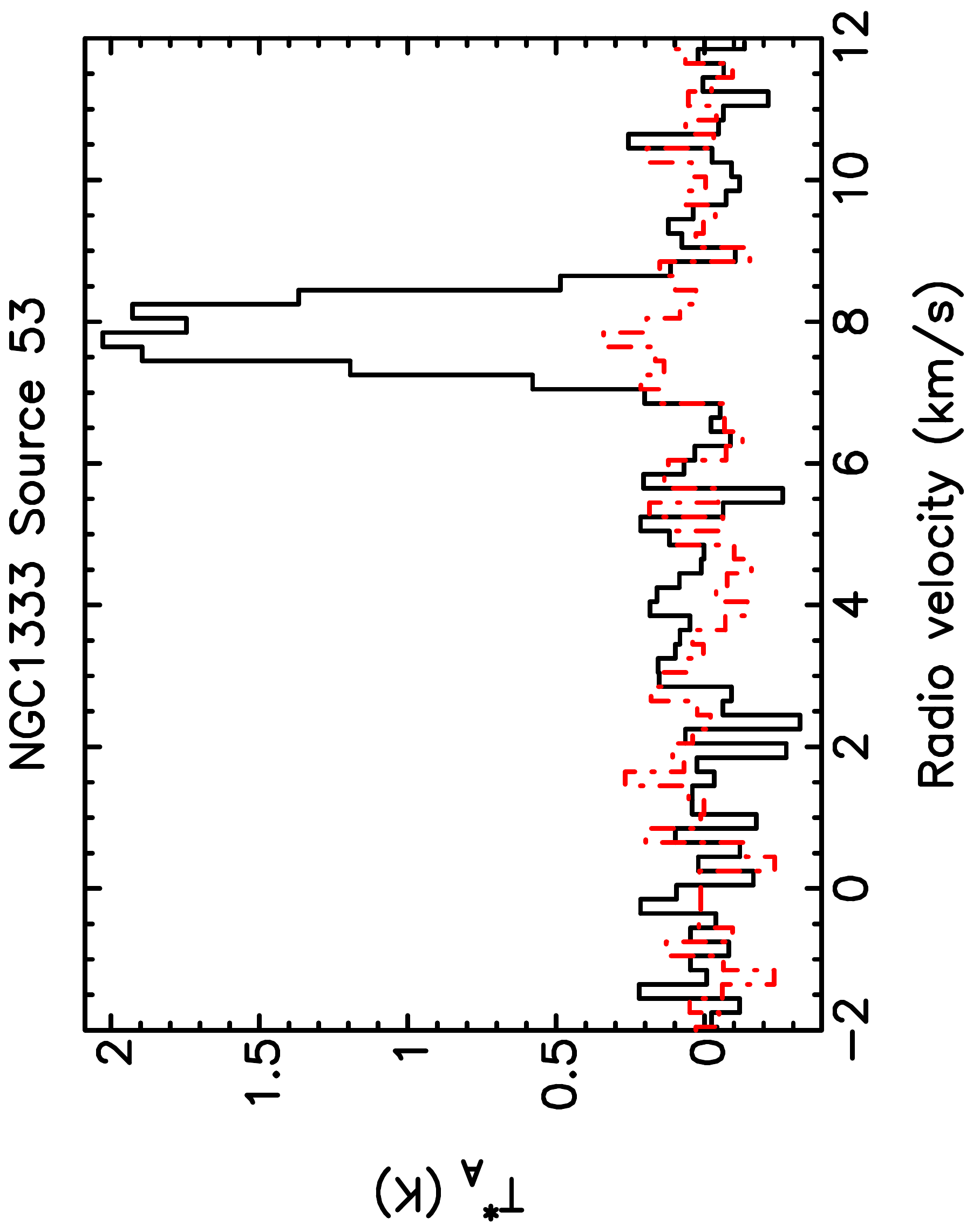}{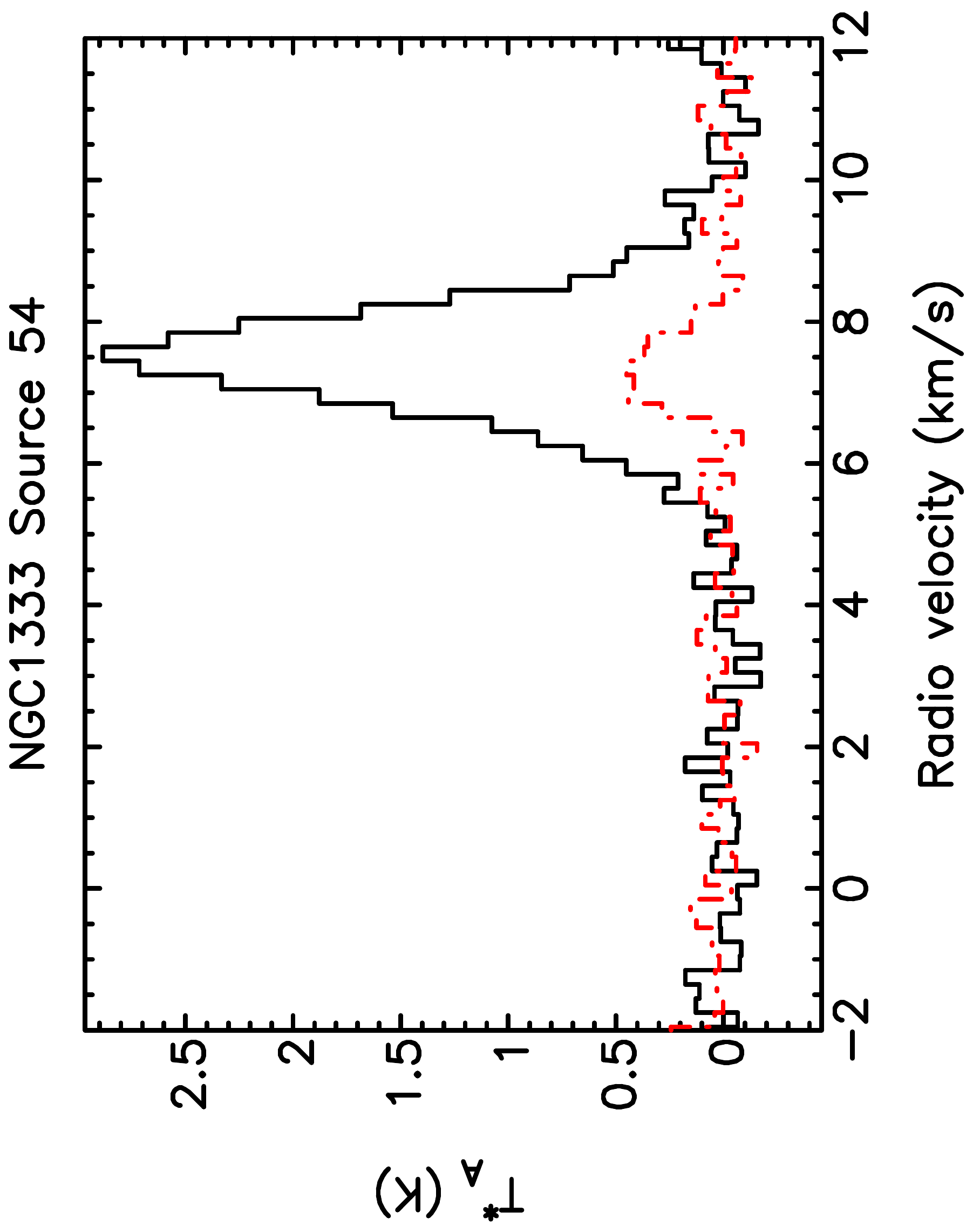}{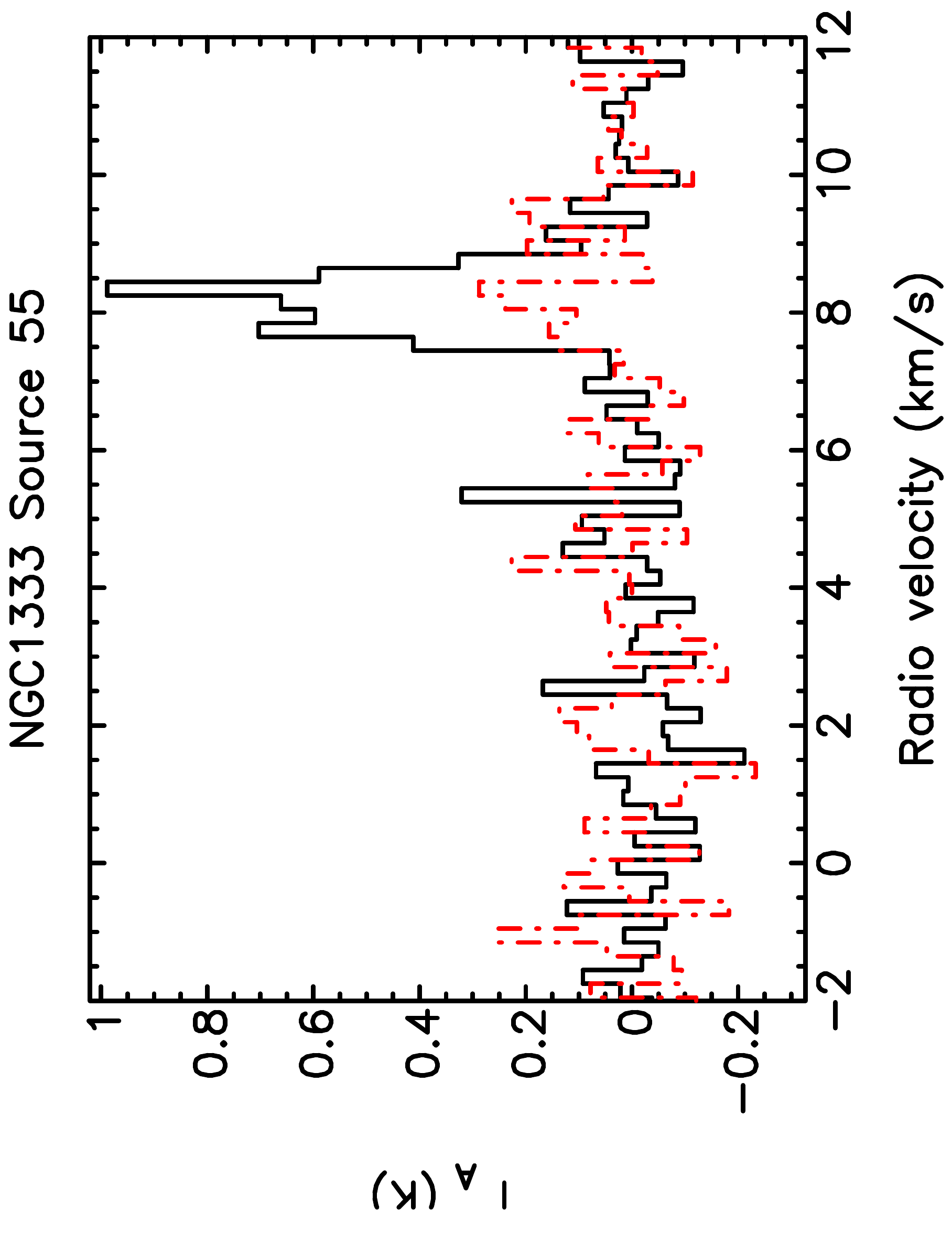}
\spectra{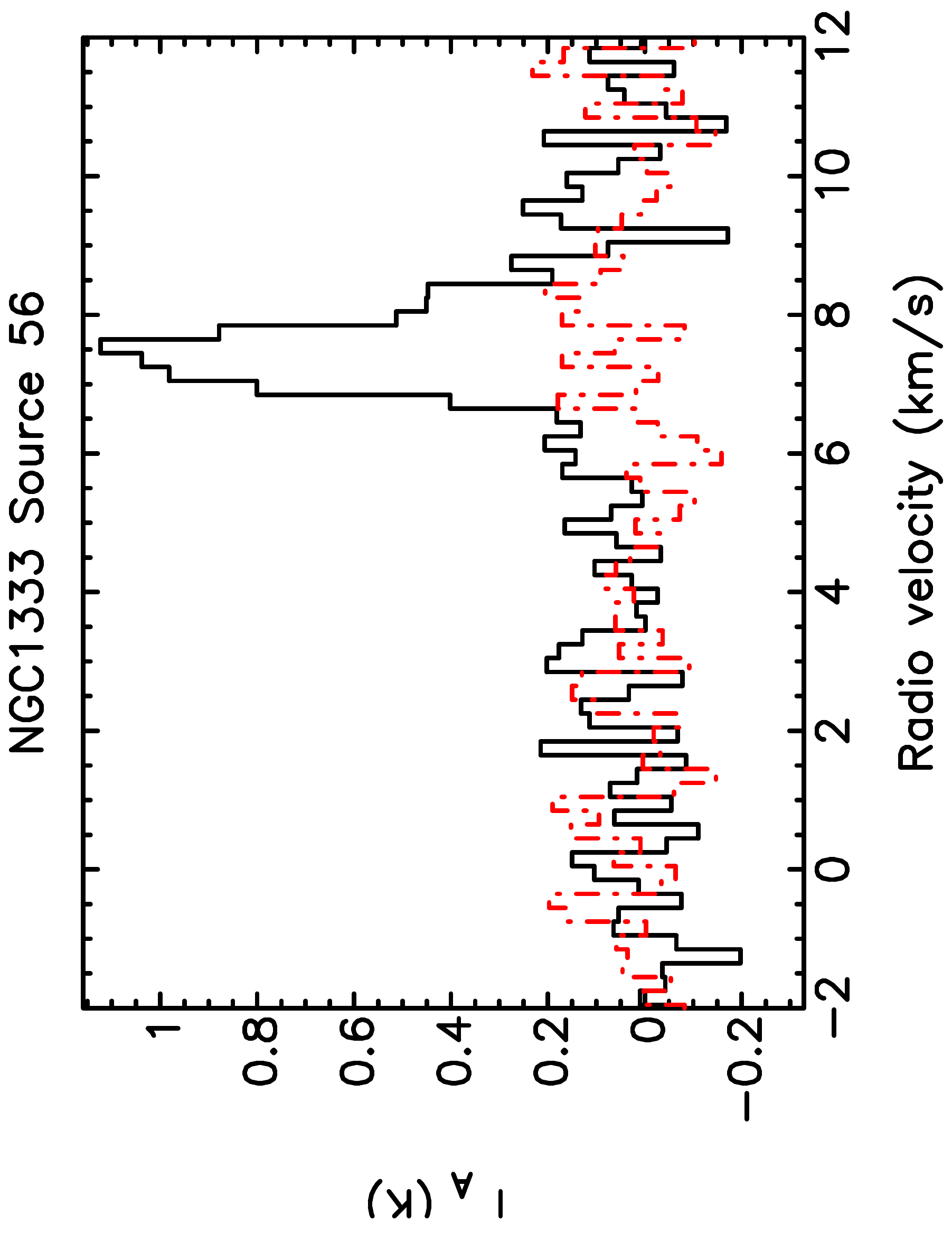}{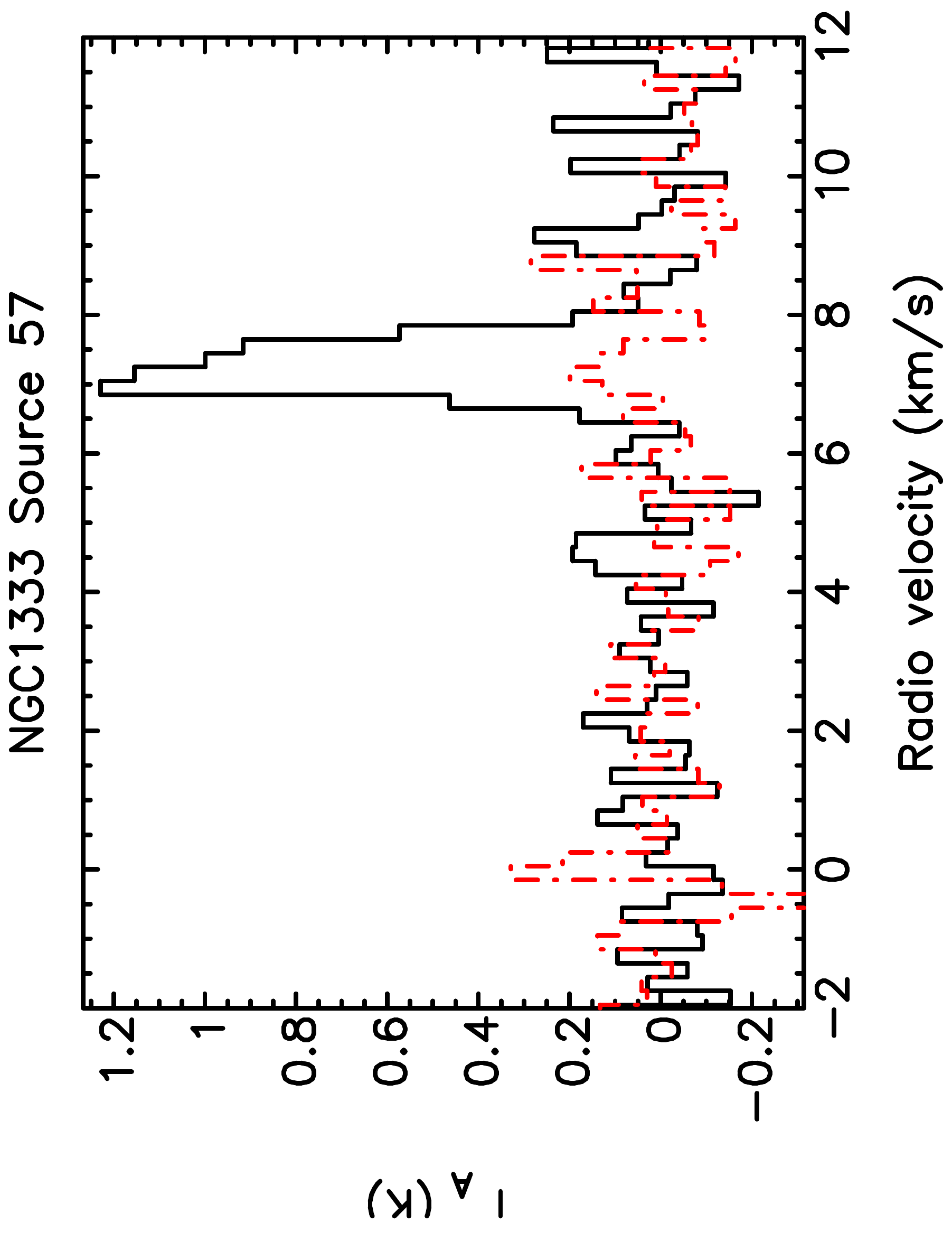}{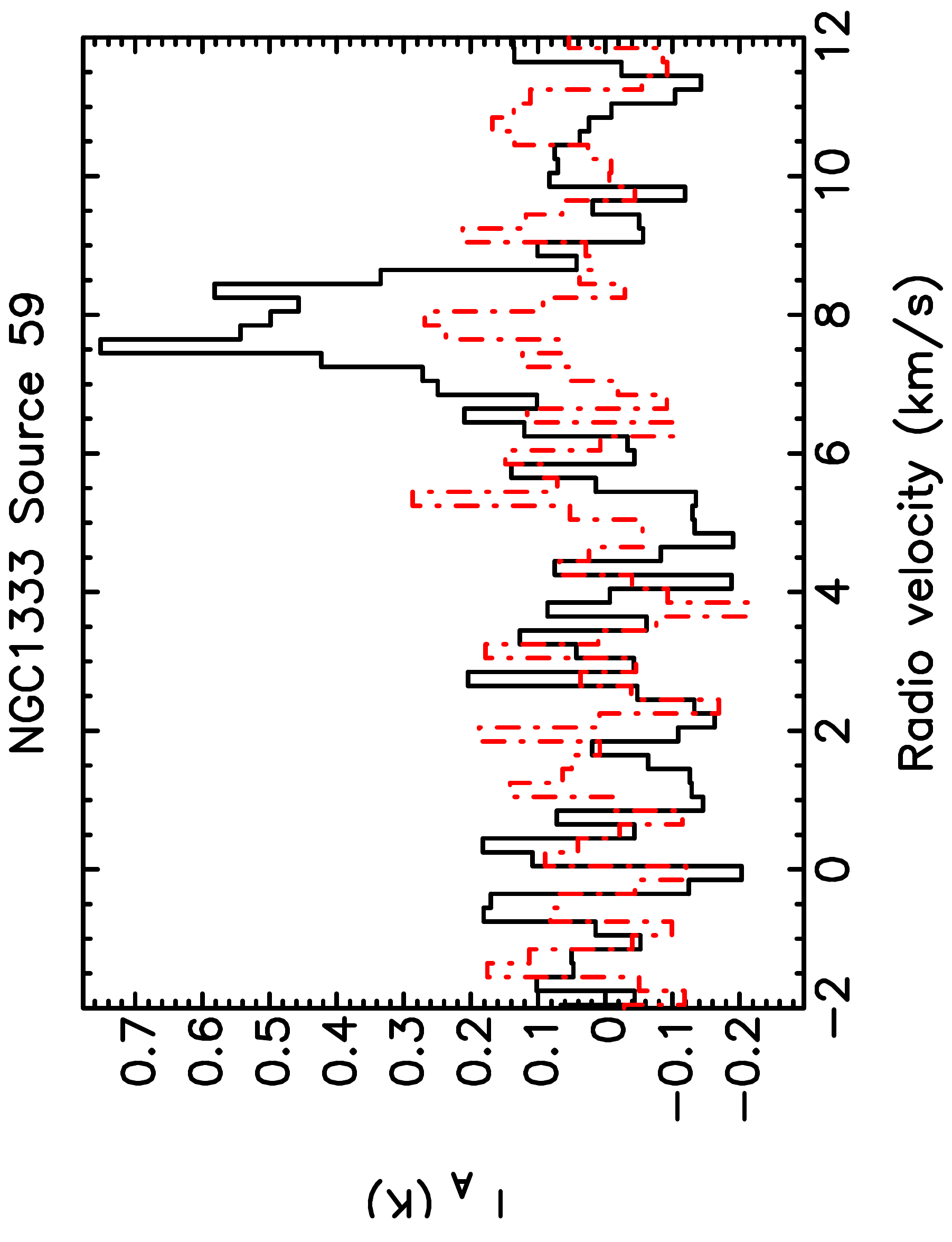}
\spectra{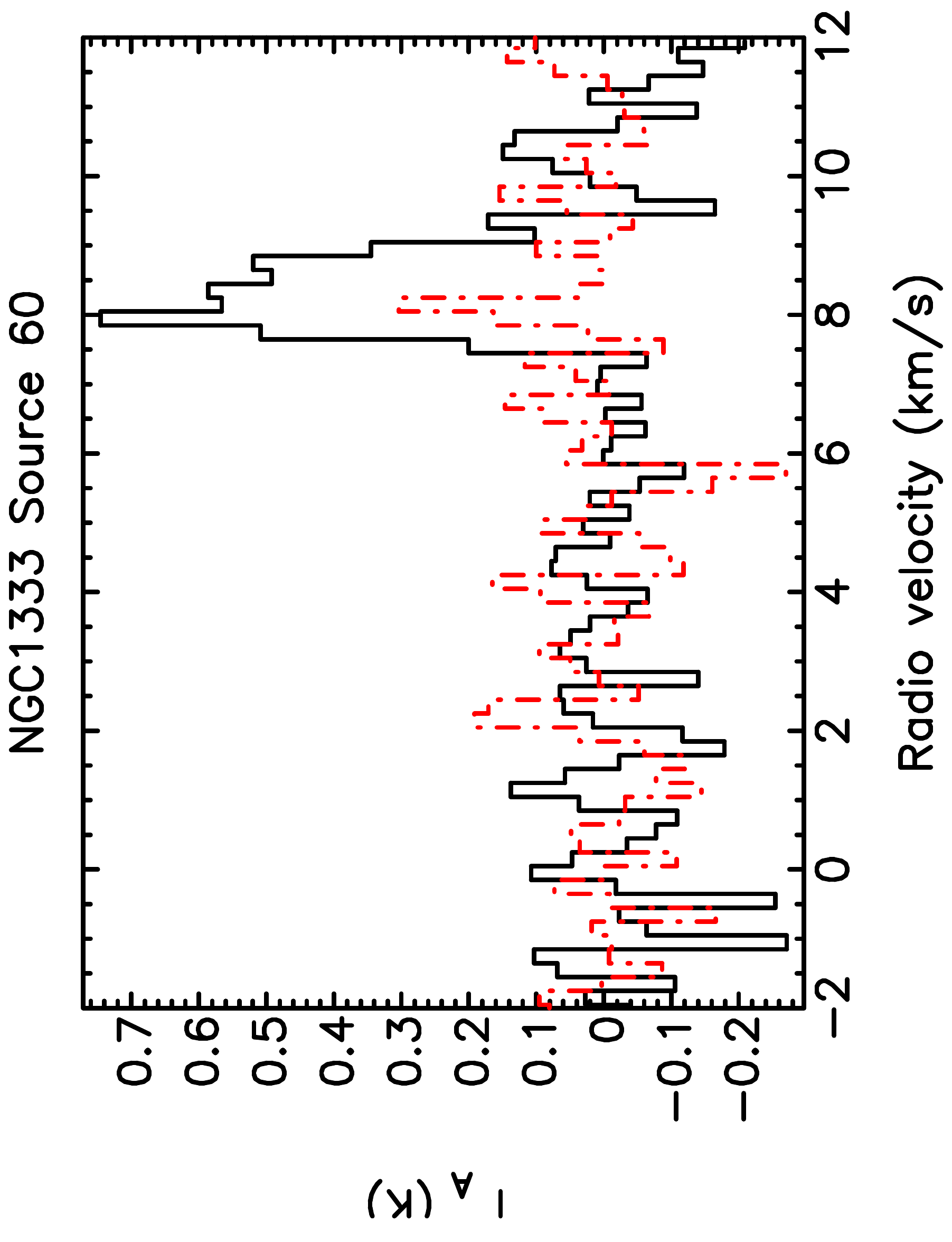}{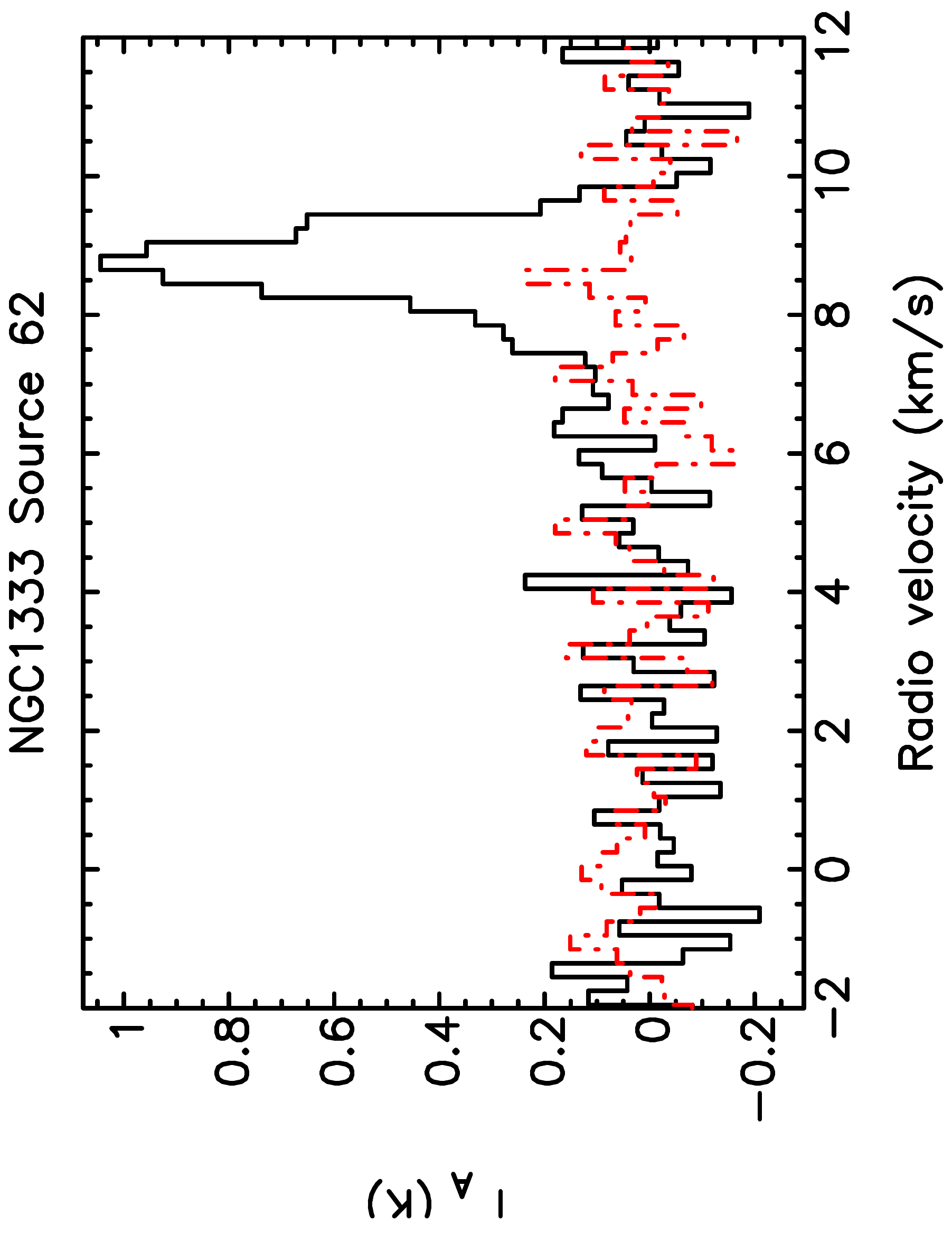}{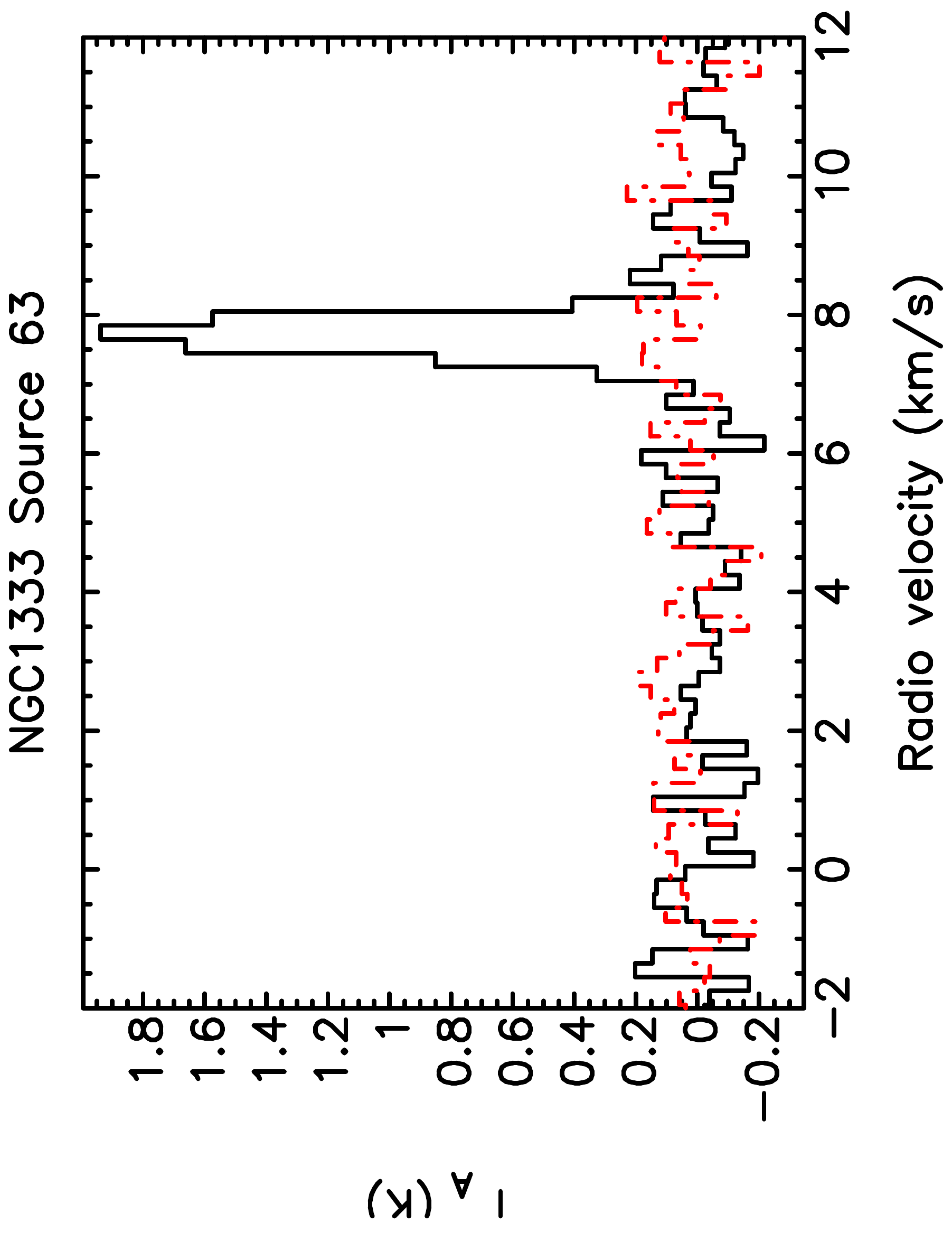}
\spectra{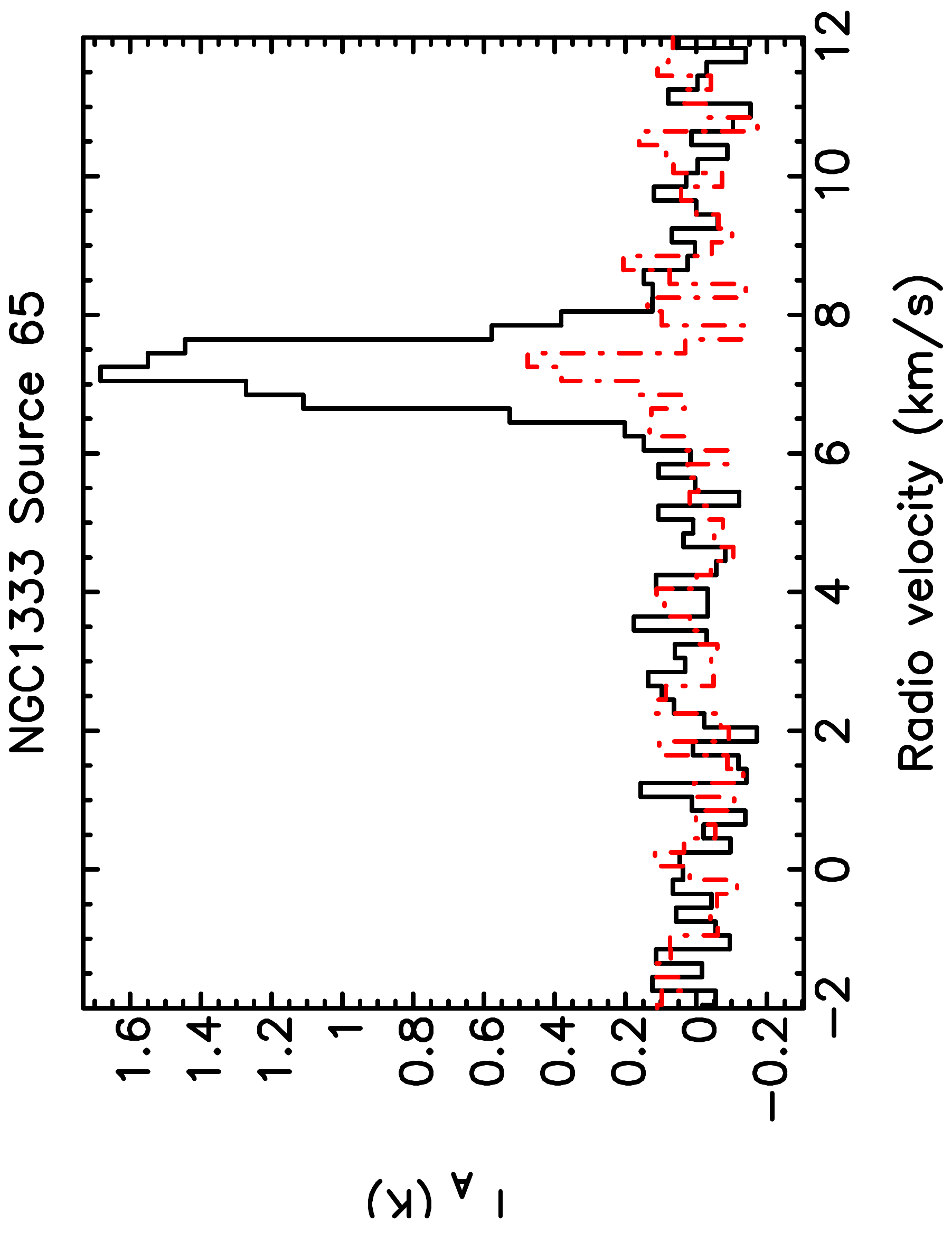}{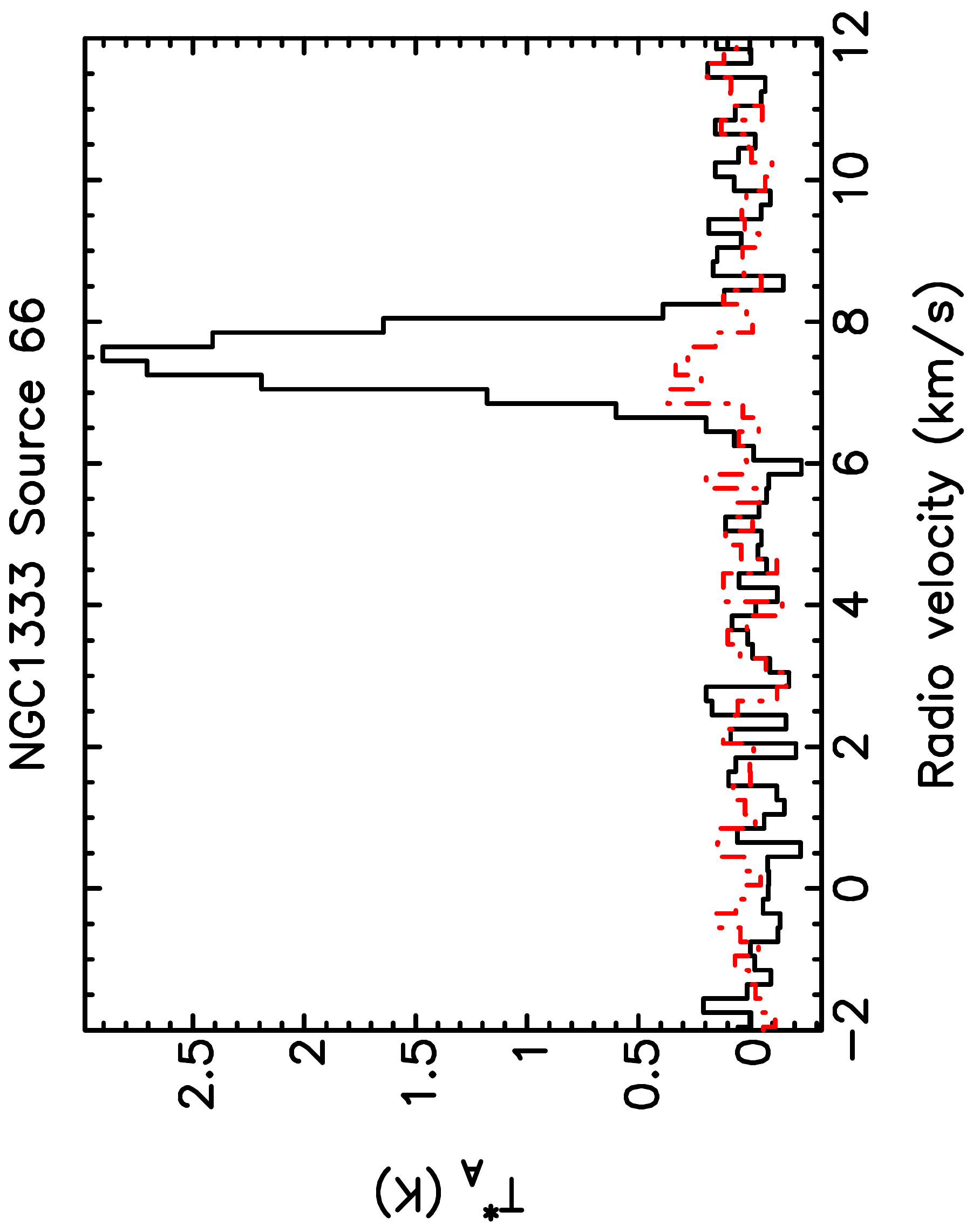}{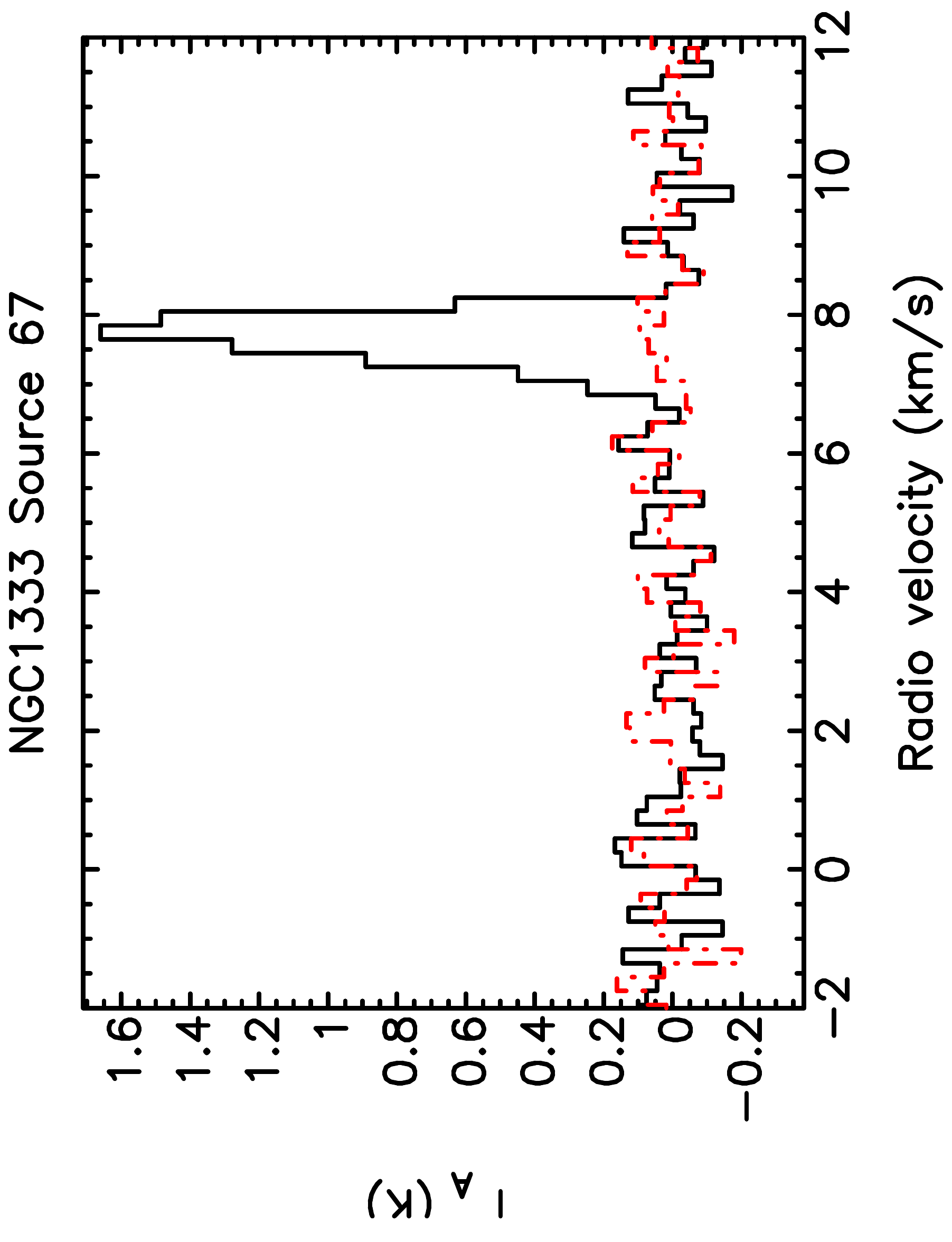}
\spectra{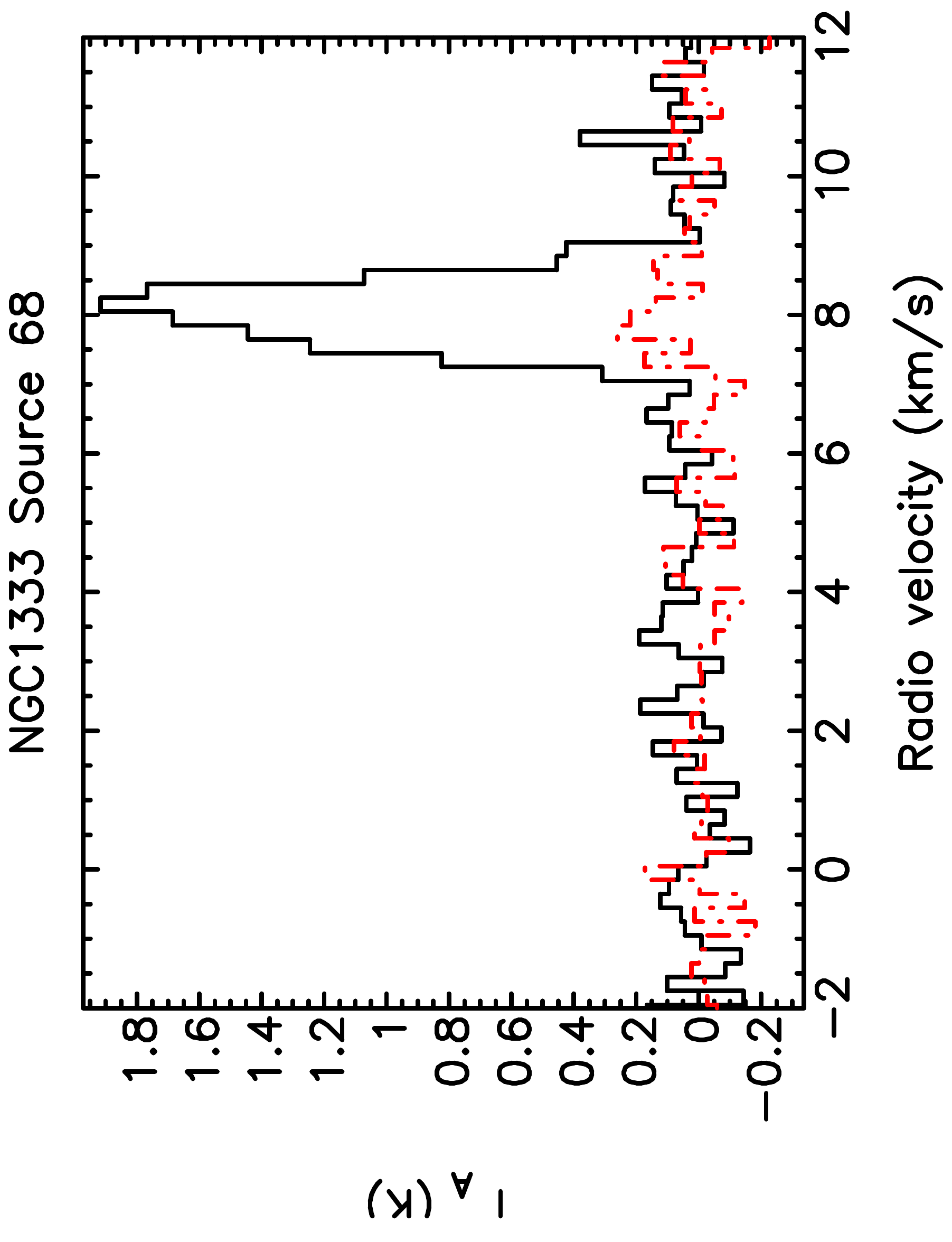}{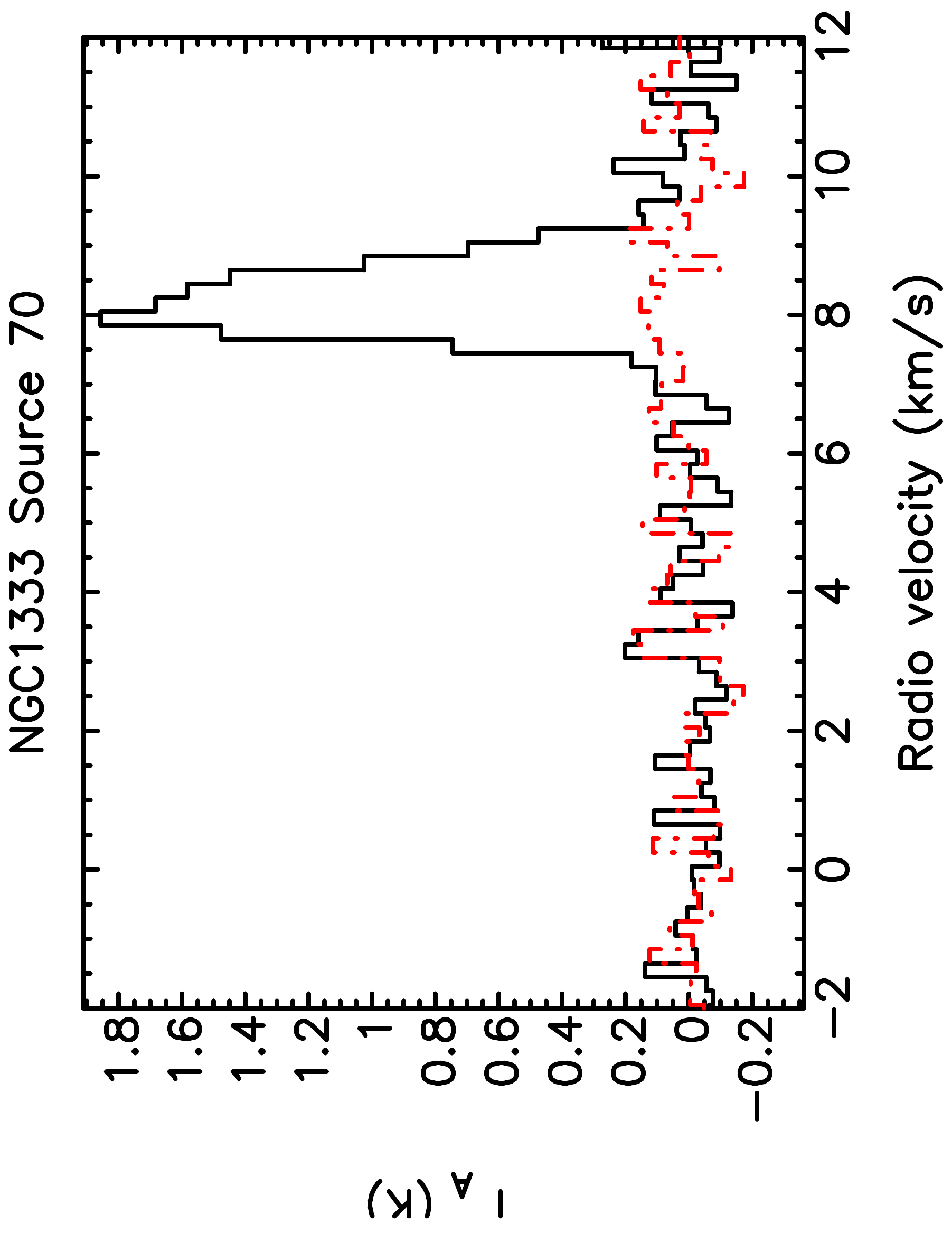}{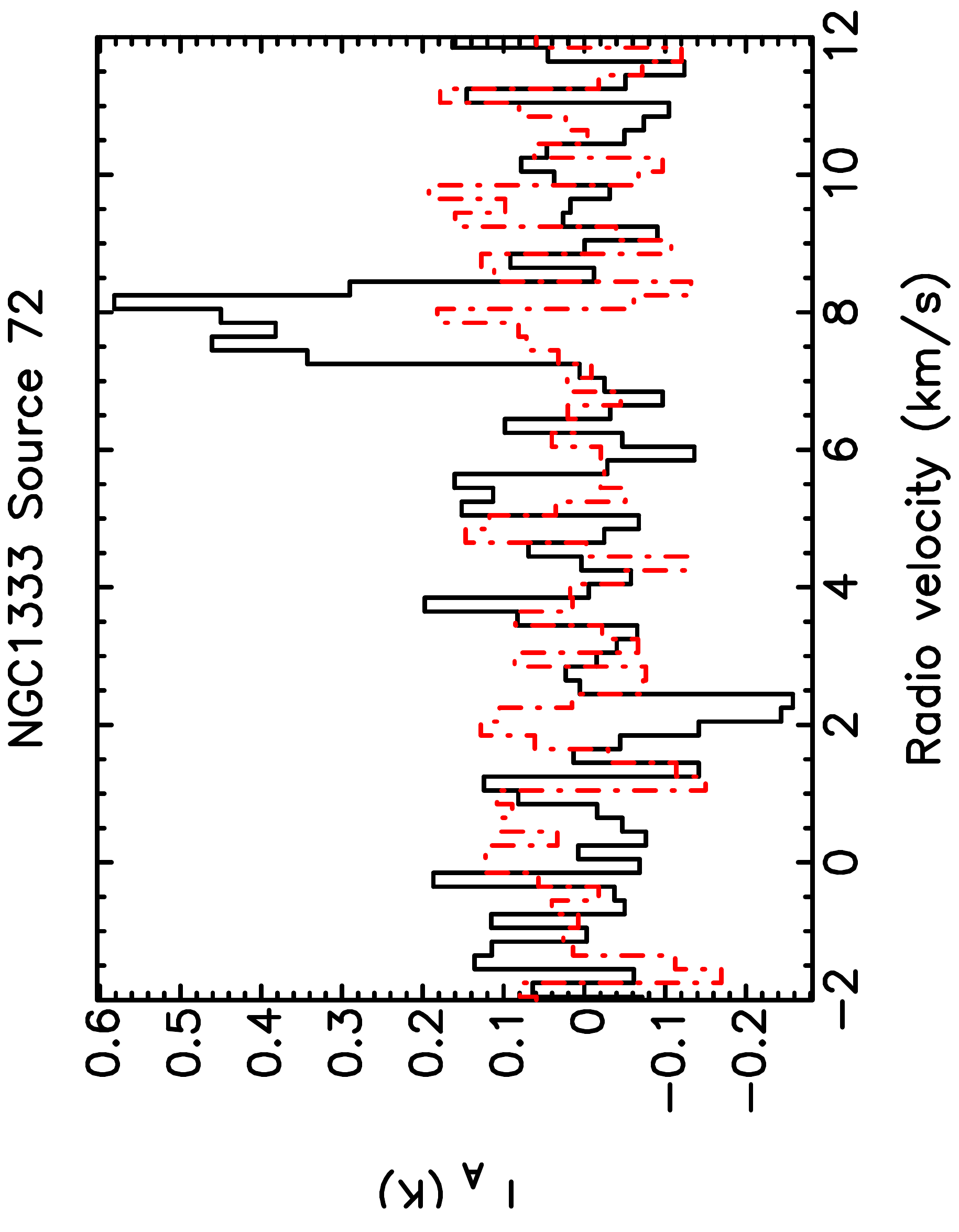}
\spectra{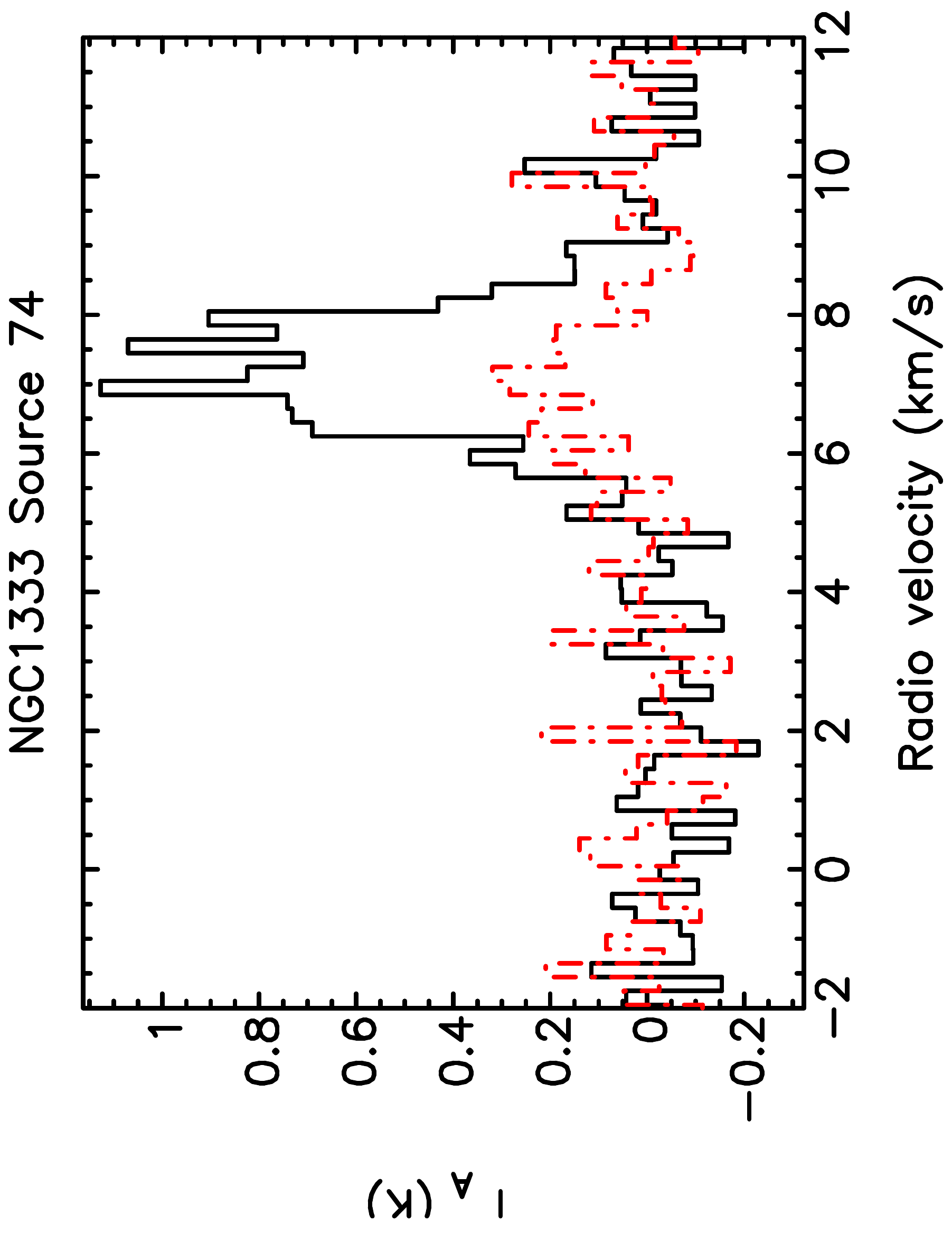}{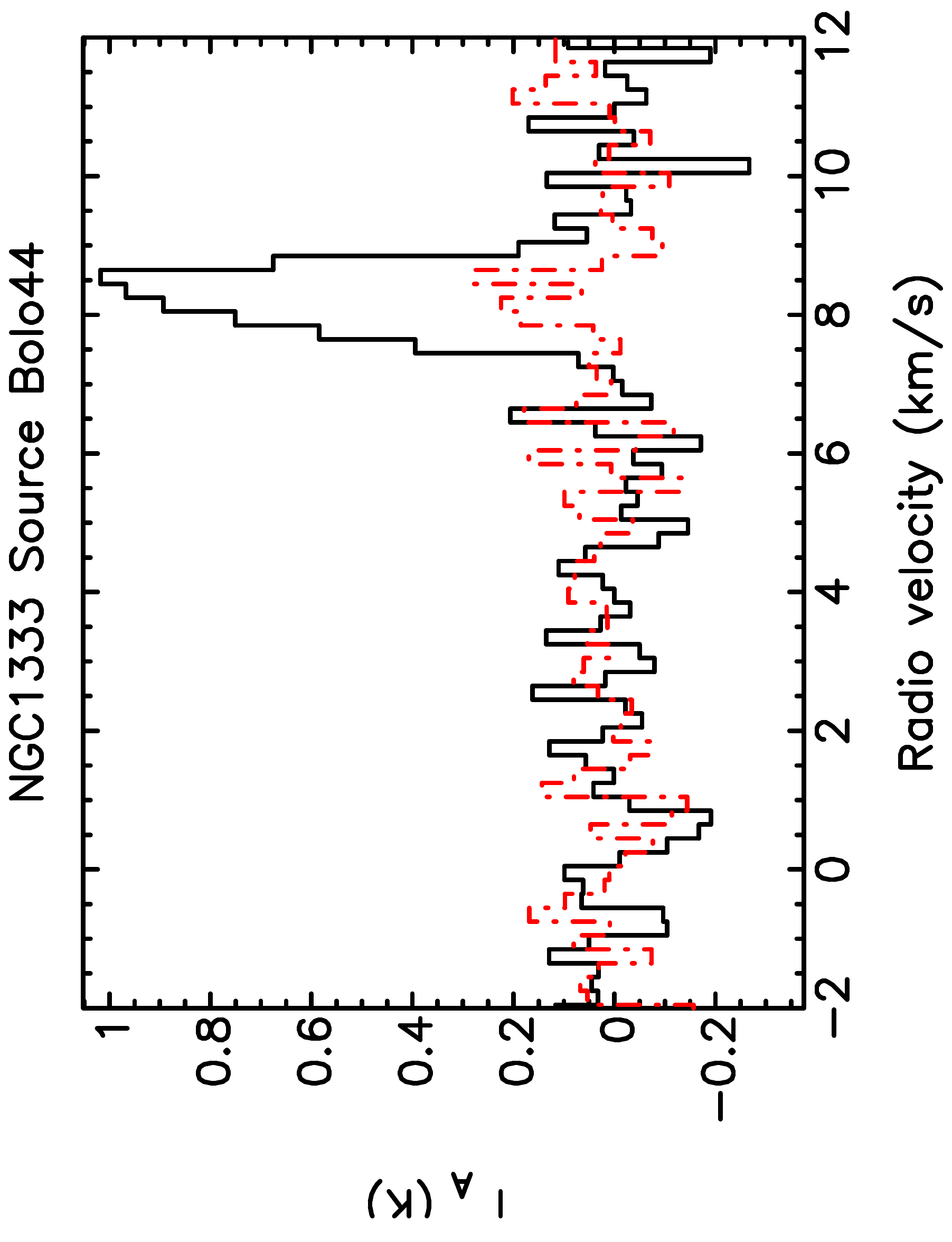}{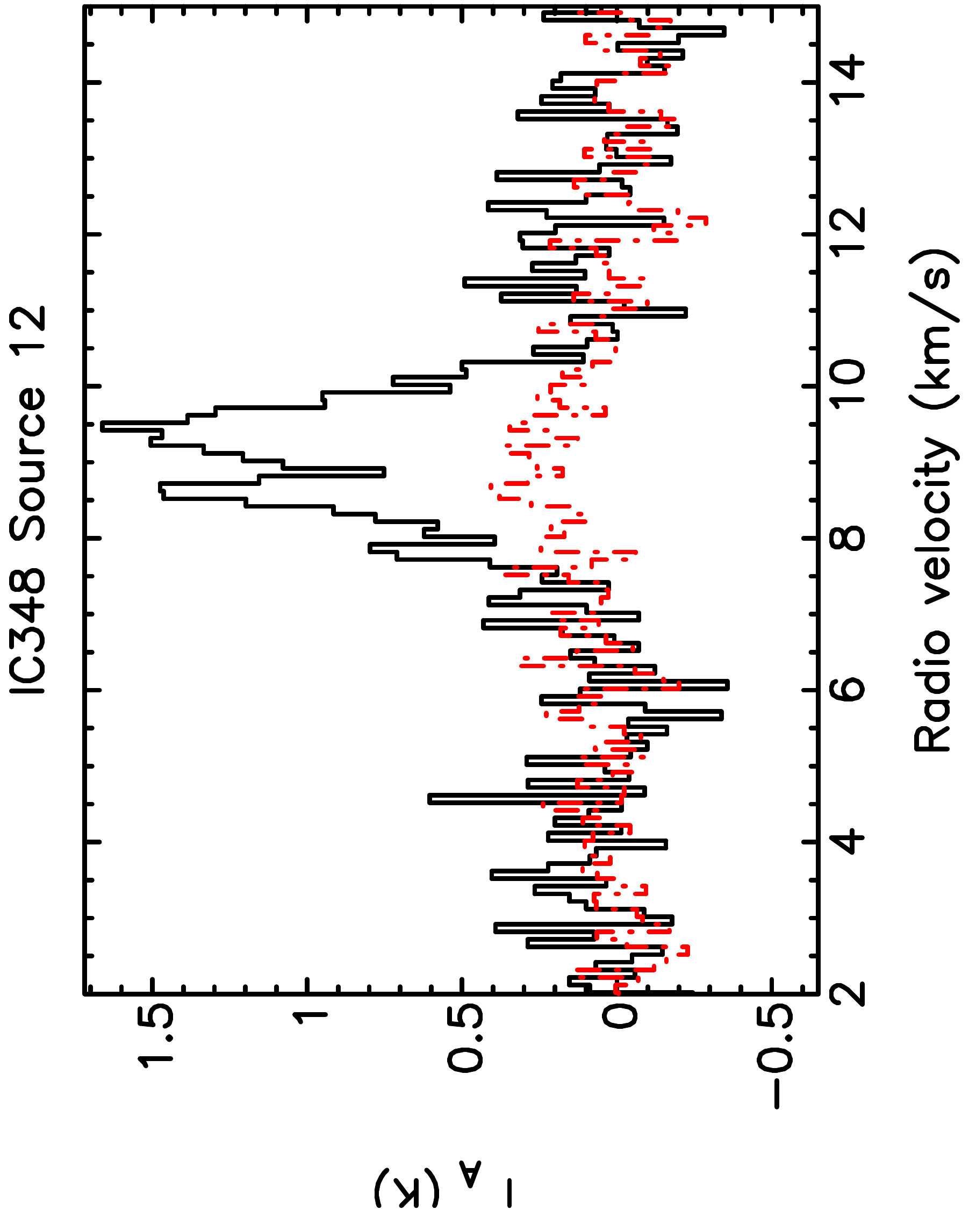}
\spectra{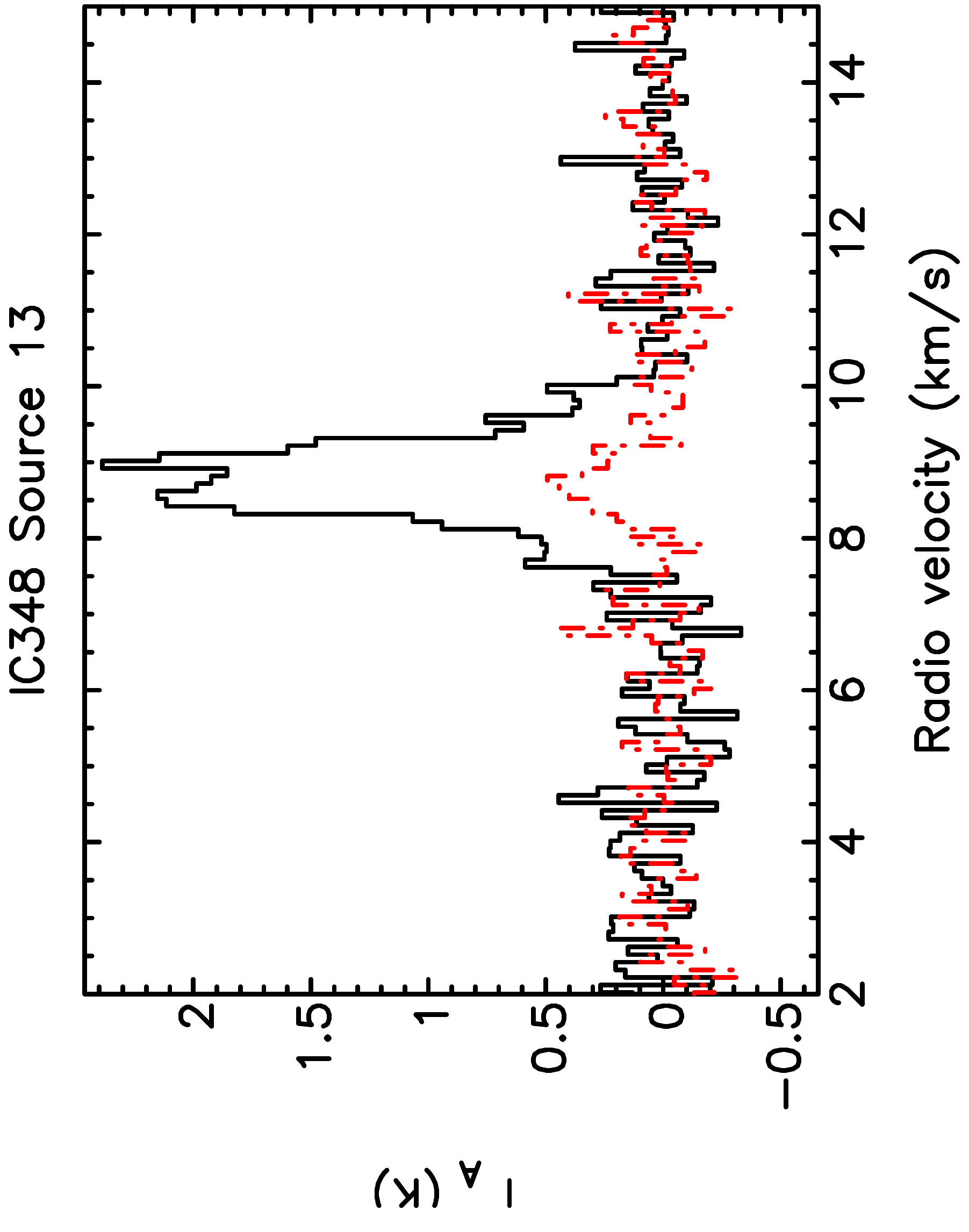}{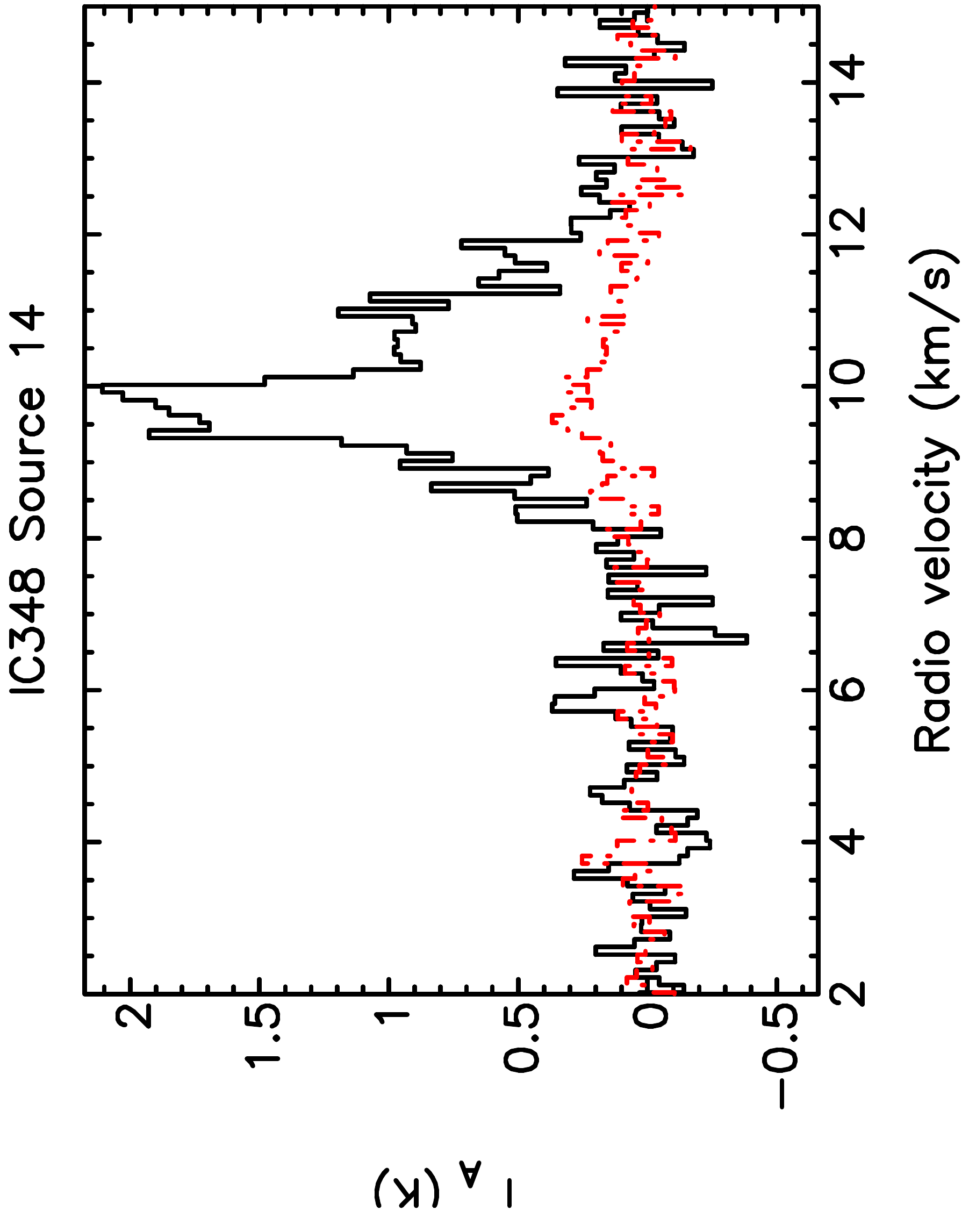}{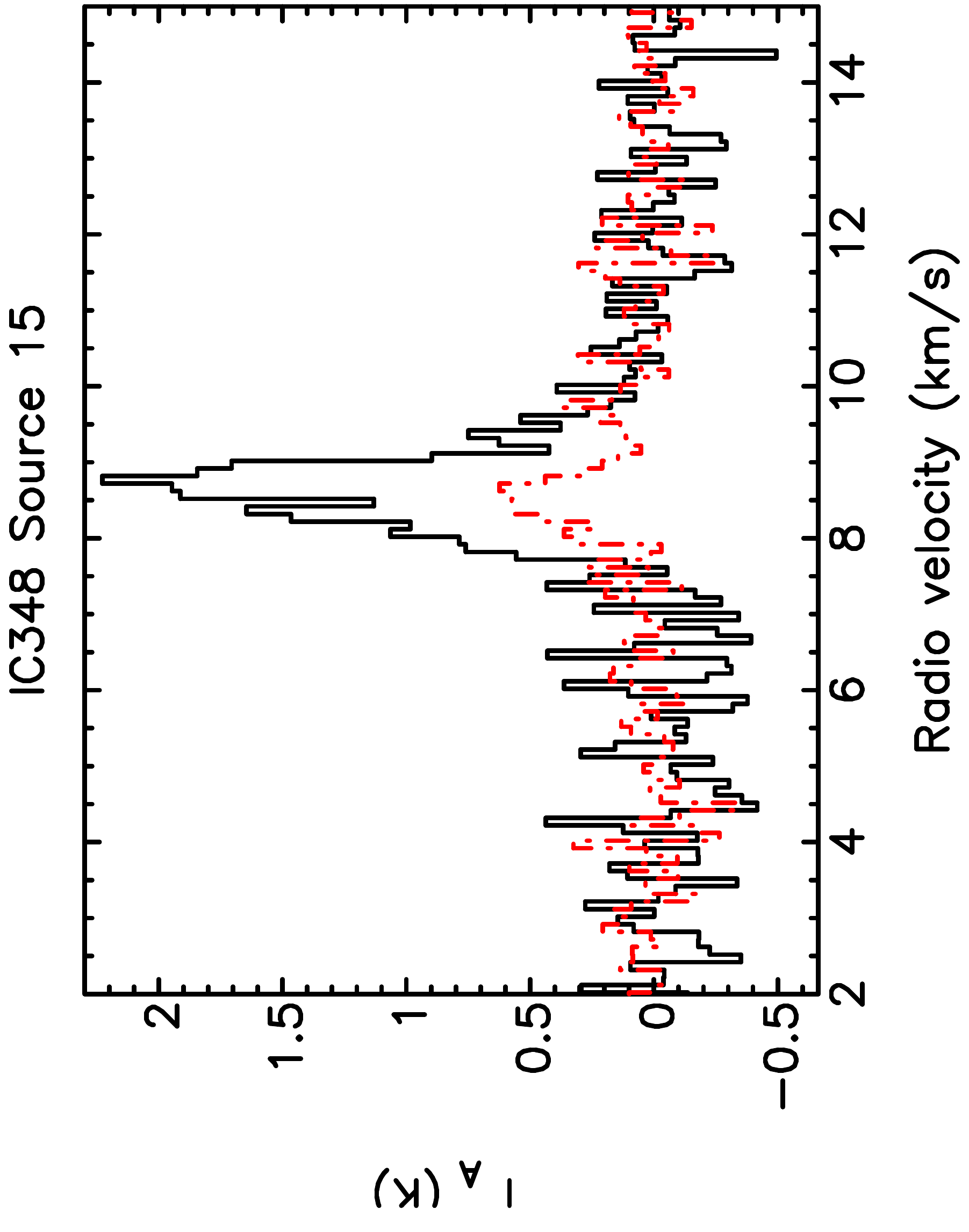}
\spectra{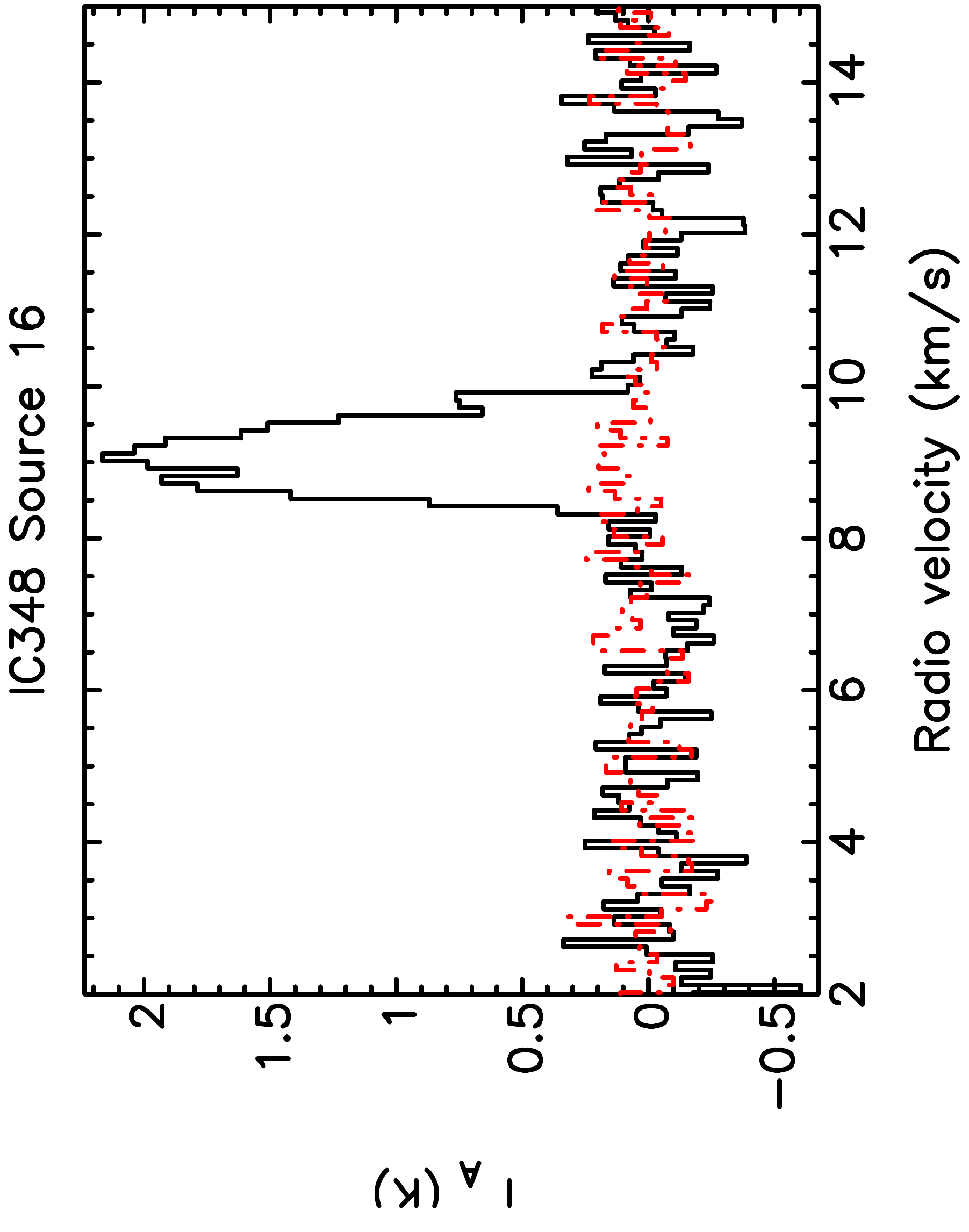}{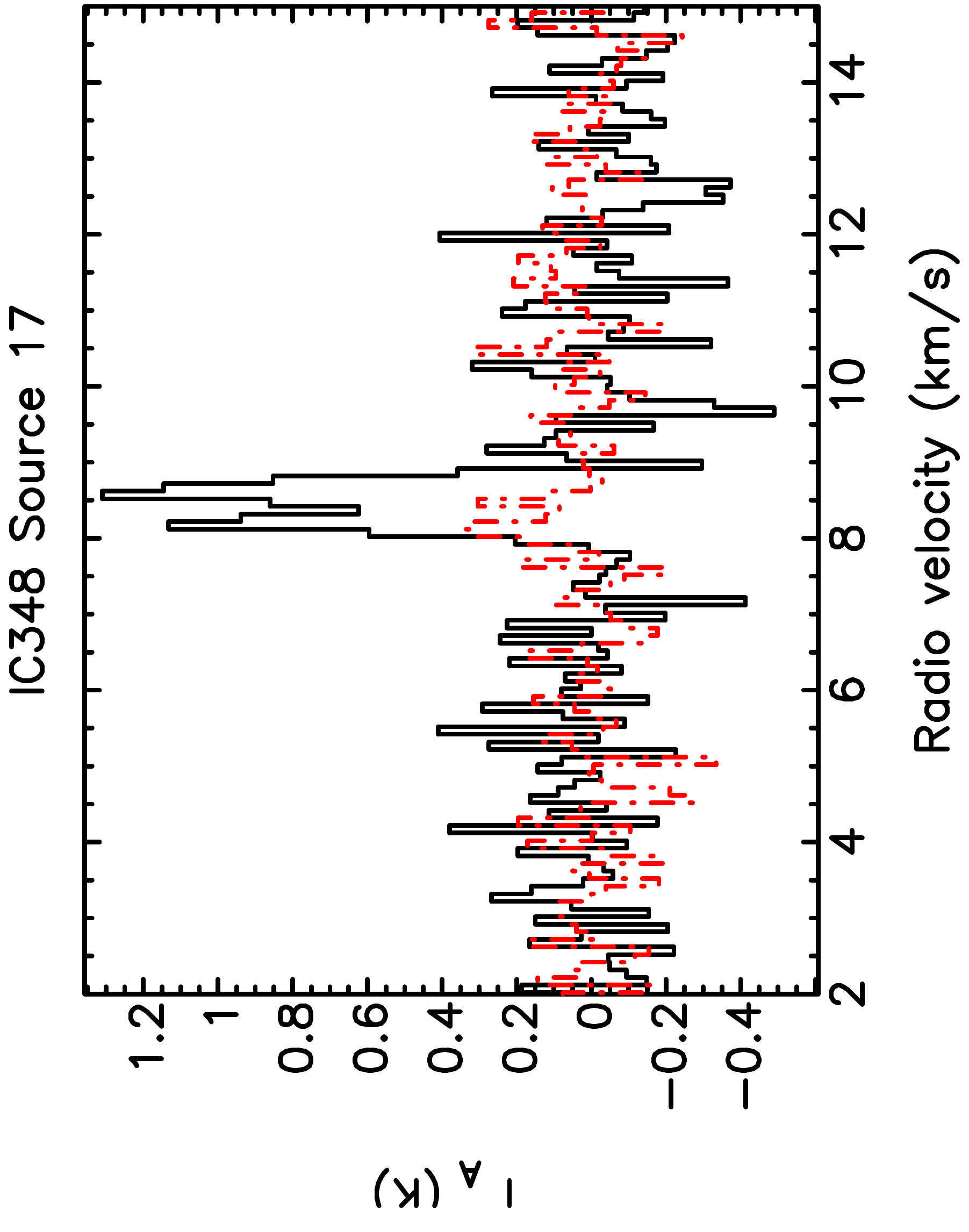}{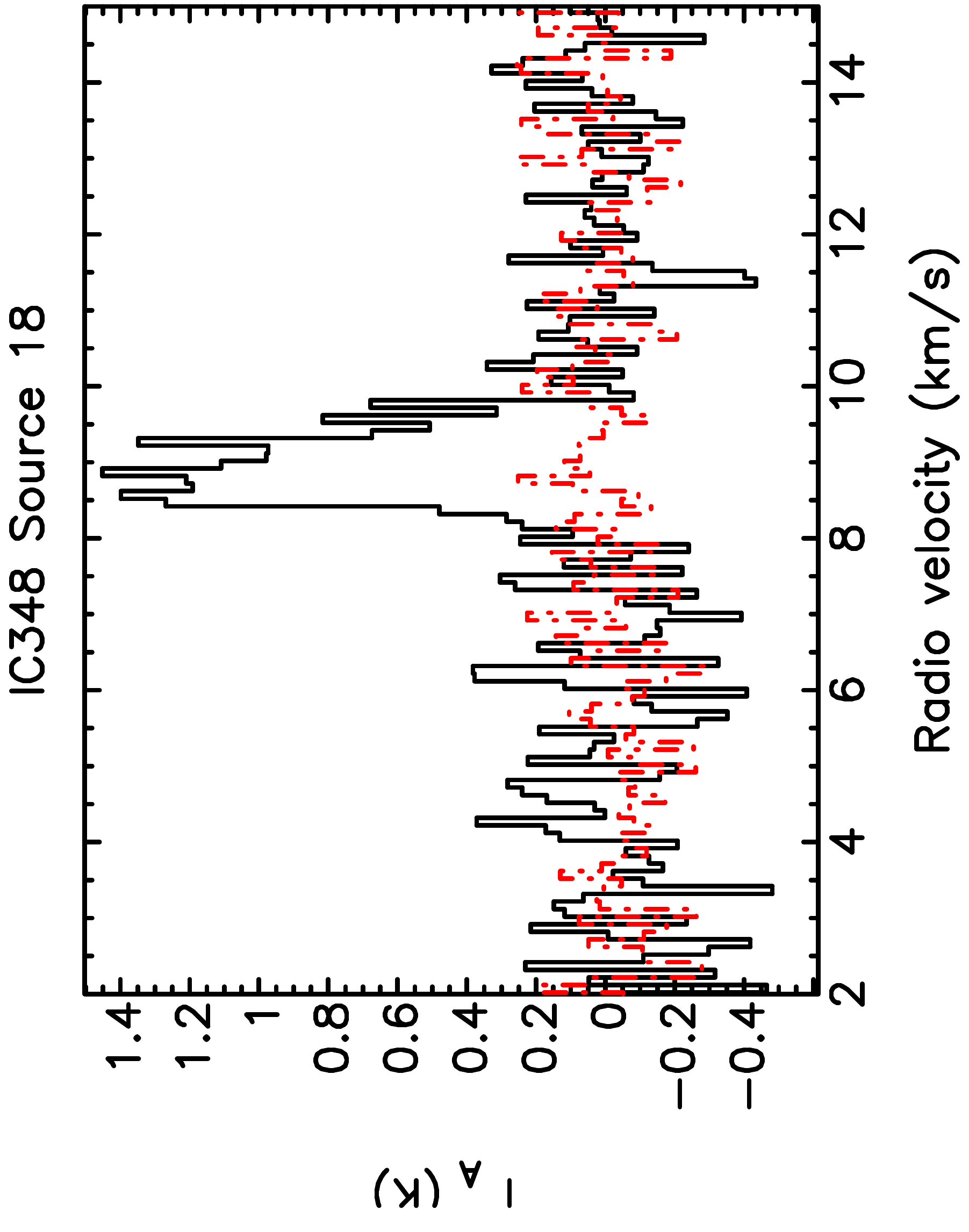}
\spectra{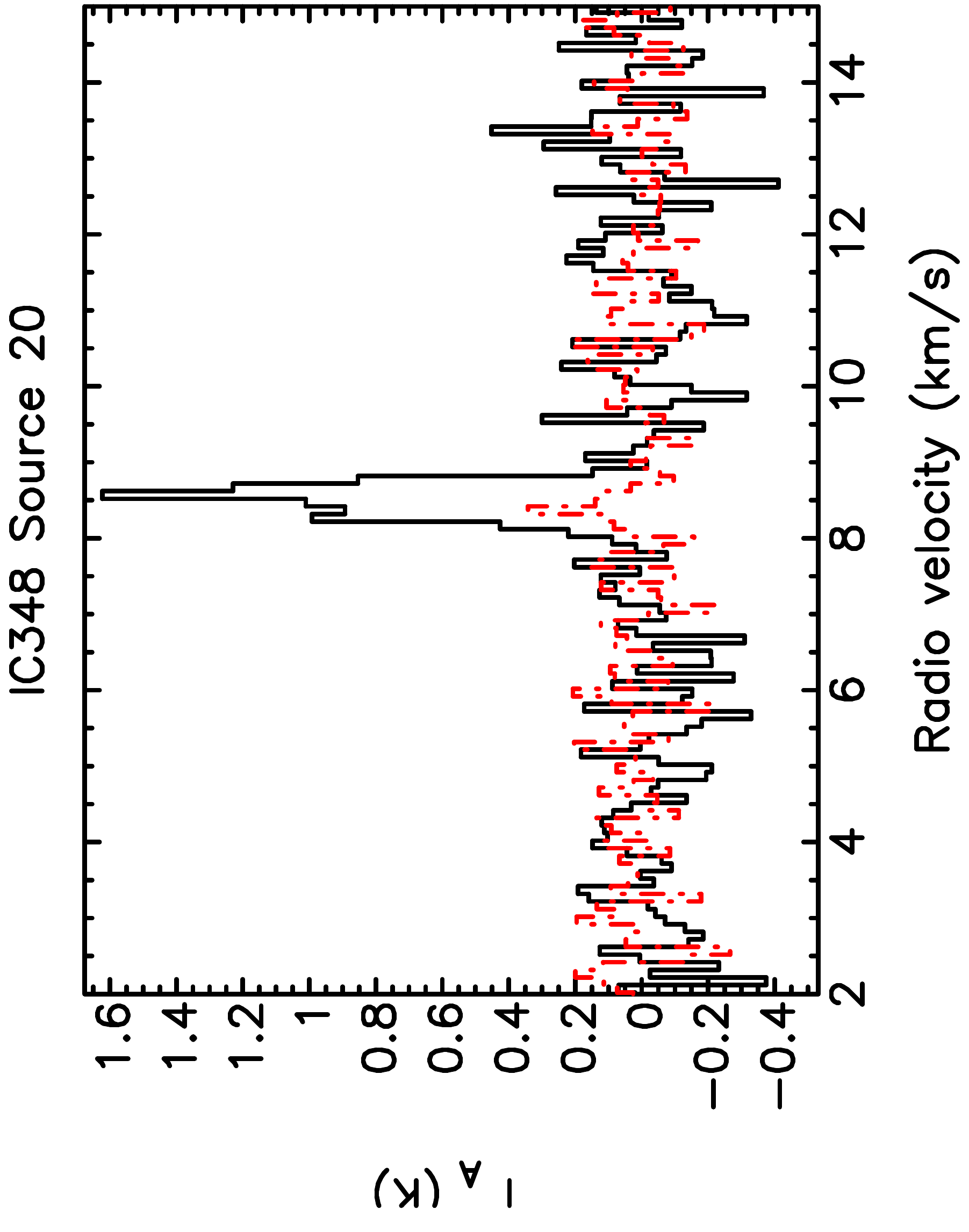}{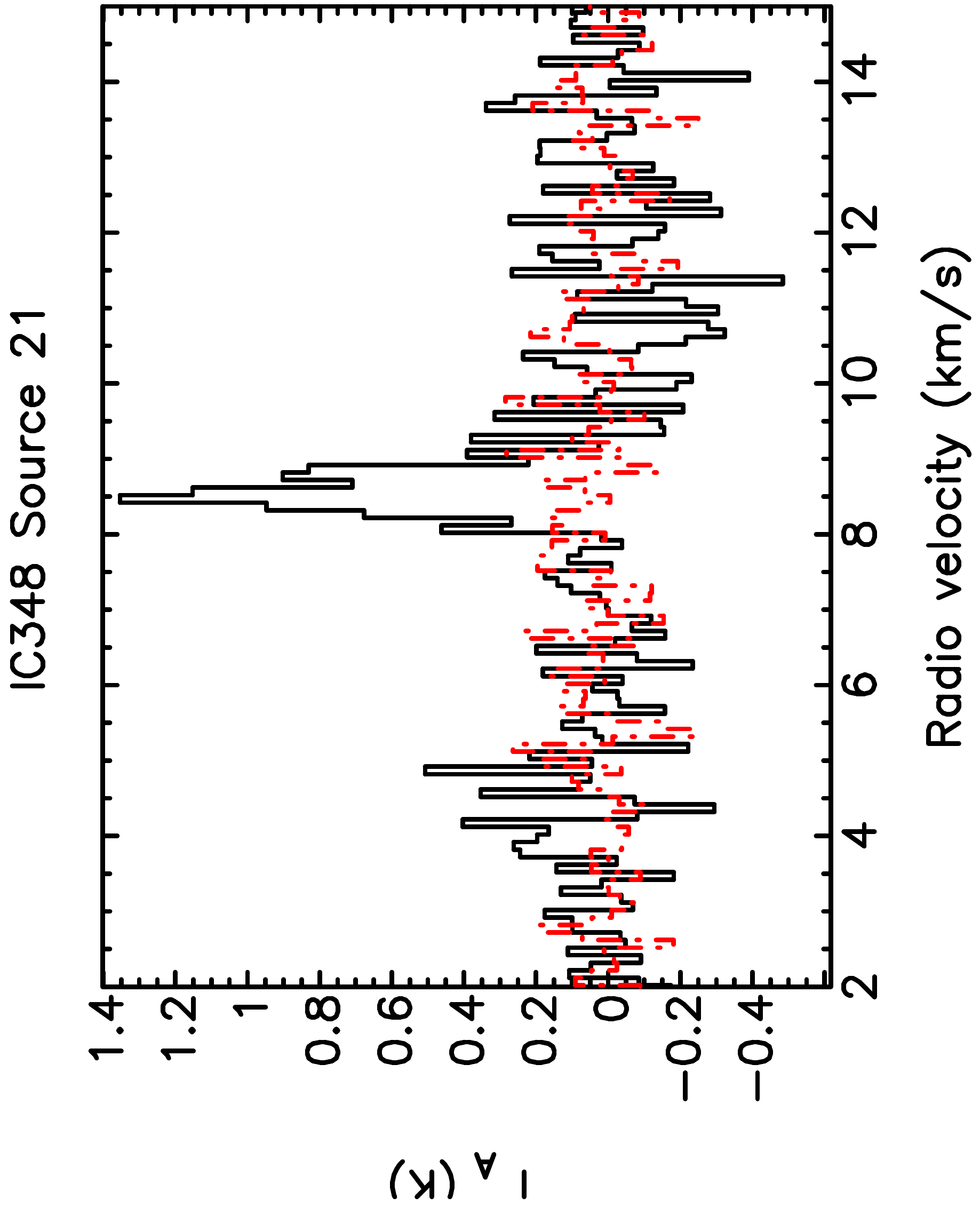}{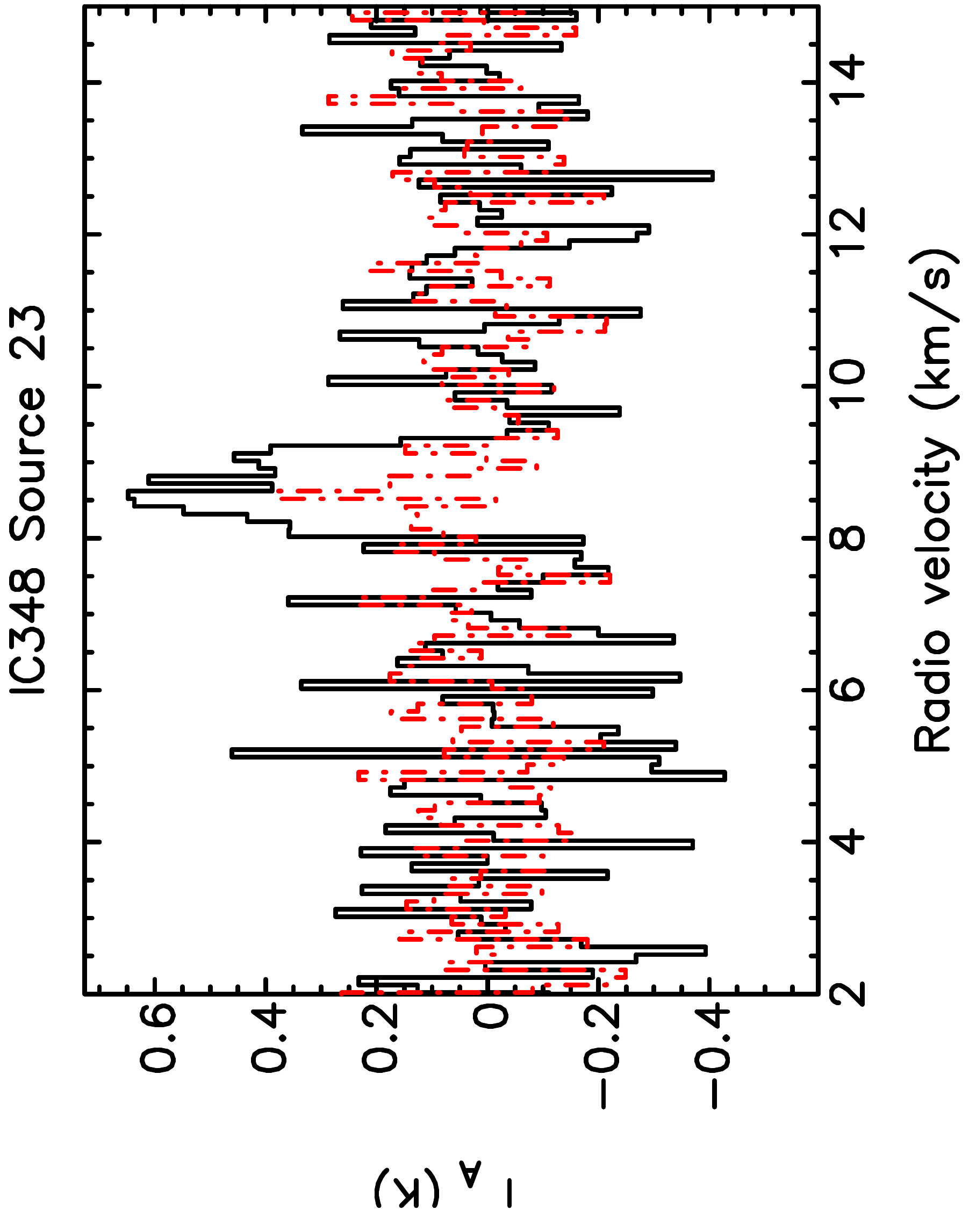}
\spectra{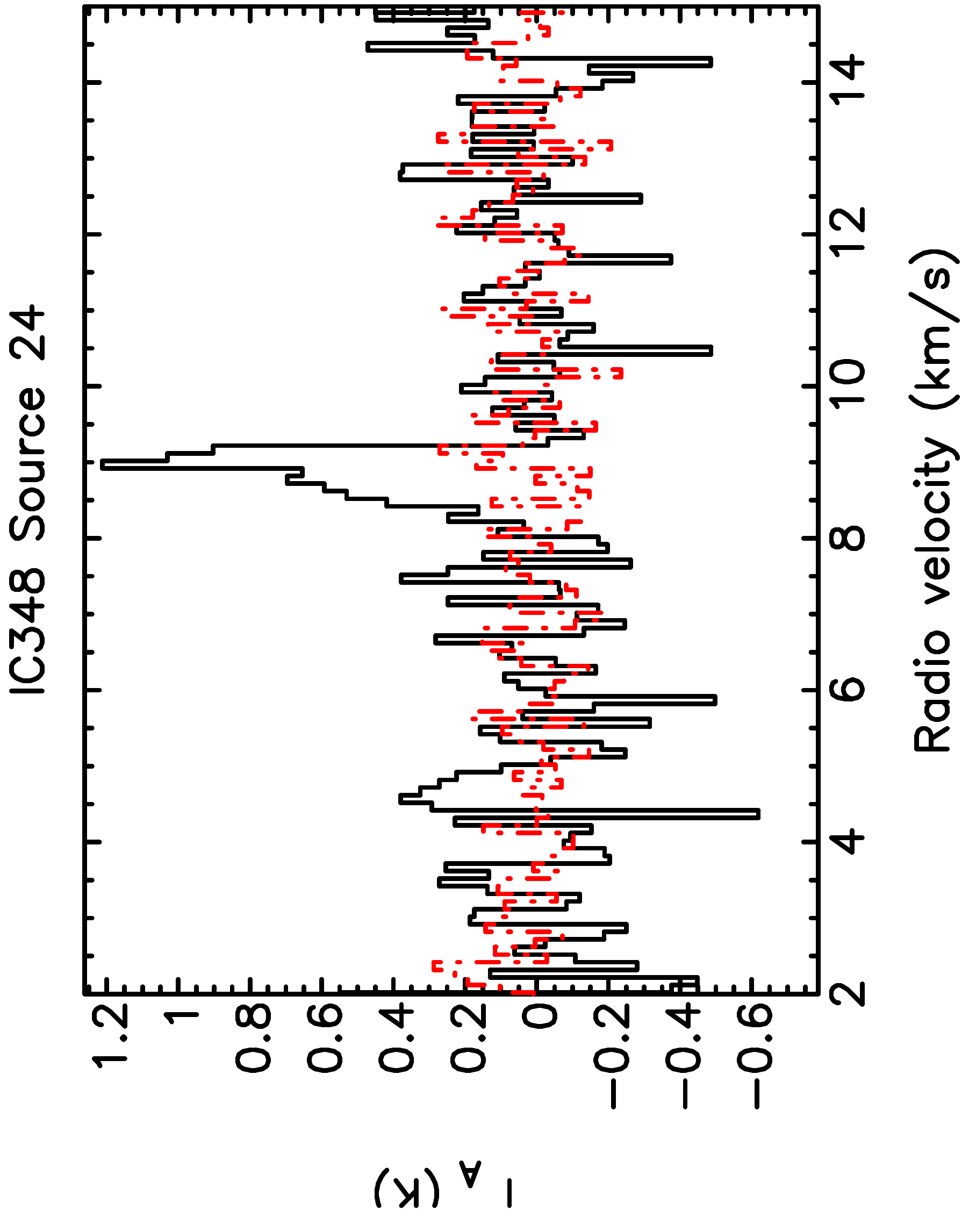}{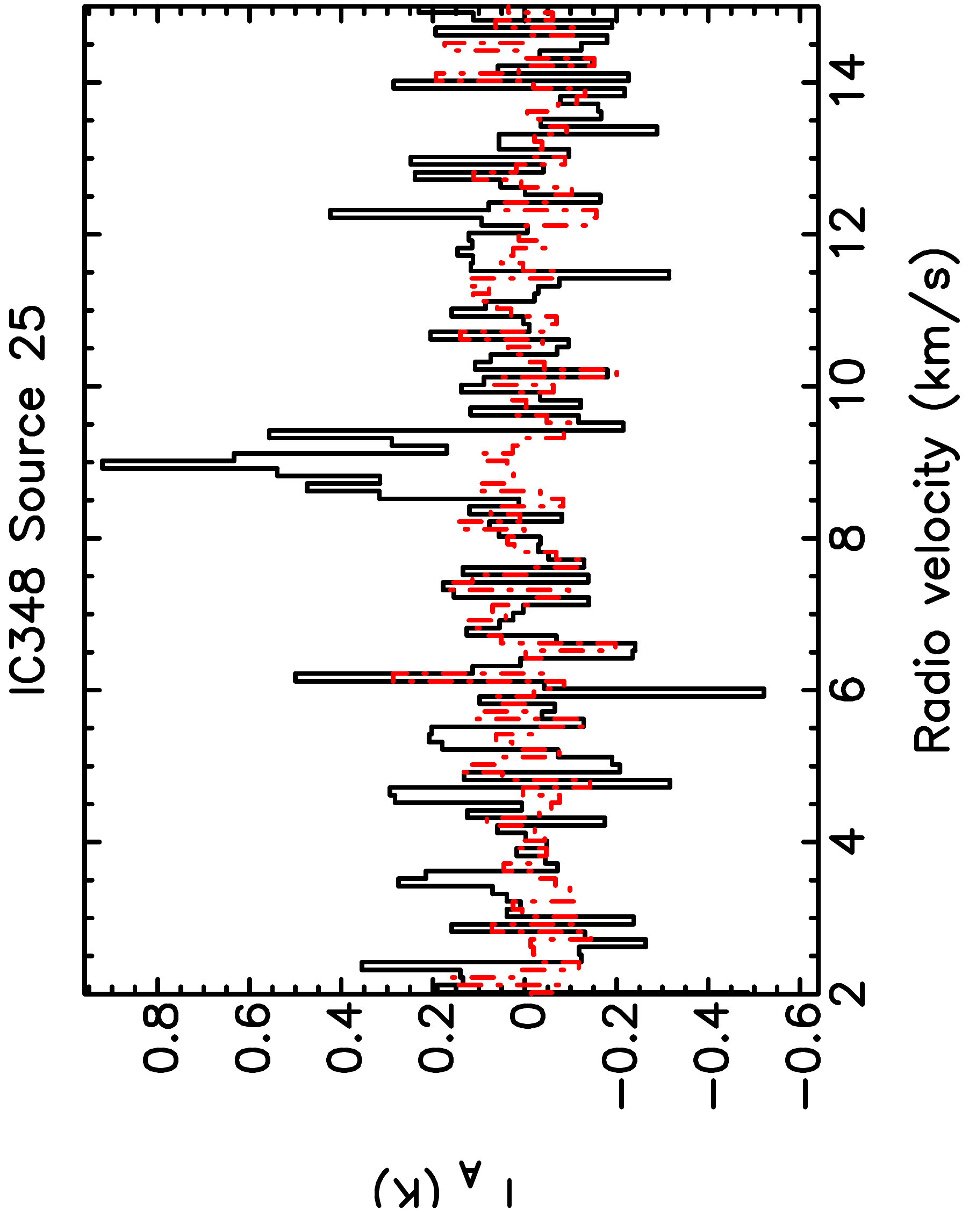}{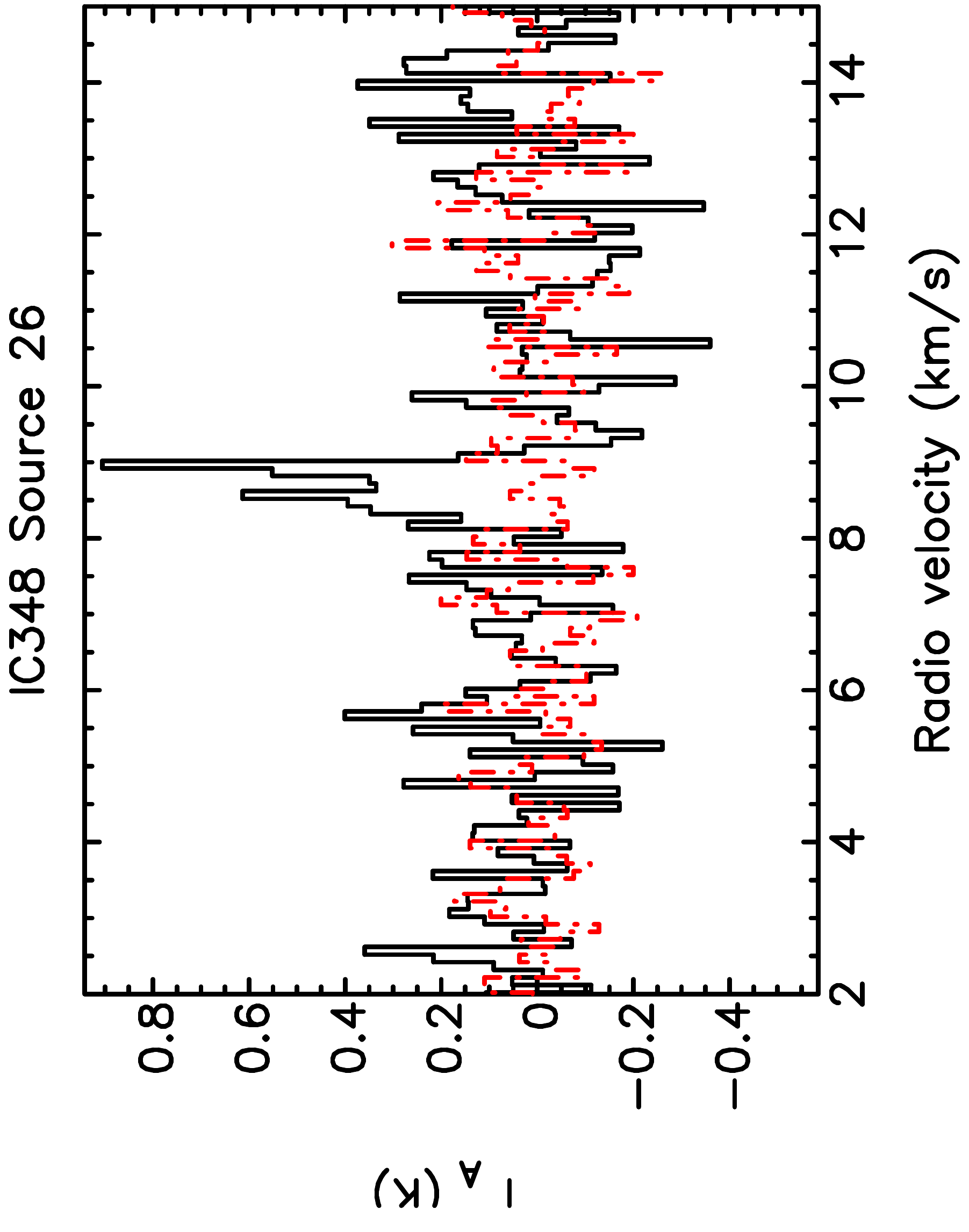}
\spectra{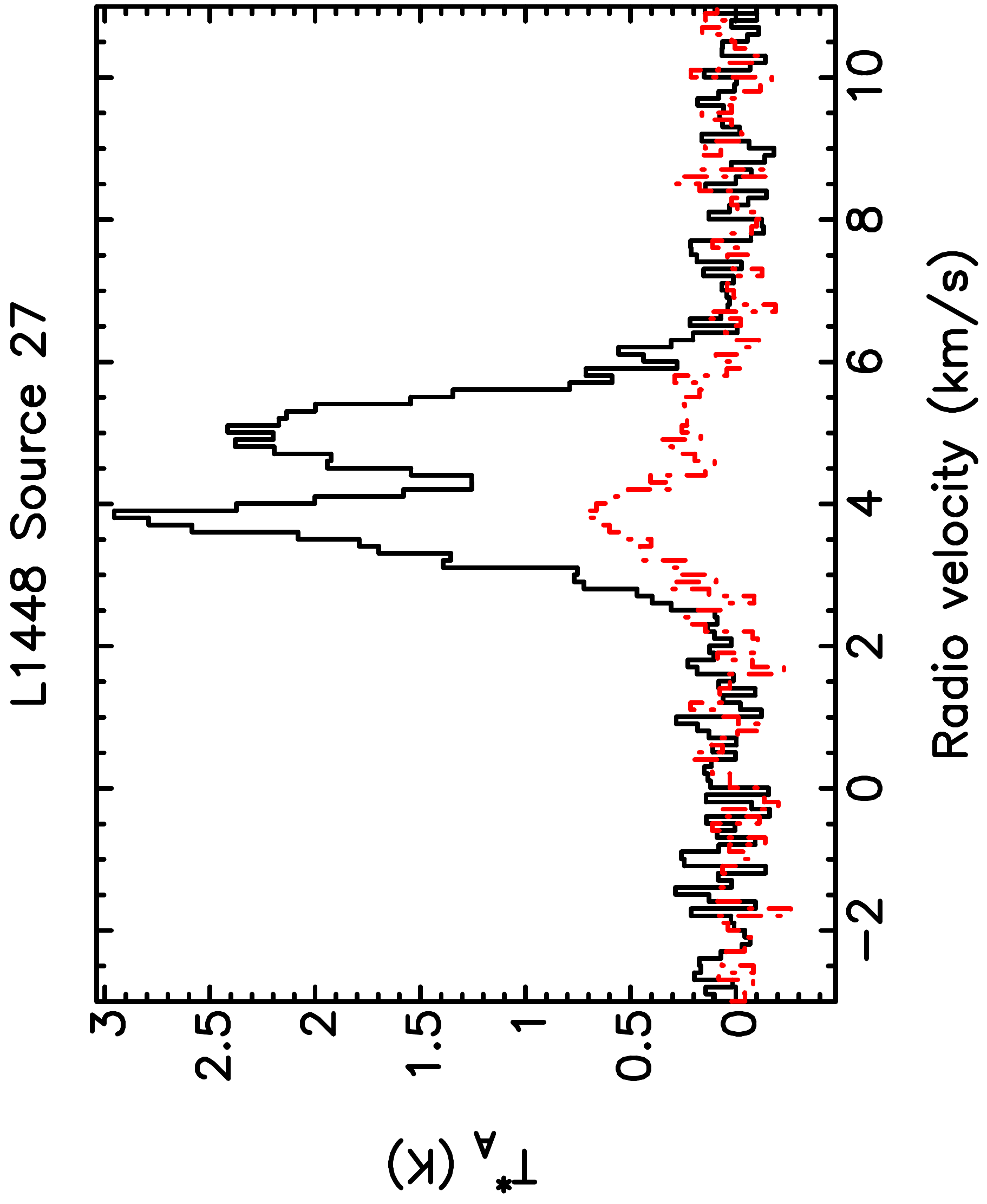}{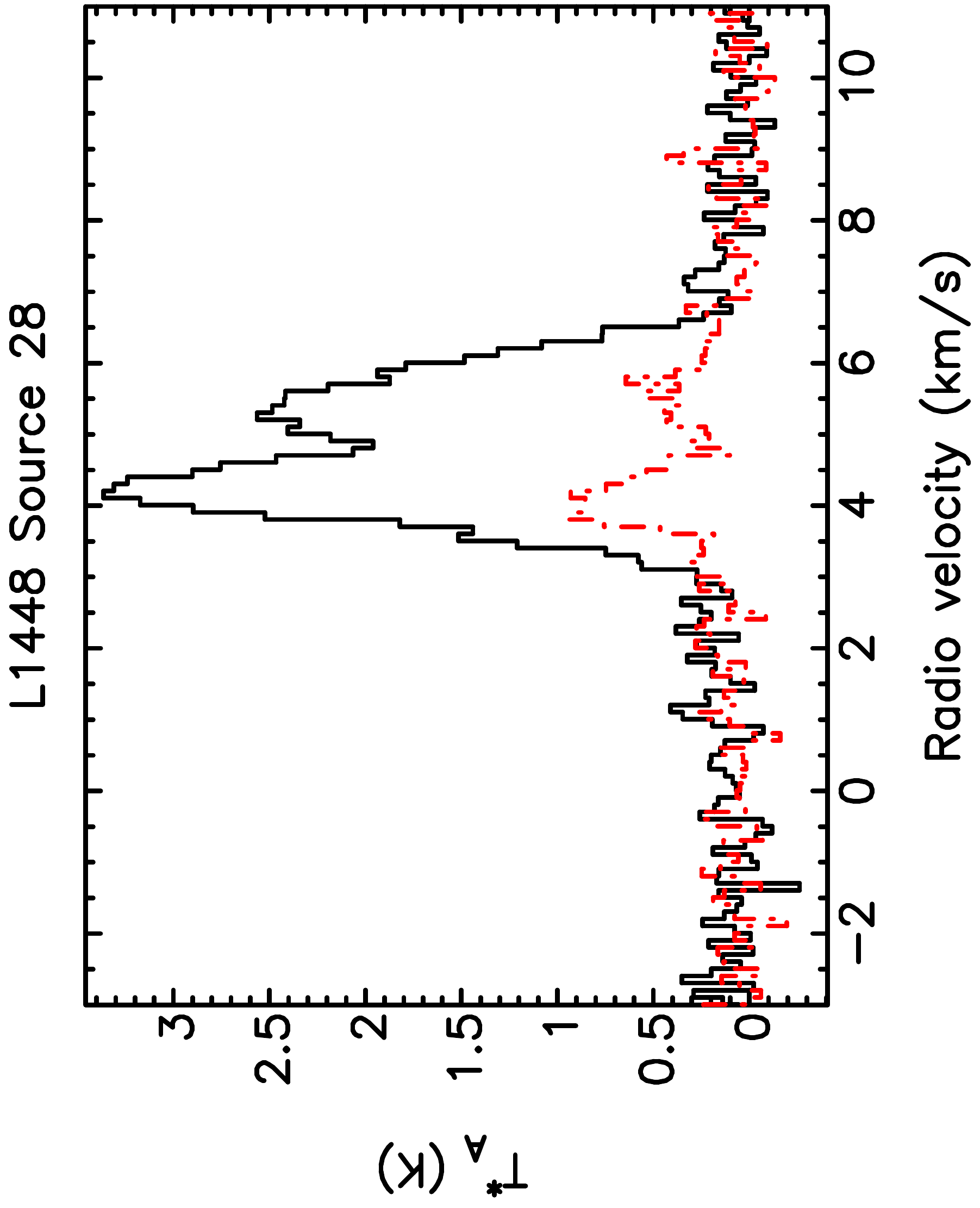}{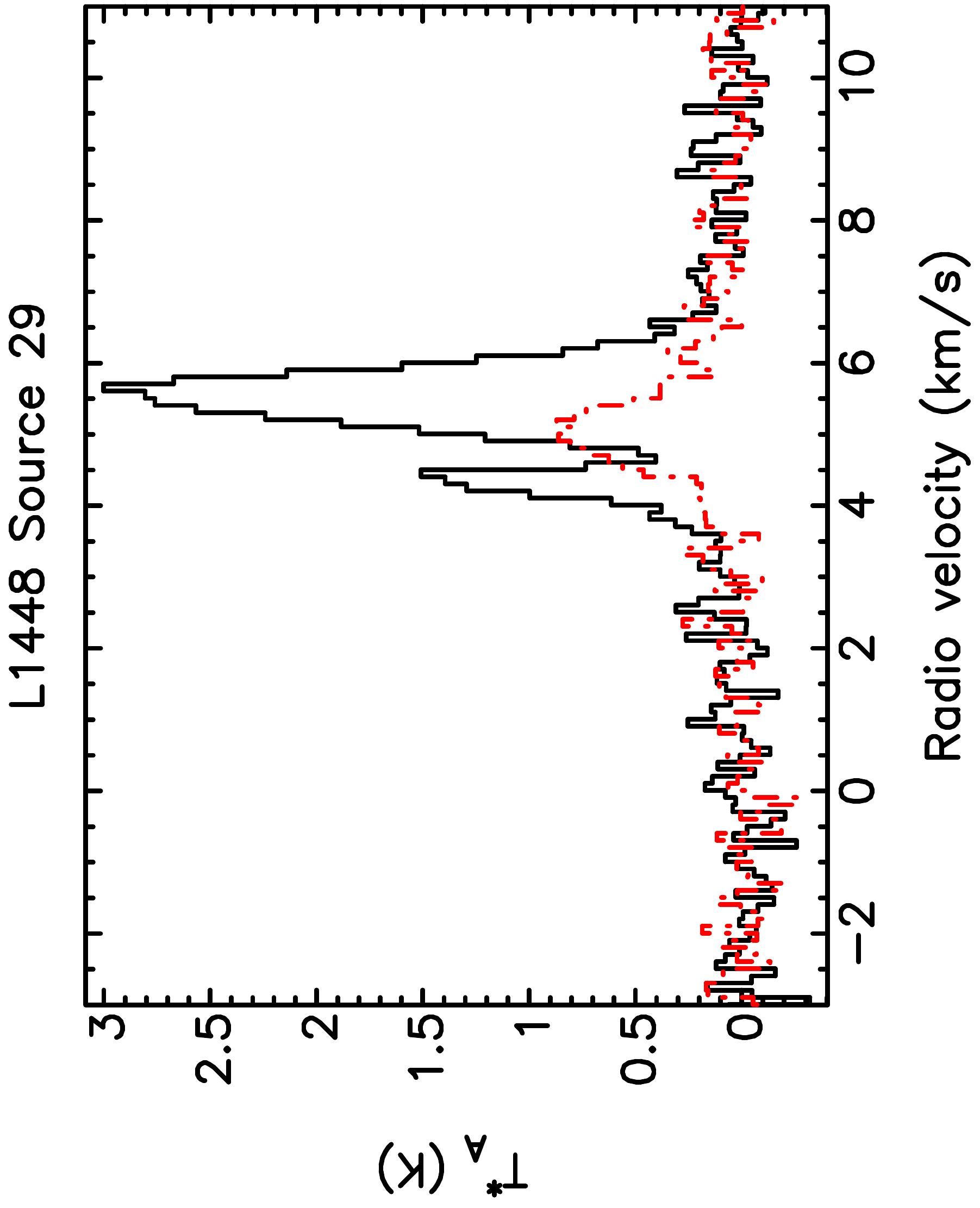}
\spectra{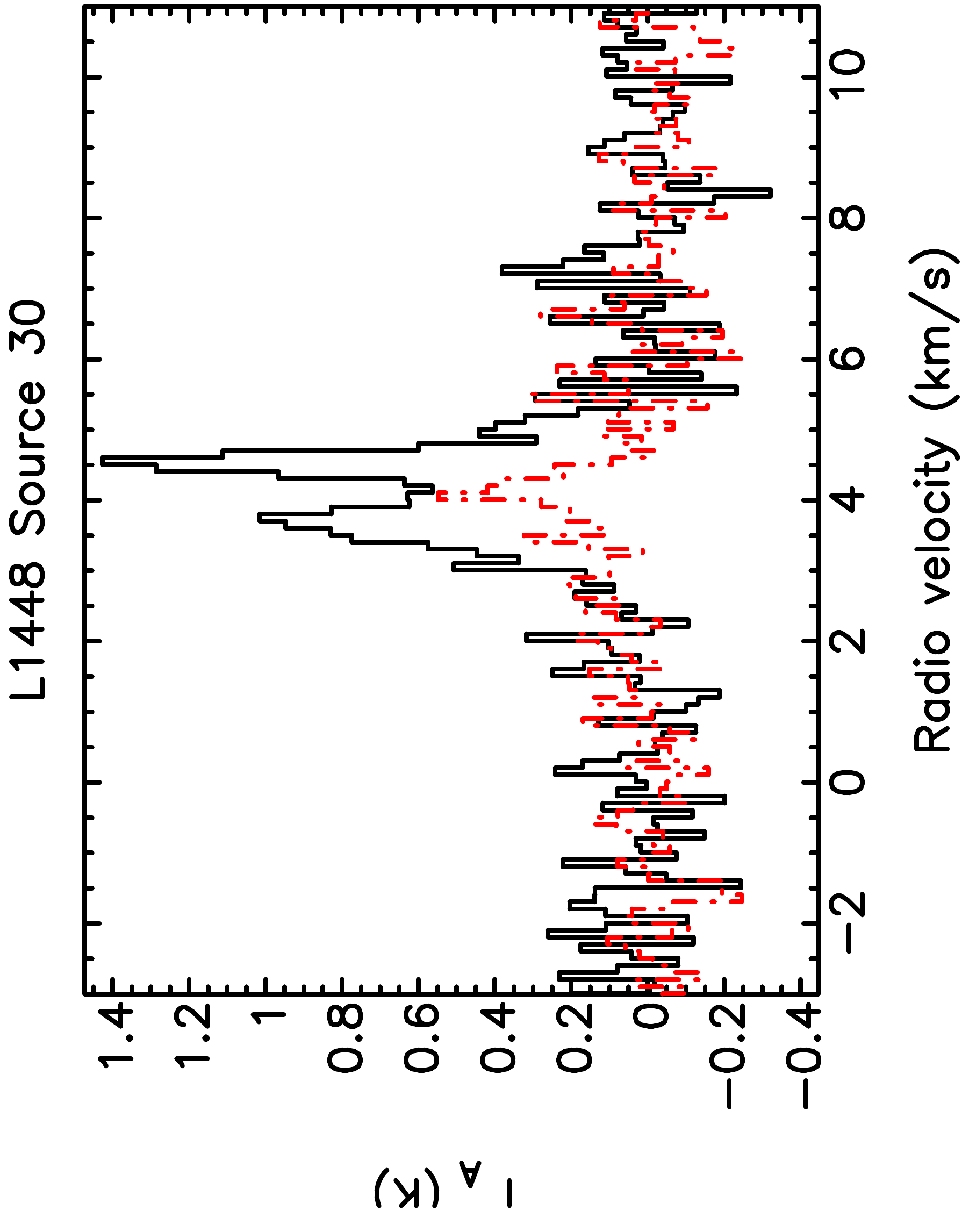}{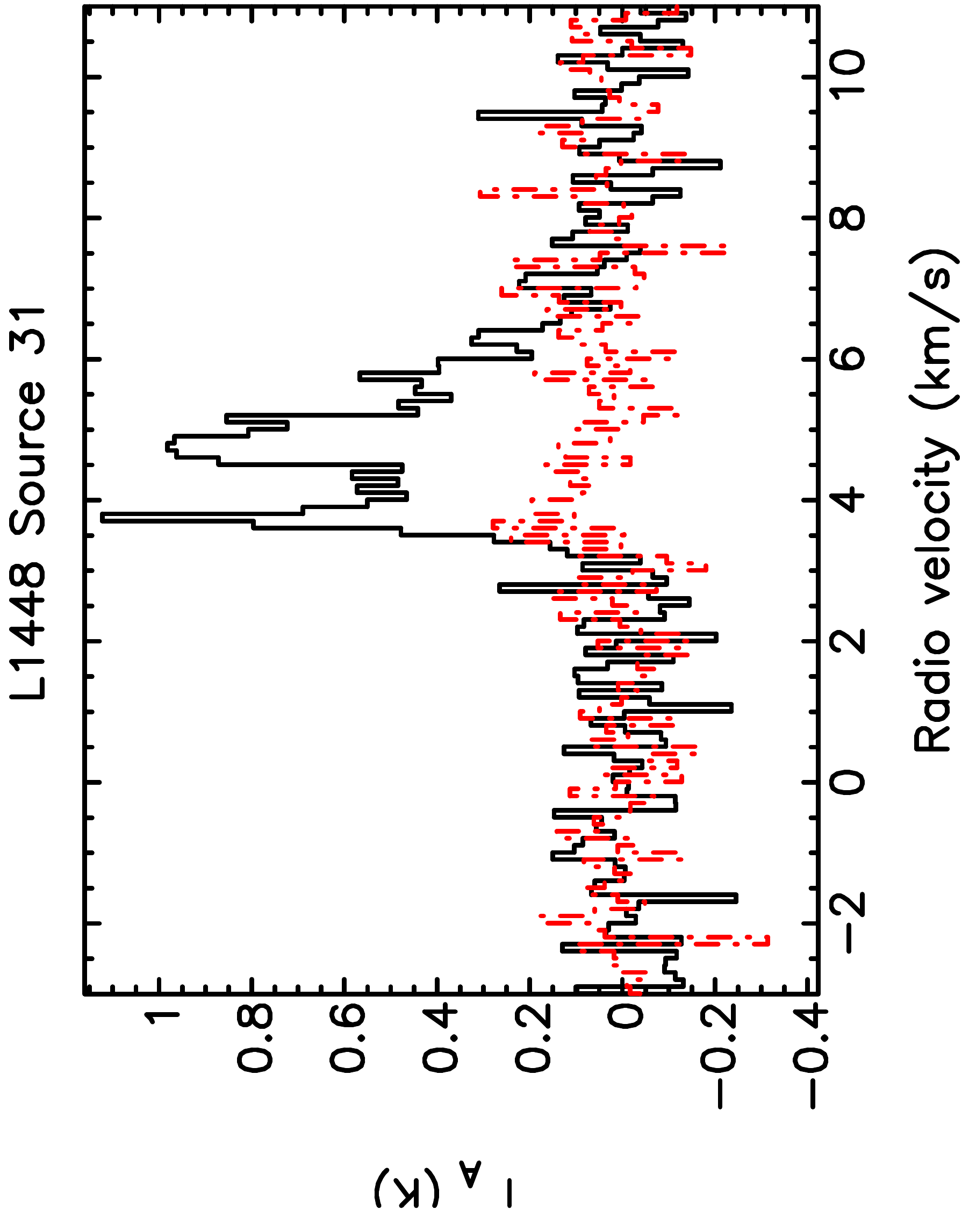}{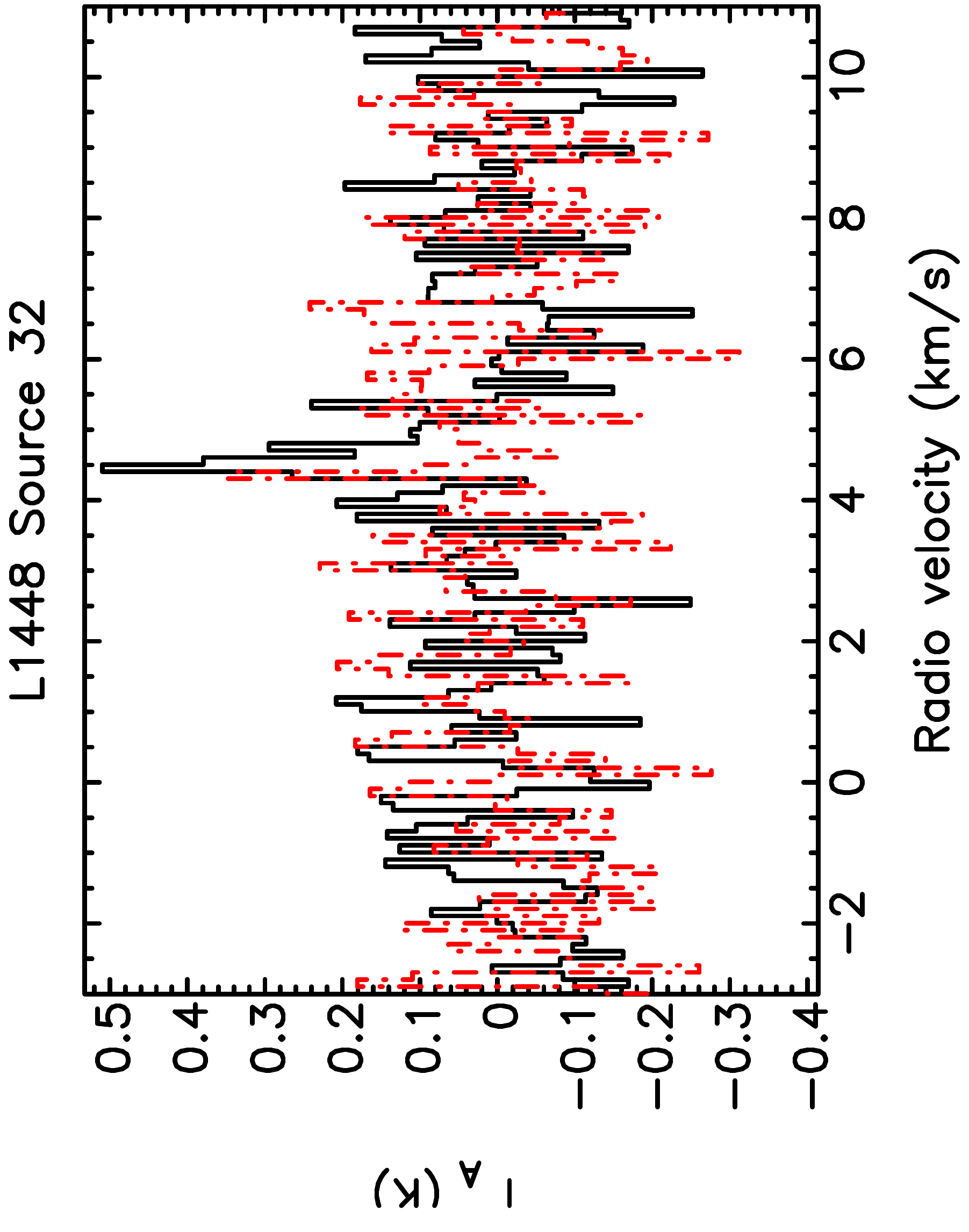}
\spectra{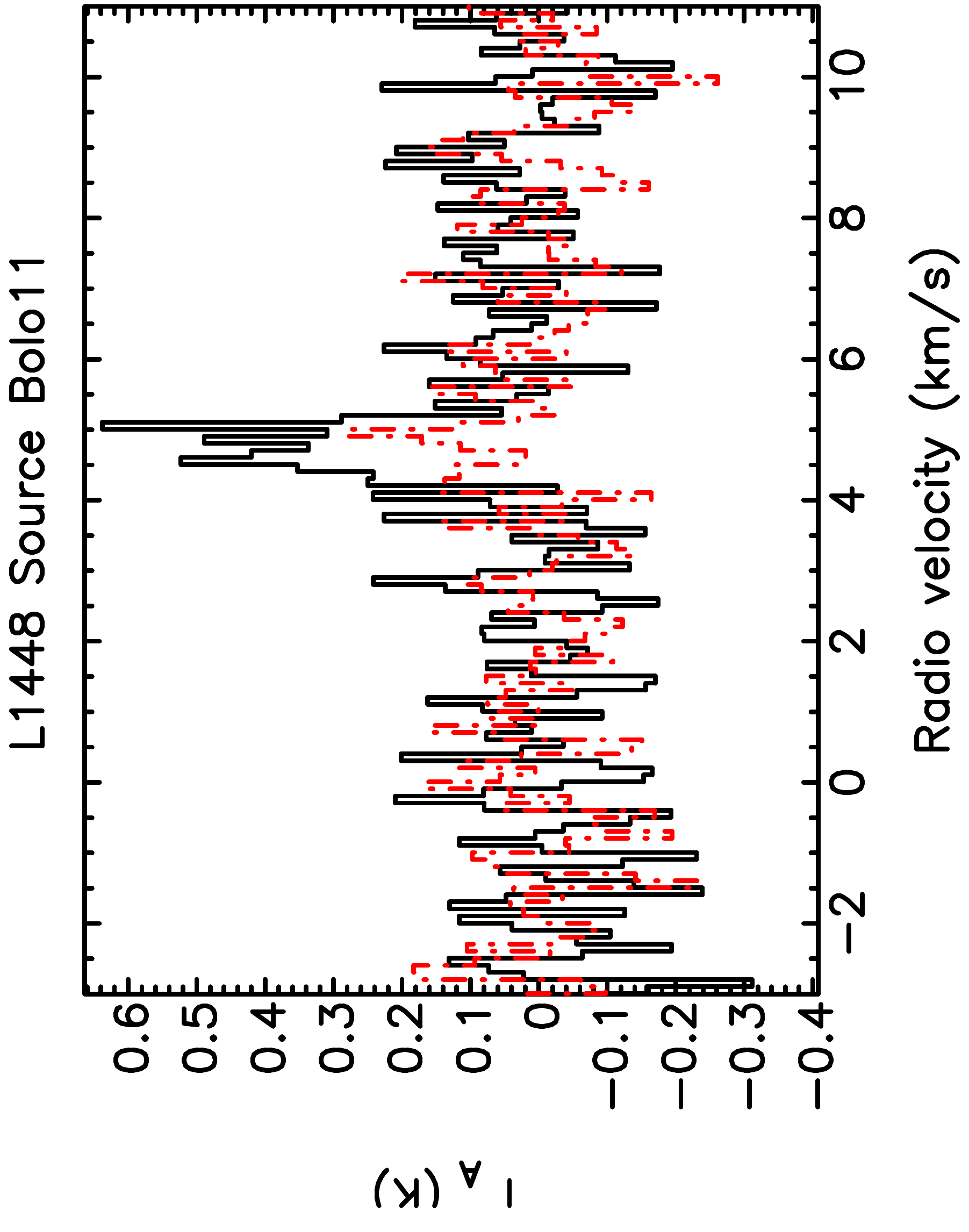}{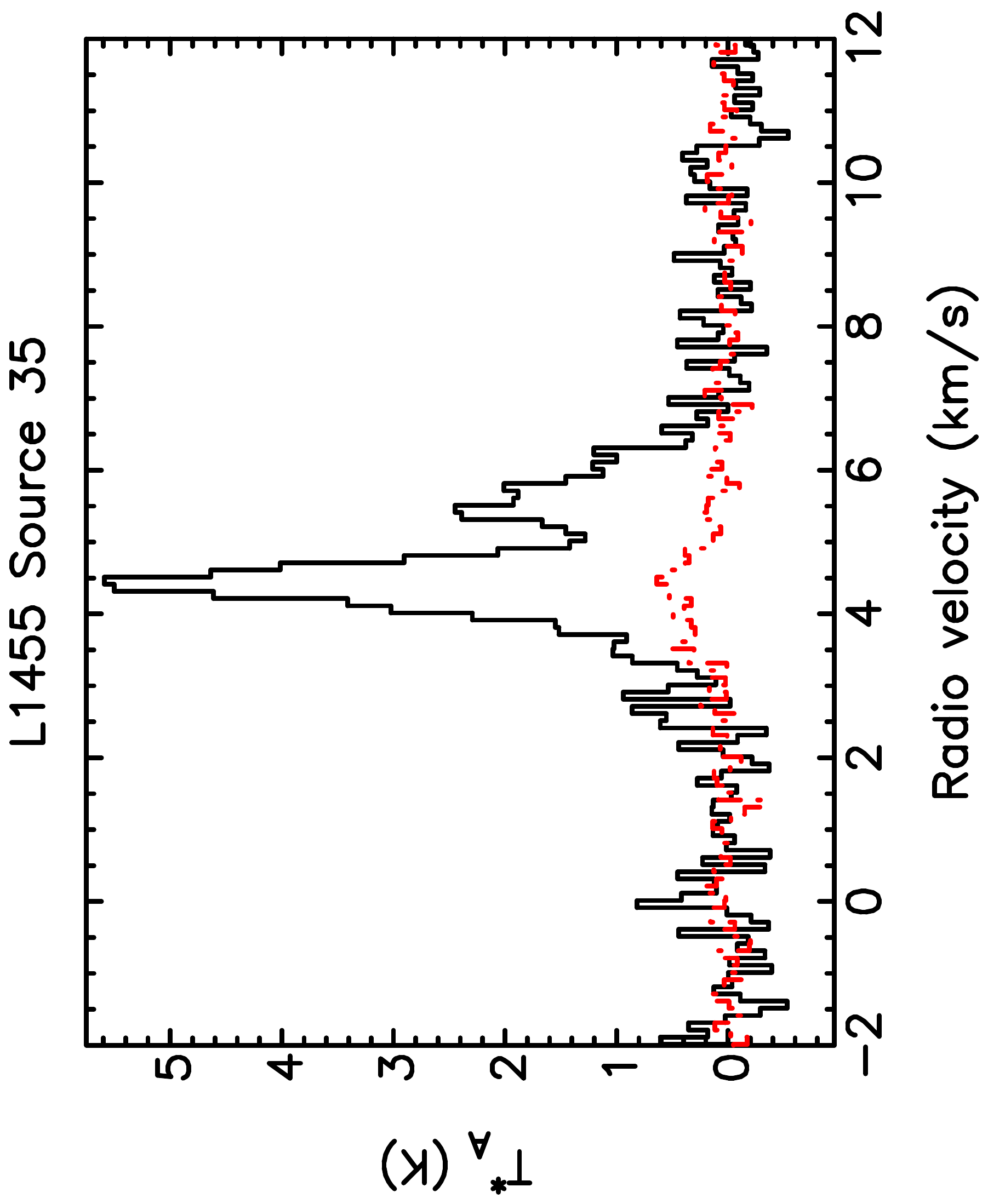}{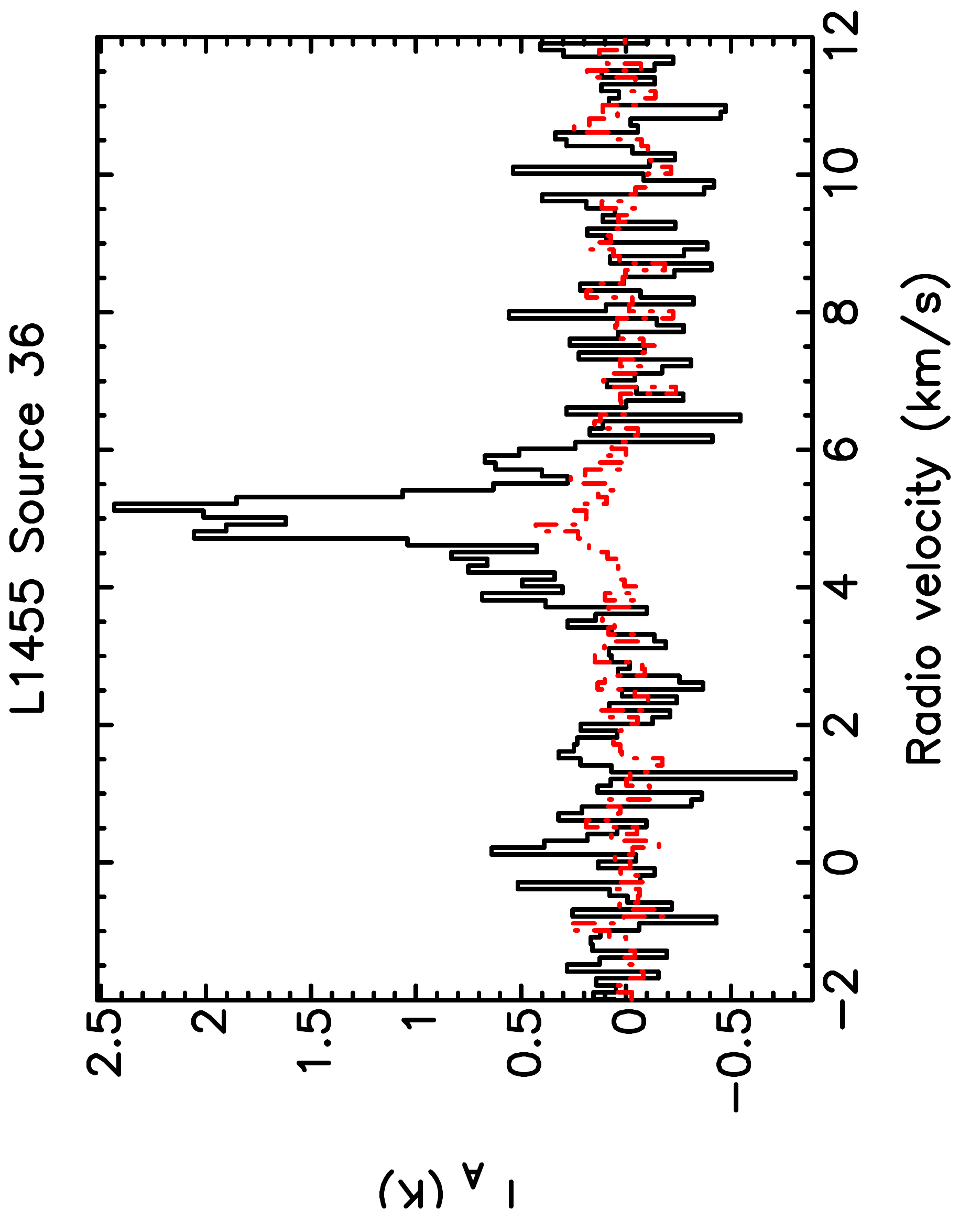}
\spectra{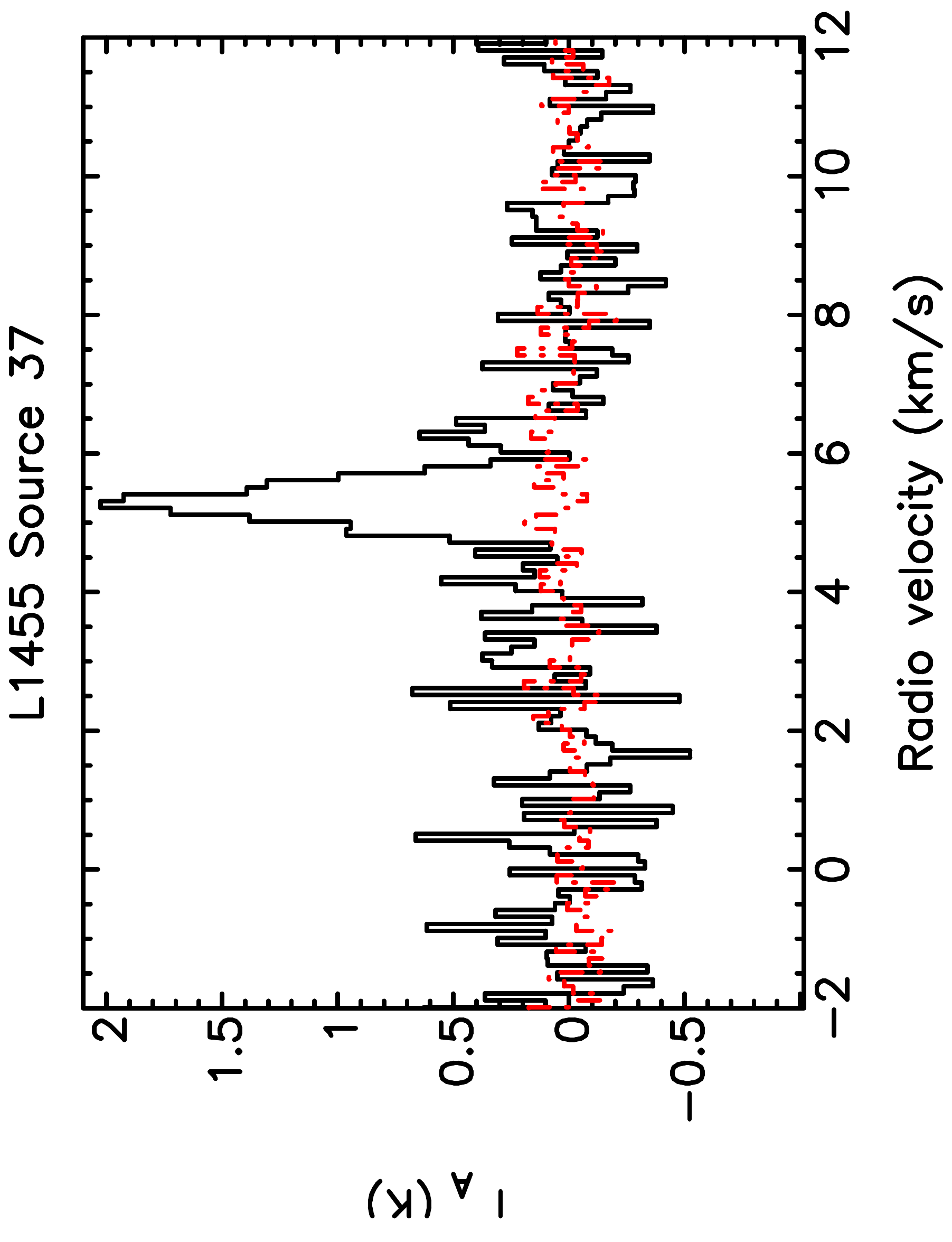}{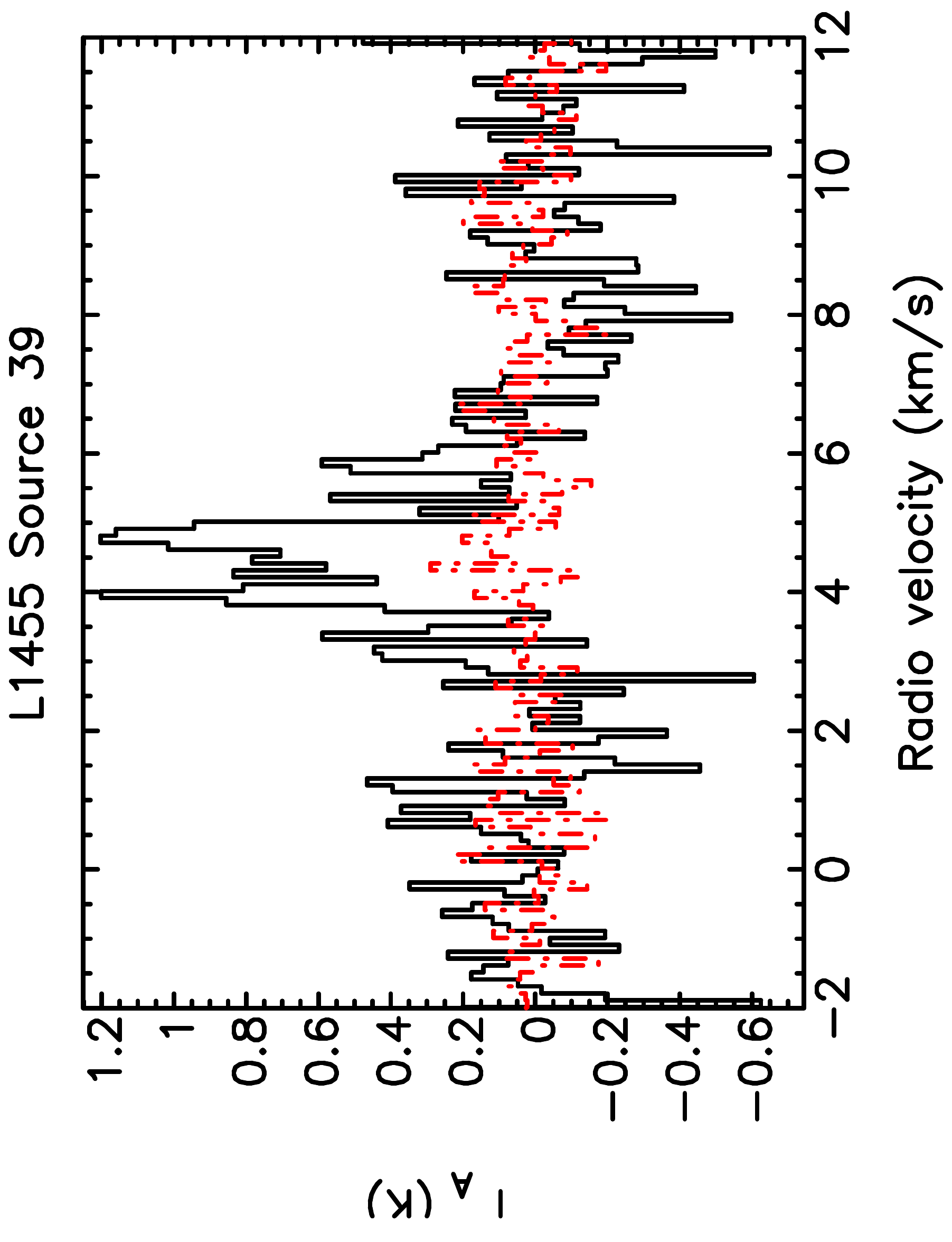}{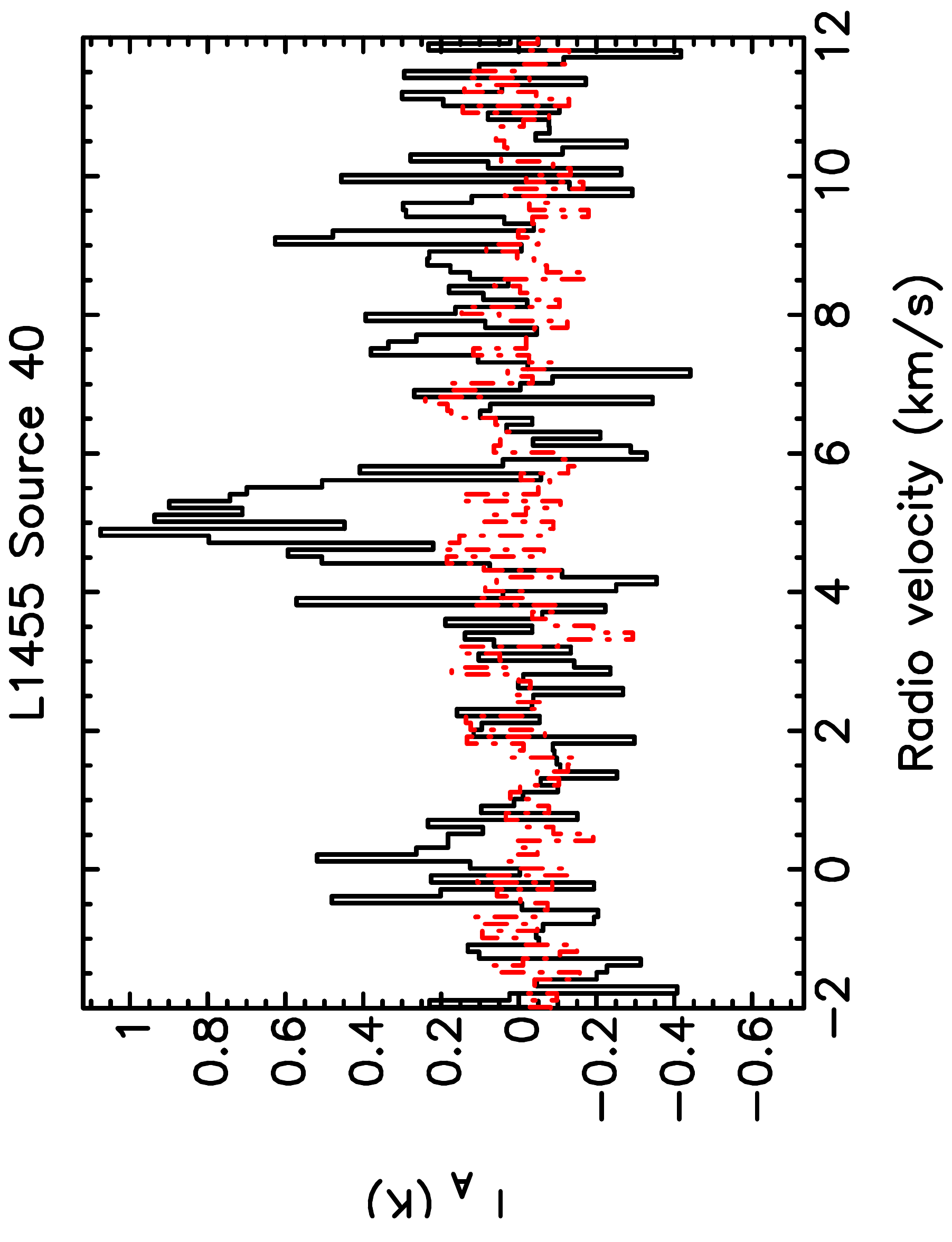}

\label{lastpage}

\end{document}